\documentclass[aps,prl,twocolumn,groupedaddress,showpacs]{revtex4}%fleqn to left %revtex4

\usepackage{amsfonts} % needed for bold Greek, Fraktur, and blackboard bold
\usepackage{amssymb}
\usepackage{braket}
\usepackage{graphicx} % needed for figures
\usepackage{cancel}

 \usepackage{lstdoc,longtable}
 \usepackage{tabularx}

\usepackage{stix} %Problem był z tym pakitem w CJP

\usepackage{amsmath}  % needed for \tfrac, \bmatrix, etc.
\usepackage{rotating} % for \rotatebox

\begin{document}

\title{The photon position operator (non-commuting) and its string-like eigenstates}

\author{Grzegorz M. Koczan}%\footnote{grzegorz\_koczan1@sggw.edu.pl}}

\affiliation{Warsaw University of Life Sciences (WULS-SGGW), Poland \\ grzegorz\_koczan1@sggw.edu.pl}

\date{\today}

\begin{abstract}
The paper provides three main definitions of the Cartesian photon position operator based on: the boost generator, the transversality condition, and the helicity operator.
In each case, the correctness of the definition and Hermitianness of the operator in the domain of physical states are proven. All considered definitions lead to the same form of the Cartesian position operator in the domain of physical states. Radial photon position operators were also defined, but they turned out to be non-equivalent. Nevertheless, the two most useful radial operators turned out to be twin counterparts in the sense of the transformation to the helical representation, which is an alternative positional representation. The components of the photon position operator do not commute, but its eigenstates do exist in analogy to the problem of eigenangular momentum.

The eigenstates of the two components of the position operator (including the radial component) are called photon strings. Photon strings on a straight line and a circle have been studied. The usual, previously known photon string states were named electric strings, but new magnetic strings were also introduced. Exact helical photon strings on a straight line and hybrid helical photon strings on a circle were also considered. On the other hand, infinitely short photon magnetic strings turned out to be flat photon vortices on planes with a normalisation formula that looks like point-localised particles.

\


\end{abstract}


\maketitle % title page is now complete

%\tableofcontents

\section{1. Introduction}
\subsection{1.1. Historical Overview of the Photon Position Operator}
The position operator in quantum mechanics was proposed in 1925, when Schr\"odinger derived his famous equation published in 1926. At a similar time, the position-momentum noncommutativity relationship and the resulting Heisenberg uncertainty principle were discovered.

However, the case of the photon (quantum of electromagnetic field) created significant problems in providing a probabilistic interpretation for the wave function and in providing the position operator. One of the first results was the definition of the wave function by Landau--Peierls in 1930 \cite{Landau}. This function was non-locally dependent on the electromagnetic field. In 1935, Born and Infeld \cite{Born--Infeld} proposed an energy-mass center operator that could be used to calculate the photon position operator. This calculation was made by Pryce in 1948 \cite{Pryce}. The Born--Infeld--Pryce operator for the photon had noncommutative components, so he was treated with caution. To make matters worse, Newton and Wigner wrote the paper \cite{Wigner} in 1949, which showed that there are no strictly localised states for photons. The work was mathematically rigorous, but it did not establish the non-existence of the photon position operator. 

In 1967, the approach to particle localisation was generalised to the so-called weakly localizable ones by Jauch and Piron \cite{Jauch Piron}. The notion of weakly localizable is based on a measure of the area of space and not explicitly on a position operator.
The further development of the issue was made more clearly by Bacry \cite{Bacry 2} in the title of Chapter~4: {\it Localizability. The photon scandal. Quantization helpless!} As a consequence, Bacry proposed a quite simple photon position operator \cite{Bacry 1988, Bacry 2}. Due to the equivalence in representational precision to Pryce's calculations for photons, this result is here called the Pryce--Bacry operator. This form of the photon position operator is confirmed, e. g. in Mourad's 1993 paper \cite{Mourad}. However, the existence of single-photon states strictly localised in a finite volume was confirmed in \cite{Skaar}, without using the tool of the position operator. Using the Schr\"odinger-type equation in the preprint \cite{Kraisler}, the evolution of localised photon states was studied. In contrast, in the work of \cite{Beige} the localisation of multiphoton packets was investigated.

The history and context of the photon position operator are long, turbulent, unclear, and above all, still alive. In fact, the problem concerns not only the photon position operator, but also other particles in the relativistic description. This is well reflected in Jancewicz's doctoral thesis \cite{Jancewicz 1974} from 1974, the main results of which were published in \cite{Jancewicz 1975} (can be compared with independently written master's thesis Koczan 2002 \cite{Koczan 2002}). Jancewicz considered three different position operators, two of which he took from Fleming \cite{Fleming}. The first was the energy-mass center operator. The second was original and called the center of inertia operator, but may have been the rest mass (see \cite{Koczan 2021,Germano}). The third was called the center of spin or the local position operator. Only the third position operator had commuting components -- if it existed -- and for the photon, only the first position operator existed.

The problem of the general position operator is also described in contemporary works by Lev from 2015 \cite{Lev} and by Zou et al. from 2020 \cite{Zou}. The description of the position operator is undertaken in the latest preprint from 2024 in the matrix representation \cite{Song}. Paradoxically, the problem of the photon position operator may be simpler than that of other particles, because it is reduced by the parameter of zero mass, and at the same time, it has the generality of non-zero spin. But the problem of photon localisation is still complex and multifaceted, as described in detail in the monograph \cite{Keller}.

The controversy over the existence (or nonexistence) and the controversy over the noncommutativity (or commutativity) of the photon position operator will be discussed after the subsection showing the experimental and observational motivations for this notion. If we accept the wave-particle duality of light and photons, then we should consistently require the existence of a positional description. However, the properties of this description should result from the nature of the object being studied, and not from \textit{ad hoc} assumptions.

\subsection{1.2. Experimental and Observational Motivations}

Attempts to theoretically deny the existence of the photon position operator or the existence of well-localised states must be confronted with experimental arguments and observations. Since the photoelectric effect and its explanation by Einstein in 1905, we know that the photon has a corpuscular nature \cite{AdP}. It was for this explanation that Einstein received the Nobel Prize in 1921. It is therefore worth quoting his words from page 133 of this key work \cite{AdP}: {\it According to this picture, the energy of a light wave emitted from a point source is not spread continuously over ever larger volumes, but consists of a finite number of \textbf{energy quanta that are spatially localized at points} of space, move without dividing and are absorbed or generated only as a whole.}

\begin{figure}[h!]
\centering
\includegraphics[width=8.5cm]{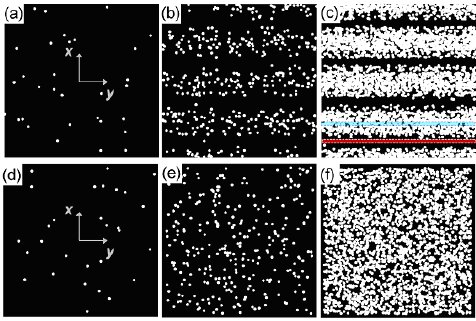}
\caption{The interference pattern produced by a single-photon source with (a) 30, (b) 300, and (c) 3000 photons registered. In contrast, the decoherent distribution of (d) 30, (e) 300, and (f) 3000 photons lacks the dark fringes. The illustration and description come from the publication \cite{Afshar}}
\label{single photon}
\end{figure}

Even Newton had already postulated the corpuscular nature of light and had a dispute with Huygens, who argued for a wave nature. Today, Newton's corpuscular theory of light and the refraction of light are considered incorrect. In fact, bullets falling into water refract differently than light. However, based on modern knowledge, it is possible (and could not be otherwise) to improve Newton's corpuscular theory into a correct theory. Buenker made such at least partially successful attempts \cite{Corpuscular 2015, Corpuscular 2022}.

During university physics course lectures, an experiment is presented that simultaneously demonstrates the particle and wave nature of a single photon. The author observed such an experiment around 2000 at the Faculty of Physics, University of Warsaw. A similar experiment conducted at Brown University can be viewed online (https://www.youtube.com/watch?v=\_MpvDAQrKbs). This is a version of the famous double-slit experiment. Contrary to appearances, this experiment does not violate the complementarity principle of particle-wave duality, which consists in the impossibility of determining which slit a photon passes through if we want to preserve the interference pattern -- like in Fig.\ref{single photon} (a),(b),(c). Simply put, if we covered one slit, the interference pattern would disappear -- like on the Fig.\ref{single photon} (d),(e),(f). 

Unfortunately, the author of the publication \cite{Afshar} and photos in Fig.\ref{single photon} claims that the principle of complementarity was violated in his experiment. This is difficult to explain, but perhaps the experiment observed single-slit diffraction rather than double-slit interference. Most important here is that single photons were found in specific positions with a particular probability distribution (Fig.\ref{single photon}).

\begin{figure}[h!]
\centering
\includegraphics[width=8.5cm]{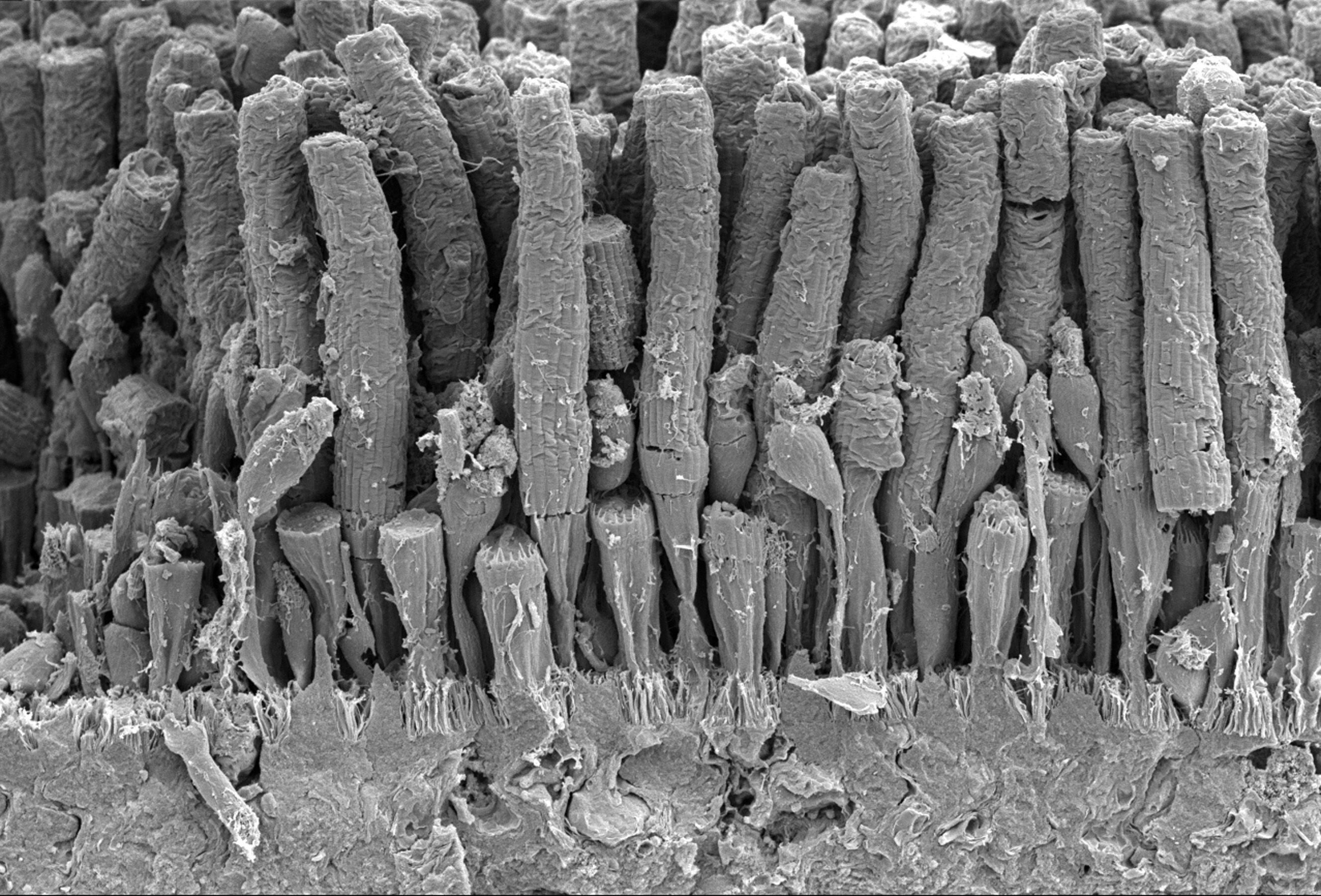}
\caption{Scanning electron micrograph of an amphibian eye retina's rods (and cones) structure. Rods about 1~\textmu m in diameter and 20~\textmu m long can absorb single photons of weak light. The photocurrents of similar natural photocells (rods) were studied in the work \cite{Pugh}. The illustration was purchased from Alamy}
\label{rods}
\end{figure}

At the most minor scale, this corpuscular nature manifests in the hydrogen atom's absorption of photons. We are talking about sizes on the order of 0.1 nm and smaller. One could say that small atoms catch very large delocalised photons, which is not convincing. Let us note that already on a scale of 1000 times larger (in the sense of micrometers and nanometers), there are biological structures capable of absorbing single photons.

The best example is the rods in the eye's retina (Fig.~\ref{rods}). In the direction perpendicular to the light capture, the rods have a diameter of the order of 1~\textmu m, so it is difficult to deny the possibility of localising photons of this order (apart from the parallel direction order 20~\textmu m). 

In the context of the position operator, we should also consider the optical/wave resolution of the image obtained using a specific type of waves: electromagnetic (visible light, radio range, X-rays), electron matter waves and even ultrasonic waves. The most popular is the following fairly general formula for "optical" resolution:
\begin{equation}
\label{resolution}
r_d=\frac{0.61 \lambda}{n \sin \theta},
\end{equation}
where: $r_d$ -- resolution distance, $\lambda$ -- wavelength, $n$ - refractive index of the medium, $\theta$ -- half cone angle of the observed wave. Although the above formula has a wave nature, the image formation is discrete, based either on pixels or quanta (photons). There is no contradiction in this, but it expresses the complementarity of the particle-wave duality, already discussed in the context of Fig. \ref{single photon}. Even for ultrasound, there is a kind of quantum-wave duality related to the existence of phonons in solid-state physics.

Based on formula 1, we can approximately assume that the resolution of optical microscopes is of the order of the wavelength of green light (500 nm), i.e. about 0.5 ~\textmu m. 

\begin{figure}[h!]
\centering
\includegraphics[width=8.5cm]{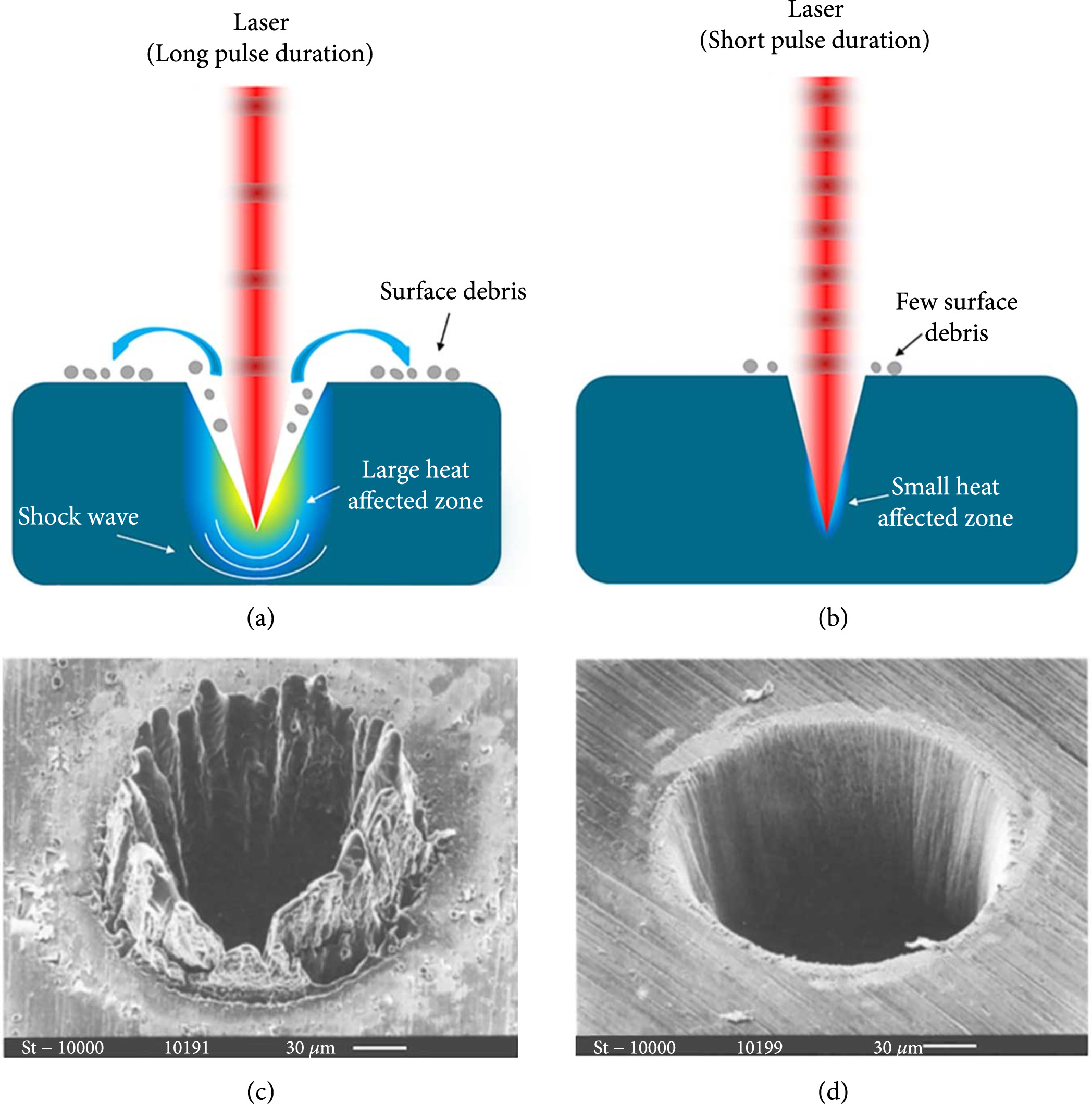}
\caption{Interaction of a 780 nm laser with 100 ~\textmu m thick steel foil under different pulse durations: (a) scheme for long pulse duration, (b) scheme for short pulse duration, (c)  electron micrograph of the ablation hole at nanosecond laser pulses of 3.3 ns, (d) electron micrograph of the ablation hole at femtosecond laser pulses of  200 fs (scale bar 30 ~\textmu m). Graphics (a), (b) are from \cite{Femto 2021}, and original micrographs (c), (d) are from \cite{Femto 1996}}
\label{Lasers}
\end{figure}

Other than detection and microscopic observations, experimental evidence of photon localisation is the precise operation of femtosecond lasers (see Fig. \ref{Lasers}). It can be easily estimated that the laser pulse in Fig. \ref{Lasers}(b),(d) took the shape of a slice with a diameter of 180 ~\textmu m and a thickness of 60 ~\textmu m. Of course, there was a huge number of photons in such a slice, which can even be calculated based on the pulse energy and the wavelength of light. However, this does not constitute a significant limitation from the point of view of a single photon, because photons are bosons and can exist in the same state. In other words, photons after leaving the laser have almost no interaction with each other. In quantum electrodynamics (QED), the interaction of photons in a vacuum is described by a four-vertex Feynman diagram. The probability of such processes is negligible compared to more typical processes with two-vertex diagrams.

At the 49th Congress of Polish Physicists in Katowice in 2025, the author posed a question to Anne Huillier, winner of the 2023 Nobel Prize for attosecond pulses. At the Congress, the author presented, among other things, the issue of the Lorentz group \cite{Poster 49 ZFP}. The question was officially posed after the lecture and concerned the existence versus nonexistence of the photon position operator. The Nobel laureate did not deny the existence of this operator but stated that she was not a theoretician. She also suggested that the laser field is classical in nature.

This suggestion, however, may be somewhat misleading, as the essence of laser operation is based on the quantum phenomenon of stimulated emission, introduced by Einstein in 1916 \cite{Einstein 1916}. However, in the late 1980s, the idea emerged that stimulated emission could supposedly be stimulated classically \cite{Fain}. In a sense, the laser field, as a coherent state of the field, is more classical than other states, but it is still quantum. Furthermore, there is a view that Maxwell's equations written using the Riemann--Silberstein vector are the Schr\"odinger equation in which the Planck constant has been canceled out or absorbed by another quantity. The existence of the above views, within the framework of the wave-particle duality of photons, only proves that quantum issues still require interpretation and ordering.

\subsection{1.3. Photon Position Operator Controversy}

Despite this, some scientists (e.g. Jordan or Białynicki--Birula marriage) had previously consistently denied the existence of the photon position operator \cite{Jordan 1978, Birula III2012, Birula VIII2012}. This was probably due to the lack of commutation of the components of such an operator and excessive faith in Wigner's authority (see \cite{Wigner} and \cite{Jordan 1980}).

However, other researchers (Hawton, Dobrski et al. and Jadczyk) even postulate the existence of a photon position operator with commuting components \cite{Hawton 1999,Hawton 2001,Dobrski,Jadczyk 2024}. Unfortunately, such operator(s) may not have the desired properties. According to the preprint \cite{Jadczyk i S}, which disputes the work of \cite{Hawton 1999}, such a photon position operator with commuting components concerns either a triplet of spinless particles or is devoid of rotational symmetry. Despite this, the lead author of the preprint published a paper \cite{Jadczyk 2024} in which the axial symmetry of the commuting components operator is allegedly not a problem. 

A slightly different approach is presented by the authors of the paper \cite{Dobrski}, who draw attention to the string discontinuity of the commuting components operator. The lack of rotational symmetry, axial symmetry or string discontinuity of the commuting components operator comes down in practice to the choice of the Cartesian coordinate system. To avoid this, in work \cite{Dobrski} a certain modification or extension of the photon wave function was proposed. Even if this does not expand the photon's degrees of freedom, it still modifies the photon's quantum theory. However, Dobrski, the main author of the paper \cite{Dobrski}, in the latest publication \cite{Dobrski 2} withdraws from the operator with commutating components and finally opts for the Pryce operator (here the Pryce–Bacry operator).

The above controversies regarding the existence or non-existence of the photon position operator based on the noncommutation property of its components (or the commutation property) show that the properties of physical quantities should be studied and not postulated \textit{ad hoc}. This remark applies not only to commutation rules.

In this subsection, it is worth commenting on the  alleged violation of the five conditions on the basis of which the photon position operator is criticised: (i) relativistic covariance, (ii) gauge arbitrariness, (iii) helicity definiteness, (iv) commuting of components, (v) domain in Hilbert space. 

Condition (i) in relation to quantum mechanics is understood too literally and too naively, when in reality it is more subtle. Based on Oziewicz's ideas, it has been shown that in the orthodox Special Relativity even the concept of relative velocity of motion has a ternary character (a body in motion, a reference body and an external observer) \cite{Koczan 2023}. Paradoxically, this was done within the framework of the manifestly covariant formalism. Results of a similar nature, also using the relativistic formalism, were also obtained for the concept of relativistic acceleration \cite{Koczan 2021,Koczan ACTA}. Therefore, one should not demand rigorous naive covariance from the notion of localisation on a three-dimensional hypersurface in quantum mechanics. Such a demand is a simple misunderstanding of the concept of relativity of simultaneity. However, the naively simple example of a covariant conservative probability current is just a distractor here.
The gauge arbitrariness condition (ii) is sometimes unreliable, too. The choice of gauge does not always imply that physical quantities depend on this choice or violate covariance. For example, in \cite{Koczan 2002} a covariant gauge was used (Gupta--Bleuler quantisation \cite{Greiner}) and results were obtained that are consistent with the present work.

However, the condition of definite helicity (iii) does not follow directly from the quantisation of Maxwell's equations -- it is an additional condition (compare \cite{Jauch Piron} vs \cite{Amrein,Galindo}). It is not entirely {\it ad hoc}, but if we treat right-handed and left-handed quantum photons separately, why do we not separate electric from magnetic photons? So condition (iii) is related to the Galindo--Amrein theorem, but is not related to the weak localizability of the photon according to Jauch and Piron. 

Even more conventional is condition (iv) of commuting components. What if particles with spin structure have the property of fundamentally lacking commuting components? The Born--Infeld operator being the center of energy (and mass -- see \cite{Koczan 2021,Germano}) operator implies that such particles (with spin) have no commuting position components -- this property applies not only to photons. Even before World War II, in 1935, Pryce found the commutating coordinates (not applicable to the photon) \cite{Pryce 1935}. However, after the war in 1948, Pryce essentially proposed new noncommutative coordinates \cite{Pryce}. Both of Pryce's works preceded the publications of Newton and Wigner in 1949 \cite{Wigner} and Foldy and Wouthuysen in 1950 \cite{Foldy 1950}. Thus, the Newton--Wigner and Foldy--Wouthuysen states or operators seem to be just particular isolated solutions based on rest values of the field, which make no sense for massless particles with spin. Photons do not have rest states, and their null momentum four-vector does not uniquely indicate a direction in three-dimensional space (which would be independent of the choice of reference frame). In other words, the momentum four-vector of a photon does not uniquely distinguish a two-dimensional space-time null hyperplane, sometimes called a flag (there is no polarisation yet, i.e. spin). Quite simply, the Born--Infeld center-of-energy (and mass) operator is a more general solution to the problem (compared to Newton--Wigner). Applying the Born--Infeld operator to the photon leads to the noncommutative Pryce--Bacry operator.

The violation of condition (v) considers generalised states described by Dirac deltas. This procedure has long been justified in the form of the Rigged Hilbert Space \cite{Gelfand}, sometimes called the Gelfand Triplet \cite{Maurin}. 
This is related to the mathematical theory of unbounded operators in Banach space and Hilbert space \cite{Dereziński}. Examples of physical applications of this theory have been developed in textbook \cite{Grabowski}.

This work has no space to explain all the complex mathematical tools of quantum theory in the context of condition (v). Therefore, a transparent approach, which is used in theoretical physics, is used, which is mathematically rigorous from the point of view of distribution theory, Rigged Hilbert Space, and theory of unbounded operators. However, one should not confuse the Rigged Hilbert Space extended by Dirac deltas with the extension of the state space by non-physical longitudinal states of photons (mode zero). This work will make this aspect of the extension of Hilbert space more precise. This clarification was made possible, among other things, by distinguishing identity equality (in extended space) from ordinary equality (in physical space) for position operators. The identity equality allowed for the disambiguation of the position operator with respect to terms that are divergence derivatives. 

The alleged violation of conditions (i--v) is sometimes the basis for unfounded claims about the nonexistence of the photon position operator. In reality, only condition (iv) commuting of components is definitely violated (apart from the works on the axially symmetric operator). Condition (iii) is controversial, but in the light of this work, it is not completely violated. On the other hand, conditions (i), (ii) and (v) are not violated, they are simply complicated mathematically.

\subsection{1.4. Scope, Aim and Formalism of the Work}

The aim of this work, however, is not a historical review, but the author wants to present own results in the most transparent and the most precise synthetic way as possible. Some of these results are an extension of the Jadczyk and Jancewicz publication from 1973 \cite{Jadczyk}. Part of the methodological approach is taken from an independent master's thesis from 2002 \cite{Koczan 2002}, which was written by the author of the present work (Koczan G. M.). 
  Some results are new and original -- see the next subsection about contributions.
  
The work has two basic tasks. The first task is to provide decent equivalent definitions of the photon position operator, so that there is no doubt about the existence, uniqueness and good specificity of such a definition(s). The second task is to provide the eigenstates of the photon position operator for two components. An additional goal turned out to be radial photon position operators and alternative position representation (that is also an alternative probability density amplitude).

This work on quantum mechanics of photons is not written in a manifestly relativistic formalism like other works by the author \cite{Koczan 2002, Koczan 2021, Koczan ACTA, Koczan 2023}, so only the Euclidean signature $(+++)$ for spatial coordinates is used here. Generally, the lower Latin indices $j,k,l,m,n,r,s\in \{1,2,3\}$ are used. Therefore, Einstein's summation convention also applies to lower indices repeating twice:
\begin{equation}
\mathrm{a}_j\mathrm{b}_j:=\mathrm{a}_1\mathrm{b}_1+\mathrm{a}_2\mathrm{b}_2+\mathrm{a}_3\mathrm{b}_3=:\mathbf{a}\cdot\mathbf{b}.
\end{equation}
Vectors (three-dimensional) are marked in bold font (with one exception, where an arrow had to be used), and their components are marked in non-bold font of the same type with a lower indicator. In the case of a coordinate vector, two marking standards will be used interchangeably:
\begin{equation}
\mathrm{r}_1=x,\ \ \ \mathrm{r}_2=y,\ \ \ \mathrm{r}_3=z.
\end{equation}
Operators are marked with a hat, and in the sense of acting on a vector wave function, they have matrix indices:
\begin{equation}
\big(\hat{O}\mathbf{\Psi}\big)_l=\hat{O}_{lm}\Psi_m. 
\end{equation}
A simple standard form (let's call it nominal) of the scalar product is assumed:
\begin{equation}
\label{skalarny}
(\Phi |\Psi)
=
\int{\Phi_l^*(\mathbf{r}) \Psi_l(\mathbf{r})d^3\mathrm{r}}
=
\int{\Tilde{\Phi}_l^*(\mathbf{p}) \tilde{\Psi}_l(\mathbf{p})\frac{d^3\mathrm{p}}{(2\pi\hbar)^3}},
\end{equation}
however, this form is not completely {\it ad hoc}, but results from the construction called {\it heterovariant} representation in \cite{Koczan 2002} (compare with \cite{Jauch Piron}). More details about this representation (photon case) are given later (see sections 3 and 9).

\subsection{1.5. Contributions}

The work presented results from many years (although discontinuous) of research, both review and original, principally by the author himself. The extent of assistance obtained from outsiders is mentioned in the Acknowledgements. The author's documented contribution to the subject so far includes: a master's thesis from 2002 \cite{Koczan 2002}, a poster at the Polish Physical Society conference in 2023 \cite{Poster} and a preprint \cite{arXiv}, version v1 of which was published in the month of the conference and then improved over two years to version v7, which is close to the present work. The second author of the poster \cite{Poster} is Arkadiusz Jadczyk, mentioned in the Acknowledgements, who co-authored an important reference paper from 1973 \cite{Jadczyk}.

Therefore, the work is a broad theoretical research in the form of a meta-analysis, leading to several original results. Contributions to this work can be listed separately according to three criteria: Review (R), Meta-analysis (M), and completely Original contribution (O). The division criterion within the meta-analysis is understood here as various decisions within the active approach in review studies. Seven elements were listed below in each of these three criteria for dividing the contributions.

As part of the division criterion based on the review results,
the following list R of main contents of this type can be given:
\begin{enumerate}
\item[R1.] Boost generator and the Born--Infeld operator for photon.
\item[R2.] Pryce's position operator for the photon.
\item[R3.] Bacry's position operator for the photon.
\item[R4.] Jancewicz and Jadczyk's definition of the photon position operator is based on projective operators onto helicity states.
\item[R5.] Eigenstates of the photon's position in the form of line integrals (e.g. along a straight line or a circle) according to Jancewicz and Jadczyk (the case now called electric strings) in the momentum representation. These results are consistent with the weakly localizable photons according to Jauch--Piron.
\item[R6.] Eigenstates of the position and helicity of the photon on a straight line according to Jancewicz and Jadczyk (the case now called helical strings on a straight line). The result does not contradict the Galindo--Amrein theorem, because the straight line is unbounded.
\item[R7.] The form of the photon position operator from Koczan's master's thesis is called the minimal form (in the sense of the shortest).
\end{enumerate}

The results of the review, subjected to detailed analysis, were the basis for further research of the meta-analysis type. In this context of meta-analysis, the following list M of results can be given:
\begin{enumerate}
\item[M1.] Explicit calculations of the Born--Infeld and Pryce--Bacry operator(s) for the photon.
\item[M2.] Explicit calculations of the photon position operator according to the Jancewicz--Jadczyk definition, which led to the helical definition of the operator.
\item[M3.] Explicit calculation of the photon position operator according to the Schr\"odinger idea of an operator not mixing signs of frequencies ({\it Zitterbewegung}).
\item[M4.] Explicit calculation of the transversal photon position operator -- the approach closest to Bacry's, despite the apparent similarity of his method to M2 or M3.
\item[M5.] Explicit proof of the equivalence of the result of the photon position operator obtained at points M1--M4.
\item[M6.] Research of the eigenstates of the position and helicity of photons of the Jancewicz--Jadczyk type on a circle (now called hybrid helical strings on a circle) and show that they are not eigenstates of the radial component. This result is consistent with the Galindo--Amrein theorem.
\item[M7.] Photon position eigenstates in the form of line integrals (along a straight line or a circle) of the Jancewicz--Jadczyk type (the case now called electric strings) in an explicit positional representation.
\end{enumerate}

In the course of in-depth review research and meta-analysis, several completely new original results were obtained. The list O of these results is as follows:
\begin{enumerate}
\item[O1.] Radial photon position operators.
\item [O2.] General formula of the line integral with curl for a photon position eigenstate, of the magnetic string type, in the positional representation.
\item [O3.] Explicit form of the eigenstate of a photon's position called a magnetic string on a straight line.
\item[O4.] Explicit form of the eigenstate of a photon's position called a magnetic string on a circle.
\item[O5.] An alternative helical positional representation of the photon and the invariance of the Cartesian position operator of the photon with respect to the transition to this representation.
\item[O6.] Obtaining string attributes of particles (photons) from an available theory (photon quantum mechanics) as opposed to postulating an \textit{ad hoc} abstract string theory requiring 11 or 26 dimensions.
\item[O7.] Demonstration of the disambiguation of the physical form of the photon position operator thanks to the extended Hilbert space with non-physical longitudinal states (for the domain and codomain).
\end{enumerate}

The above contribution, presented in three parts, does not constitute a table of contents, but a synthetic summary of its most essential elements from the division's point of view, which is into review, research, and authorial content.

Appendices A--F, especially C and F, can significantly help to understand the contributions better. Appendix C presents concisely the equivalent definitions of the Cartesian photon position operator, together with tables of their specific names and notations. Appendix F lists all string photon states and tables, including their names, notations, and eigenvalues. In addition, the paper presents graphical models of the most important string eigenstates of the poton position.

Regardless of the contributions, the work layout had to take a form more subordinated to the regime of derivational logic and mathematical calculations, including proofs. The reader is therefore faced with the prospect of an extended reading of a specialised mathematical text. The alternative is selective reading, in which the detailed titles of the section and subsection headings can be helpful.

\section{2. DEFINITION OF THE PHOTON POSITION OPERATOR
BASED ON THE BOOST GENERATOR}

\subsection{2.1. Boost Generator for Photon}

Two of the generators of the Poincaré group refer directly to the position vector. These are the rotation generator, which is the total angular momentum operator, and the Lorentz boost generator, which is the moment of energy operator. Apart from the spatial part that interests us, both generators also contain a spin part. In the case of photons, there may be difficulties in separating the spatial part from the spin part if we want to have operators well defined in the domain of physical states. Therefore, for each of these generators, in a sense, the spatial part and the spin part do not have to be {\it a priori} correct separately, but taken together they constitute the correct entire operator. 

Nevertheless, it is possible to isolate the physical photon position operator and the physical spin operator from these generators. However, the position operator will not be a simple coordinate multiplication, and the physical spin operator will not be a set of ordinary antisymmetric matrices. Despite this, we formally use the "ordinary" matrix photon spin operator. However, such an operator has a correct physical meaning only as a projection on the direction of momentum (i.e. as helicity). Similarly, simple multiplication of the wave function by the coordinates makes sense only in the context of a matrix element (dot product) of physical states (see later).

The orbital angular momentum, which is the spatial part of the rotation generator, depends on the vector product of position and momentum. This product disappears for the component of the position vector parallel to the momentum, so we lose information about this component. Therefore, it is better to base the photon position operator not on the rotation generator, but on the boost generator.

We assume the standard form of the $\hat{\mathbf{N}}$ boost generator (energy moment operator), only slightly adapted to the photon (see \cite{Foldy} for $m=0, t=0$ which is equivalent to the formula in \cite{Jadczyk}):
\begin{equation}
\label{N}
\hat{\mathbf{N}}:\equiv\frac{c}{2}(\hat{\mathrm{p}}\ \hat{\mathbf{r}}+\hat{\mathbf{r}}\ \hat{\mathrm{p}})+c\ \hat{\mathrm{p}}^{-1}\hat{\mathbf{p}}\times\hat{\mathbf{S}},
\end{equation}
where: $c$ is speed of light, $\hat{\mathrm{p}}=|\hat{\mathbf{p}}|=\hbar\sqrt{-\Delta}$ (or $\hat{\mathrm{p}}=|\mathbf{p}|=\mathrm{p}$) is absolute value of momentum operator $\hat{\mathbf{p}}=-i\hbar\nabla$ (or $\hat{\mathbf{p}}=\mathbf{p}$), and $\hat{\mathbf{r}}=\mathbf{r}$ (or $\hat{\mathbf{r}}=i\hbar\partial/\partial\mathbf{p}$) is a coordinate vector of "ordinary" position, and $\hat{\mathbf{S}}$ is the "ordinary" spin matrix for a vector field $\mathrm{S}_{k,lm}=-i\hbar\varepsilon_{klm}$. Lack of order symmetrization in the first term of the boost generator occurs when the nominal form of scalar product is not used (see \cite{Beckers} and first of all \cite{Jadczyk 2024}, where the generator formula was derived strictly for the photon).

Interestingly, there are critical voices about certain properties of the representation of Poincaré generators (especially for the massless case, like a photon). Dragon denies the Hermitianity of these operators \cite{Dragon2}. The correctness of these generators, including the invariance of the Hilbert space domain, is questioned based on the requirement of total smoothness of the operators by Dragon and Oppermann \cite{Dragon1}. The requirement of complete smoothness for unbounded operators is absurd and contradicts the mathematical theory of such operators \cite{Grabowski, Dereziński}. Based on this flawed requirement, the authors of previously cited papers on abstract strings in fantastic 26 dimensions deny the foundations of photon quantum mechanics (quantum electrodynamics QED) without interactions. 

For example, in this way, these authors deny the existence of a position operator for massless particles. The procedure presented is incorrect four times. First, the position operator of a scalar massless particle is not a problem (its components commute) \cite{Wigner, Born--Infeld, Koczan 2002}. Newton and Wigner in \cite{Wigner} wrote: \textit{has been carried out also for the equations with zero mass.  In the case of spin 0 and 1/2, we were led back to the expressions for localised systems.} Moreover, the commutator components of the Born-–Infeld operator vanish for zero spin, independently of the value of the mass \cite{Born--Infeld}. Second, using the same arguments, one could refute the usual nonrelativistic Schr\"odinger position operator (e.g. under the pretext of the lack of smoothness of the $r=|\hat{\mathbf{r}}|$ operator used to draw the hydrogen atom orbitals). Third, the \textit{ad hoc} assumption of commutativity of the components of the position operator is a preference for the abstract Heisenberg group over the physical Poincaré group and Maxwell's equations. Fourth, a translation on a mass shell or a cone in the massless case (formula "(40)" in Dragon and Oppermann) is never an affine translation and must be based on an appropriate parallel translation connection (see e.g. \cite{Jadczyk 2024}). An ordinary affine translation on the light cone in momentum space immediately violates the transversality condition (on which the next section 3 is based).

As for the issue of alleged violations of invariance, this issue should be considered in terms of explicit or implicit covariance, not invariance (see \cite{Koczan 2023}). It is then easier to simultaneously investigate the freedom of gauge. Furthermore, it is worth mentioning in this context that there is an explicitly covariant method of quantising the electromagnetic field, which is called the Gupta-Bleuler quantisation \cite{Greiner}. So the problems with the representation of Poincaré generators for photons, despite the difficulties, seem to be solvable. For example, in the publication \cite{Ruscy}, there were no problems with such generators in higher dimensions, and in the preprint \cite{Sazdovic}, Maxwell's equations were derived based on group theory.

Once written out, the indicator form of the boost generator in the momentum representation takes the following shape (for $\hbar:=1$, $c:=1$):
\begin{equation}
\hat{\mathrm{N}}_{k,lm}\equiv\Big(\mathrm{p}\frac{i\partial}{\partial\mathrm{p}_k}+\frac{i\mathrm{p}_k}{2\mathrm{p}}\Big)\delta_{lm}+\frac{i}{\mathrm{p}}\delta_{km}\mathrm{p}_l-\frac{i}{\mathrm{p}}\delta_{kl}\mathrm{p}_m.
\end{equation}
This operator does not satisfy the condition of transversality, in the sense of identity:
\begin{equation}
\mathrm{p}_l\hat{\mathrm{N}}_{k,lm}\equiv\mathrm{p}\frac{i\partial}{\partial\mathrm{p}_k}\mathrm{p}_m-\frac{i\mathrm{p}_k}{2\mathrm{p}}\mathrm{p}_m\nequiv 0.
\end{equation}
However, for physical states satisfying the transversality condition, the value of the boost operator also satisfies this condition:
\begin{equation}
\mathrm{p}_l\hat{\mathrm{N}}_{k,lm}\Psi_m=\mathrm{p}\frac{i\partial}{\partial\mathrm{p}_k}\mathrm{p}_m\Psi_m-\frac{i\mathrm{p}_k}{2\mathrm{p}}\mathrm{p}_m\Psi_m= 0
\end{equation}
Each of the two zeroing terms of this expression results from both the spatial and spin parts of the boost operator. Therefore, on separate parts (spatial and spin), the condition would not be met.

\subsection{2.2. Born--Infeld and Pryce--Bacry Operators}

Therefore, the definition of the photon position operator should not and cannot eliminate the spin part from the boost generator. On the contrary, as will be explained later, such a definition actually absorbs the entire spin part. Therefore, it is enough to provide the simplest formula for the Hermitian (self-adjoint) operator with the dimension of position:
\begin{equation}
\label{q}
\hat{\mathbf{q}}:\equiv\frac{1}{2c}(\hat{\mathrm{p}}^{-1}\ \hat{\mathbf{N}}+\hat{\mathbf{N}}\ \hat{\mathrm{p}}^{-1}).
\end{equation}
A general version (not only for photons) of such a position operator was defined in 1935 by Born and Infeld \cite{Born--Infeld}. This definition, thanks to the appropriate commutation condition, can be written slightly shorter:
\begin{equation}
\hat{\mathbf{q}}\equiv
\frac{1}{\sqrt{\hat{\mathrm{p}}c}}\hat{\mathbf{N}}
\frac{1}{\sqrt{\hat{\mathrm{p}}c}}.
\end{equation}
The Born--Infeld position operator was and is well defined for virtually every particle, including the photon, which is massless. Nevertheless, Born and Infeld did not develop {\it explicite} their operator for the photon. 

The first known such expansion was given in 1948 by Pryce \cite{Pryce}, who used a rather peculiar Kemmer-type matrix formalism for Maxwell's equations (similar to the matrix in the Dirac equation). As a result, the Pryce form of the Born--Infeld operator for a photon differed formally in one term from the forms considered in this work (see below). A simplified form of the photon position operator (without this additional term) was given in 1988 by Bacry \cite{Bacry 1988}. However, Bacry did not explicitly use the Born--Infeld definition in 1988 to derive the given formula. 

Nevertheless, in a sense, the compatibility of the two approaches has been shown. Bacry's earlier paper from 1981 \cite{Bacry 1981} included the Born-Infeld definition, but did not give a particular result. Ultimately, it is not known how Bacry obtained the formula given in 1988, because he did not provide a detailed derivation or proof. Formally, however, he wrote down the (purely) transversal definition of the photon operator, which is not identically equal, only on physical states, in the formula given (see section 3). Bacry knew this and therefore wrote down information about a certain insignificant ambiguity under the formula. This ambiguity was not refined, as in the present work, in terms of identity equality "$\equiv$" or equality across states "$=$". At the same time, Bacry also wrote about a frequency-nonmixing position operator and, in a sense, suggested a helicity-nonmixing position operator (see section 4). 

Due to the above, the obtained formula of the photon position operator will be called the Pryce--Bacry definition (or Pryce--Bacry operator):
\begin{equation}
\hat{\mathbf{x}}:\equiv \hat{\mathbf{r}}+
\frac{\hat{\mathbf{p}}\times\hat{\mathbf{S}}}{\hat{\mathrm{p}}^2}.
\end{equation}

In any case, strictly applying the definition of (\ref{q}) to the operator (\ref{N}) leads to the following fact.
\\
\\
STATEMENT 1 (Born--Infeld position operator for photon)
\textit{The photon position operator computed from the boost generator has the Pryce--Bacry vector form (see \cite{Pryce, Bacry 1988, Koczan 2002}):}
\begin{equation}
\label{qq}
\hat{\mathbf{q}}\equiv \hat{\mathbf{r}}+
\frac{\hat{\mathbf{p}}\times\hat{\mathbf{S}}}{\hat{\mathrm{p}}^2}\equiv\hat{\mathbf{x}}
\end{equation}
\textit{and the following indicator form:}
\begin{equation}
\label{qk}
\hat{\mathrm{q}}_{k,lm}\equiv \hat{\mathrm{r}}_k\delta_{lm}+\frac{i\hbar}{\hat{\mathrm{p}}^2}\delta_{km}\mathrm{\hat{\mathrm{p}}}_l-\frac{i\hbar}{\hat{\mathrm{p}}^2}\delta_{kl}\mathrm{\hat{p}}_m.
\end{equation}
\textit{Moreover, it is a physical operator, i.e. it maintains the condition of transversality in the domain of physical states. In addition, this operator is automatically Hermitian (self-adjoint) for the standard dot product} (\ref{skalarny}).
\\
\\
PROOF. To obtain the vector form of the Born--Infeld position operator, it is enough to calculate the following expression in the momentum representation (for $\hbar:=1$):
\begin{equation}
\mathrm{p}^{-1}\hat{\mathbf{r}}\mathrm{p}+\mathrm{p}\hat{\mathbf{r}}\mathrm{p}^{-1}    
\equiv
\mathrm{p}^{-1}\frac{i\partial}{\partial\mathbf{p}}\mathrm{p}+\mathrm{p}\frac{i\partial}{\partial\mathbf{p}}\mathrm{p}^{-1} \equiv
2\frac{i\partial}{\partial\mathbf{p}}
\equiv
2\hat{\mathbf{r}}.
\end{equation}
Hence, the form (\ref{qq}) follows from the definitions of (\ref{N}) and (\ref{q}).

To obtain the index form, we need to write the vector product of momentum and spin:
\begin{equation}
(\mathbf{p}\times\hat{\mathbf{S}})_{k,lm}
\equiv
\varepsilon_{kjn}\mathrm{p}_j\mathrm{S}_{n,lm}
\equiv
-i\hbar \varepsilon_{kjn}\varepsilon_{nlm}\mathrm{p}_j,
\end{equation}
and use the identity for the Levi-Civita symbol:
\begin{equation}
(\mathbf{p}\times\hat{\mathbf{S}})_{k,lm}
\equiv
\frac{\hbar}{i} (\delta_{kl}\delta_{jm}-\delta_{km}\delta_{jl})\mathrm{p}_j
\equiv
\frac{\hbar}{i} (\delta_{kl}\mathrm{p}_m-\delta_{km}\mathrm{p}_l).
\end{equation}
Therefore, the vector formula (\ref{qq}) can be written in the index version (\ref{qk}) in any representation (momentum or position).

Thanks to the index form, it is possible to demonstrate the transversality of the position operator. Let's start by checking the identity transversality of the main term of the position operator (in the momentum representation):
\begin{equation}
\mathrm{p}_l\hat{\mathrm{r}}_k\delta_{lm}
\equiv
\mathrm{p}_m\hat{\mathrm{r}}_k
\equiv
\mathrm{p}_mi\hbar\frac{\partial}{\partial\mathrm{p}_k}
\equiv
\hat{\mathrm{r}}_k\mathrm{p}_m-i\hbar\delta_{km}
\nequiv 0.
\end{equation}
The condition is not met as identity for the entire operator, although the key non-transversal term is reduced:
\begin{equation}
\mathrm{p}_l\hat{\mathrm{q}}_{k,lm}
\equiv
\hat{\mathrm{r}}_k\mathrm{p}_m-\frac{i\hbar}{\mathrm{p}^2}\mathrm{p}_k\mathrm{p}_m
\nequiv 0,
\end{equation}
so it can be seen that the position operator is transversal on physical states (transversal states):
\begin{equation}
\mathrm{p}_l\hat{\mathrm{q}}_{k,lm}\Psi_m
=
\hat{\mathrm{r}}_k\mathrm{p}_m\Psi_m-\frac{i\hbar}{\mathrm{p}^2}\mathrm{p}_k\mathrm{p}_m\Psi_m
= 0.
\end{equation}

What remains to be proven is the Hermitian (self-adjoint) position operator in the identity way, with the standard scalar product (\ref{skalarny}). To do this, it is enough to make a complex conjugate of imaginary units ($i^*=-i$) and rearrange the wave function indicators in places (for $\hbar:=1$):
\begin{equation}
(\hat{\mathrm{q}}_{k,lm})^{\dagger}
\equiv 
\hat{\mathrm{q}}^*_{k,ml}
\equiv
\hat{\mathrm{r}}_k\delta_{ml}+\frac{i^*}{\hat{\mathrm{p}}^2}\delta_{kl}\hat{\mathrm{p}}_m-\frac{i^*}{\hat{\mathrm{p}}^2}\delta_{km}\mathrm{\hat{p}}_l
\equiv
\hat{\mathrm{q}}_{k,lm}.
\end{equation}
Thanks to the above, the calculated forms of the position operator have all the requirements of the real physical quantity operator. Q.E.D.

\subsection{2.3. A Discussion of Two Slightly Different Representations}

As mentioned, Pryce did not receive a photon operator form identical to (\ref{qq}). His form contained an additional element $-i\hbar\hat{\mathbf{p}}/(2\hat{\mathrm{p}}^2)$, which could give the impression of a non-Hermitian element. It resulted from the specificity of the Kemmer matrix representation used by Pryce. Currently, this can be explained by an asymmetric order of operators or a non-standard form of the dot product. For example, dividing the boost generator by energy on the right leads strictly to the Pryce form \cite{Pryce}:
\begin{equation}
\hat{\mathbf{N}} \frac{1}{\hat{\mathrm{p}} c}
\equiv
\hat{\mathbf{r}}-\frac{i\hbar\hat{\mathbf{p}}}{2\hat{\mathrm{p}}^2}+\frac{\hat{\mathbf{p}}\times\hat{\mathbf{S}}}{\hat{\mathrm{p}}^{2}}
\equiv: \{\hat{\mathbf{q}} \}_{\phi}.
\end{equation}
The form of the position operator defined in this way will be Hermitian if the dot product in the momentum representation of the wave function takes the form:
\begin{equation}
\label{skalarny/p}
\int{\Tilde{\phi}_l^* \frac{1}{\mathrm{p}} \Big(\hat{\mathbf{N}}\frac{1}{\mathrm{p}c} \tilde{\phi}_l\Big) \frac{d^3\mathrm{p}}{(2\pi\hbar)^3}}.
\end{equation}
A similar form of the dot product is characteristic of relativistic field theories. Using an appropriate transformation, you can go to the representation with a standard dot product:
\begin{equation}
\tilde{\Psi}_l=\frac{1}{\sqrt{\mathrm{p}}}\tilde{\phi}_l ,
\ \ \ \ \ \
\{\hat{\mathbf{q}} \}_{\Psi}=\frac{1}{\sqrt{\mathrm{p}}}\{\hat{\mathbf{q}} \}_{\phi}\sqrt{\mathrm{p}}.
\end{equation}
After this procedure, the Born--Infeld operator in Pryce form would take the Pryce--Bacry form (\ref{qq}). In \cite{Koczan 2002}, the representation of the above type is called {\it heterovariant} representation, due to its non-local properties and non-local transformation laws (see also \cite{Jauch Piron}). 

Because the Born--Infeld operator in Pryce form contained an additional term, the photon position operator in the literal form (\ref{qq}) was probably given for the first time by Bacry in 1988 \cite{Bacry 1988}. However, Bacry used a slightly different derivation method, and it is only a coincidence that he did not obtain the additional Pryce term. The point is that Bacry, when working in the Riemann--Silbertein vector representation, should work with the dot product (\ref{skalarny/p}), and not the standard dot product (\ref{skalarny}). If that were the case, Bacry would have received the same additional term as Pryce.
  
Since we finally have the desired photon position operator, we can use it to express the energy moment operator. To do this, we calculate $\hat{\mathbf{r}}$ from (\ref{qq}) and insert it into (\ref{N}), obtaining:
\begin{equation}
\label{pq+qp}
\hat{\mathbf{N}}\equiv\frac{c}{2}(\hat{\mathrm{p}}\ \hat{\mathbf{q}}+\hat{\mathbf{q}}\ \hat{\mathrm{p}}),
\end{equation}
which can also be transformed into:
\begin{equation}
\label{pqp}
\hat{\mathbf{N}}\equiv
\sqrt{\hat{\mathrm{p}}c}\ \hat{\mathbf{q}}\sqrt{\hat{\mathrm{p}}c}.
\end{equation}
The above formulas constitute both a Born--Infeld type photon position operator and strengthen the interpretation of the boost generator as an energy moment operator, without the spin term, which is troublesome to interpret. 

In addition, the operator $\hat{\mathbf{N}}$, up to the factor $c^2$, can be treated as the moment of relativistic mass (see \cite{Koczan 2021}). Therefore, the position operator is the center of relativistic mass or, more generally, the center of mass-energy. This does not contradict the fact that the rest mass of a photon is zero.

\section{3. Definition of the photon position operator based on the condition of transversality}

\subsection{3.1. Vector Wave Function in Coulomb Gauge}

In the previous section 2, a very significant impact of the transversality condition on the quantum theory of the photon was revealed. As we shall see in this section 3, the transversality condition itself:
\begin{equation}
\nabla \cdot
\mathbf{\Psi}(\mathbf{r})=0 \ \ \ \Leftrightarrow \ \ \ 
\mathbf{p}\cdot\tilde{\mathbf{\Psi}}(\mathbf{p})=0
\end{equation}
suffices to derive the photon position operator correctly.

What is the interpretation of the transversality condition of the photon wave function? The simplest answer to this question can be said that a photon is a transversely polarized particle -- it is a transverse electromagnetic wave. In the language discussed in the next section 4, we would say that a photon can have a spin either strictly in the direction of momentum (helicity +1) or opposite to it (helicity $-1$), but cannot have a spin perpendicular to the momentum (helicity 0). That is, the disappearance of states with helicity 0 is equivalent to the disappearance of longitudinal polarisation and is therefore equivalent to the transversality condition.

In the quantum theory of the electromagnetic field, the transversality condition has a slightly different origin. The easiest way to describe the electromagnetic field is not using the electric and magnetic fields, but using the four-potential, i.e. the so-called scalar potential and the vector (three-dimensional) potential taken together. However, the four-potential has five troublesome properties: (i) it is defined ambiguously to the gradient of any function, (ii) the scalar potential is not a dynamic quantity because its canonical momentum vanishes identity-wise, (iii) it requires meeting the Coulomb gauge condition $\nabla\cdot \mathbf{A}=0 $ or the Lorentz gauge condition $\partial^{\mu}A_{\mu}=0$ on the states, (iv) it is accompanied by non-physical gauge modes sometimes called "ghosts" (depending on the version of the description with positive, zero or even negative norm), (v) its canonical quantization (in standard Coulomb gauge) breaks the explicit covariance of the theory (but not implicite covariance or even not explicit covariance in nonstandard Gupta--Bleuler quantization \cite{Greiner}). 

Property (iii) is introduced to solve the problematic properties (i) and (ii). The point is that either the scalar potential is fixed (or zeroed), or it is referred to as "ghosts" (with a negative norm). Therefore, the introduction of property (iii) leads to the inconvenience of (iv) accompanying non-physical states ("ghosts"). Of course, we can use the domain definition to eliminate "ghosts" from the theory, but it is easier and more convenient to use a certain superdomain from which nonphysical states ("ghosts") are separated, which ensures the physicality of the theory. 

The closest method to this work is the Coulomb gauge for free  photons ($A_0=0$), in which even additional nonphysical states ($\nabla\cdot \mathbf{A}\neq 0 $) have a positive norm ("ghost" modes with longitudinal polarisation, i.e. with helicity 0). This approach is characterised by property (v), i.e. it breaks the explicit covariance of the theory -- which does not mean that the theory is not covariant (implicitly). 

In the alternative Gupta-Bleuler method \cite{Greiner}, the price for maintaining explicit covariance is the use of a nonzero potential $A_0$ relating to nonphysical states with a negative norm (so-called "evil ghosts"). Appropriately applying the Lorentz gauge condition to physical states allows the elimination of "evil ghosts" from the domain of physical states \cite{Greiner}. However, there are still null norm states ("zero ghosts") \cite{Koczan 2002}. "Zero ghosts" have no practical physical significance, but they are also physically harmless -- so they are inert rather than nonphysical. However, they describe some gauge freedom in the Lorentz gauge framework and thus allow explicit covariance to be preserved. Zero ghosts (not to be confused with the longitudinal zero mode) can also be theoretically eliminated using equivalence classes.

Nevertheless, papers are still being published with the erroneous opinions that photon quantum mechanics (or more generally QED) violates gauge freedom of type (i) or violates Lorentz covariance in scenario (v). In fact, none of these conditions (gauge freedom, explicit Lorentz covariance) need be violated -- this is achieved in the Gupta--Beuler quantisation \cite{Greiner}. However, the Coulomb gauge is an allowed simplification because the choice of gauge is allowed. The issue with the Lorentz covariance condition is more serious, because this condition cannot be violated -- and it is not violated, although it may seem so. This is shown by the correspondence with the Gupta-Bleuler quantisation \cite{Greiner, Koczan 2002}. Since this quantisation in any chosen reference frame is equivalent to the quantisation in the Coulomb gauge -- via an appropriate choice of the representative of the equivalence classes -- then the Coulomb quantisation satisfies the Lorentz covariance property, but implicitly. The last conclusion follows from the physical equivalence of theories, one of which is Lorentz covariant and explicitly so (Gupta--Bleuler quantisation). However, it is so subtle that it is easier to baselessly deny it than to understand it.

It is easy to see that within Maxwell's equations (in particular the Ampere-Maxwell equation), the vector potential for the magnetic field in the Coulomb gauge is expressed uniquely in terms of the electromagnetic field and the current distribution. Namely, from the equations:
\begin{equation}
\nabla\times\mathbf{B}=\mu_0 \ \mathbf{j}+\frac{1}{c^2}\frac{\partial \mathbf{E}}{\partial t}, \ \ \ 
\mathbf{B}=\nabla\times\mathbf{A}, \ \ \ 
\nabla \cdot\mathbf{A}=0,
\end{equation}
the formula for the vector potential follows:
\begin{equation}
\label{A}
\mathbf{A}=\frac{1}{-\Delta}\Big(\mu_o\mathbf{j}+\frac{1}{c^2}\frac{\partial \mathbf{E}}{\partial t}\Big),
\end{equation}
where the inverse of the minus Laplacian is a typical Coulomb integral (see Appendix A). Therefore, the vector potential is a unique function of the time derivative of the electric field and the charge current density $\mathbf{j}$. The entire electromagnetic field can be expressed using the vector potential and charge density $\rho$:
\begin{equation} 
\mathbf{B}=\nabla\times\mathbf{A}, \ \ \ 
\mathbf{E}=-\frac{\partial \mathbf{A}}{\partial t}-\frac{\nabla}{-\Delta}\Big(\frac{\rho}{\varepsilon_0}\Big).
\end{equation}

Therefore, the sourceless (transversal) vector potential is an initially good candidate for representing the wave function. However, three problems remained: normalisation, the dot product form, and the lack of complex numbers. In the SI system, the dimension of the vector potential is as follows:
\begin{equation} 
\big[\mathbf{A}\big]=\frac{1\ \mathrm{N}}{1\ \mathrm{A}}=\frac{1\ \mathrm{kg} \cdot 1\ \mathrm{m}}{1\ \mathrm{A}\cdot 1\ \mathrm{s}^2}.
\end{equation}
The wave function in the positional representation should have units dimension $1\ \mathrm{m}^{-3/2}$. A significant discrepancy in the dimensions of the units means that the normalising factor, in addition to the dimensional constants, must contain a fourth-degree root of the minus Laplacian:
\begin{equation} 
\Bigg[\frac{\sqrt[4]{-\Delta}\ \mathbf{A}}{\sqrt{\mu_0\hbar c}}\Bigg]=\frac{1}{1\ \mathrm{m}^{3/2}}.
\end{equation}
Due to the formula (\ref{A}), it is basically enough to know the calculation of the $(-\Delta)^{-3/4}$ operator given in Appendix A. The above normalisation of the wave function can be justified not only based on dimensional analysis, but also based on the simplified form of a dot product. It is most convenient to refer here to the commutator for creating and annihilating parts of the vector potential. In order for this commutator to be normalised to the Kronecker and Dirac deltas (or the transversal Kronecker delta), the vector potential must be multiplied by an operator proportional to $(-\Delta)^{-1/4}$:
\begin{equation} 
\vec{\mathcal{A}} \sim \sqrt[4]{-\Delta}\ \mathbf{A}.
\end{equation}
This methodology was used in the work \cite{Koczan 2002} and, as you can see, it is consistent with the result obtained based on the units analysis.

The complex nature of the renormalised vector potential is still missing from the ready representation of the wave function. Often this goal is achieved somewhat artificially using a Riemann--Silberstein vector of the form $\mathbf{E}+ic\mathbf{B}$. However, creating a complex classical field is unnecessary before quantising it. The formalism of quantum field theory will automatically give the wave function components complex values. In the method adopted here, in practice, it comes down to separating part of the field with positive frequencies \cite{Koczan 2002}:
\begin{equation} 
\label{Psi definition}
\mathbf{\Psi}(\mathbf{r},t) \sim \sqrt[4]{-\Delta}\ \mathbf{A}_{(+)}(\mathbf{r},t):=\bra{0} \sqrt[4]{-\Delta}\ \hat{\hat{\mathbf{A}}}(\mathbf{r},t)\ket{\Psi},
\end{equation}
the field operator in the sense of the second quantisation is marked here with a double hat to distinguish it from operators acting directly on the wave function. The operation of this field definition is that the annihilating part of the field (negative frequencies) disappears when it is affected by a vacuum state on the left side. The positive frequency of the field satisfying the wave equation can be easily calculated:
\begin{equation}
\label{A+}
\mathbf{A}_{(+)}=\frac{\hat{\mathrm{p}}c+i\hbar\partial/\partial t}{2\hat{\mathrm{p}}c}
\mathbf{A}=
\frac{1}{2}\mathbf{A}+\frac{i}{2\sqrt{-\Delta}}\frac{1}{c}\frac{\partial\mathbf{A}}{\partial t}.
\end{equation}
Therefore, the photon wave function in the vector potential representation has the form:
\begin{equation}
\label{Psi from A}
\mathbf{\Psi}(\mathbf{r},t)
\sim
\sqrt[4]{-\Delta}\ \mathbf{A}+\frac{i}{\sqrt[4]{-\Delta}}\frac{1}{c}\frac{\partial\mathbf{A}}{\partial t},
\end{equation}
which can be expressed directly in terms of the electric field (neglecting current and charge densities):
\begin{equation}
\label{Psi from E}
\mathbf{\Psi}(\mathbf{r},t)
\sim
\frac{1}{\sqrt[4]{-\Delta}^3}\frac{1}{c^2}\frac{\partial\mathbf{E}}{\partial t}-\frac{i}{\sqrt[4]{-\Delta}}\frac{1}{c}\mathbf{E},
\end{equation}
or also using the magnetic field:
\begin{equation}
\label{Psi from B and E}
\mathbf{\Psi}(\mathbf{r},t)
\sim
\frac{\nabla\times\mathbf{B}}{\sqrt[4]{-\Delta}^3}-\frac{i}{\sqrt[4]{-\Delta}}\frac{1}{c}\mathbf{E}.
\end{equation}
This form is not identical to the renormalised Riemann--Silberstein vector. However, for states with a specific helicity, the relationship with this vector appears, but for different helicities it differs, in addition to normalisation, also in complex coupling.
Thus, the transversal complex wave function of the photon has been constituted by one of the last three equations.

Since the domain of physical wave functions is traversal vector fields, operators of physical quantities should also operate within this space. This condition can be ensured using the projection operator on transversal states, which is the transversal Kronecker delta:
\begin{equation}
\label{P}
\hat{P}_{lm}
:\equiv
\hat{\delta}^{\perp}_{lm}
:\equiv
\delta_{lm}-\frac{\hat{\mathrm{p}}_l\hat{\mathrm{p}}_m}{\hat{\mathrm{p}}^2}
\equiv
\delta_{lm}-\Delta^{-1}\partial_l\partial_m,
\end{equation}
where: $\partial_l=\partial/\partial\mathrm{r}_l$. Let us also introduce the projection operator on non-physical states ("ghosts" with positive norm and helicity 0):
\begin{equation}
\label{P0}
\hat{P}_0 :\equiv 1-\hat{P},
\end{equation}
however, the wave function indicators have been omitted here.
These are projection operators onto orthogonal subspaces, so they meet the conditions:
\begin{equation}
\hat{P}^2\equiv\hat{P},\ \ \ \ \hat{P}_0^2\equiv\hat{P}_0,\ \ \ \ \hat{P}\hat{P}_0\equiv\hat{P}_0\hat{P}\equiv0.
\end{equation}

\subsection{3.2. Transversal Photon Position Operator}

We can now define the photon's position operator as a position vector that transforms physical states into physical states:
\begin{equation}
\label{QPP}
\hat{\mathbf{Q}}:\equiv\hat{P}\ \hat{\mathbf{r}}\ \hat{P}+\hat{P}_0\ \hat{\mathbf{r}}\ \hat{P}_0.
\end{equation}
As we can see, this operator also includes the "ghosts" subspace (longitudinal polarisation or helicity zero) in order to maintain identity consistency with the remaining definitions. Operating on non-physical states in part of this operator does not make it a non-physical operator, because this operator by definition does not mix physical with non-physical states (and {\it vice versa}). Of course, we can consider an operator by definition truncated only to physical states -- which for physical states coincides with the more general operator (\ref{QPP}) -- a pure transversal operator:
\begin{equation}
\label{QP}
\hat{\mathbf{R}}:\equiv\hat{P}\ \hat{\mathbf{r}}\ \hat{P}.
\end{equation}

The operator in the general version equivalent to (\ref{QPP}) was defined in 1973 in the publication \cite{Jadczyk}. The technical difference was that the projection operators $\hat{P}$ and $\hat{P}_0$ were expressed using the helicity operator. However, the definition in the form (\ref{QP}) was given in 1988 by Bacry \cite{Bacry 1988}, although he did not use identity transformations, but equivalent transformations for physical states.
\\
\\
STATEMENT 2 (On transversally diagonal photon position operator)

\textit{The photon position operator} (\ref{QPP}), \textit{which does not mix physical transversal states with non-physical longitudinal states, is identity-equivalent to the Pryce--Bacry (Born--Infeld) operator for the photon:}
\begin{equation}
\label{Q=q}
\hat{\mathbf{Q}}\equiv\hat{\mathbf{x}}\equiv \hat{\mathbf{r}}+
\frac{\hat{\mathbf{p}}\times\hat{\mathbf{S}}}{\hat{\mathrm{p}}^2}\equiv\hat{\mathbf{q}},
\end{equation}
\textit{and this applies to both the vector and indicators versions. However, the purely transversal photon position operator} (\ref{QP}) \textit{on the physical Hilbert subspace is equal to the two mentioned operators:}
\begin{equation}
\hat{\mathbf{R}}=\hat{\mathbf{Q}}=\hat{\mathbf{x}}=\hat{\mathbf{q}}\ \ \ \ \text{on} \ \ \ \ \mathcal{H}_{phys},
\end{equation}
\textit{and has an indicator form equal identity to:}
\begin{equation}
\label{Q|}
\hat{\mathrm{R}}_{k,lm}\equiv 
\hat{\mathrm{r}}_k\delta_{lm}
+\frac{i\hbar}{\hat{\mathrm{p}}^2}\delta_{km}\mathrm{\hat{\mathrm{p}}}_l
-\hat{\mathrm{r}}_k\frac{\hat{\mathrm{p}}_l\hat{\mathrm{p}}_m}{\hat{\mathrm{p}}^2}
-i\hbar\frac{\hat{\mathrm{p}}_k\hat{\mathrm{p}}_l\hat{\mathrm{p}}_m}{\hat{\mathrm{p}}^4}.
\end{equation}
\textit{In this form, the photon position operator is an identity Hermitian operator with a dot product} (\ref{skalarny}) \textit{and is identity transversal on two sides.}
\\
\\
PROOF. The calculations will be performed in the momentum representation, assuming $\hbar:=1$. First, an operator limited to the physical space will be calculated:
\begin{equation}
\hat{\mathrm{R}}_{k,lm}\equiv 
\Big(\delta_{lj}
-\frac{\mathrm{p}_l\mathrm{p}_j}{\mathrm{p}^2}\Big)
\frac{i\partial}{\partial\mathrm{p}_k}
\Big(\delta_{jm}
-\frac{\mathrm{p}_j\mathrm{p}_m}{\mathrm{p}^2}\Big).
\end{equation}
After multiplying the brackets, we get the expression:
\begin{equation} 
\label{wyrazenie}
   \delta_{lm}\frac{i\partial}{\partial\mathrm{p}_k}
-\frac{i\partial}{\partial\mathrm{p}_k}
 \frac{\mathrm{p}_l\mathrm{p}_m}{\mathrm{p}^2}
-\frac{\mathrm{p}_l\mathrm{p}_m}{\mathrm{p}^2}
  \frac{i\partial}{\partial\mathrm{p}_k}
+\frac{\mathrm{p}_l\mathrm{p}_j}{\mathrm{p}^2}
  \frac{i\partial}{\partial\mathrm{p}_k}
  \frac{\mathrm{p}_j\mathrm{p}_m}{\mathrm{p}^2}.
\end{equation}
In the last two terms, we are rearranging the operator of the derivative to the beginning, calculating the necessary derivatives:
\begin{equation} 
-\frac{\mathrm{p}_l\mathrm{p}_m}{\mathrm{p}^2}
  \frac{i\partial}{\partial\mathrm{p}_k}
\equiv
  -\frac{i\partial}{\partial\mathrm{p}_k}
  \frac{\mathrm{p}_l\mathrm{p}_m}{\mathrm{p}^2}
+\frac{i\delta_{kl}\mathrm{p}_m+\mathrm{p}_l i\delta_{km}}{\mathrm{p}^2}
-\frac{2i\mathrm{p}_k\mathrm{p}_l\mathrm{p}_m}{\mathrm{p}^4},
\end{equation}
\begin{equation} 
\label{P0rP0}
\frac{\mathrm{p}_l\mathrm{p}_j}{\mathrm{p}^2}
  \frac{i\partial}{\partial\mathrm{p}_k}
  \frac{\mathrm{p}_j\mathrm{p}_m}{\mathrm{p}^2}
\equiv
  \frac{i\partial}{\partial\mathrm{p}_k}
  \frac{\mathrm{p}_l\mathrm{p}_m}{\mathrm{p}^2}
-\frac{i\delta_{kl}\mathrm{p}_m}{\mathrm{p}^2}
+\frac{i\mathrm{p}_k\mathrm{p}_l\mathrm{p}_m}{\mathrm{p}^4}.
\end{equation}
After inserting expressions for these last two terms into (\ref{wyrazenie}), we get:
\begin{equation}
\label{QI}
\hat{\mathrm{R}}_{k,lm}\equiv 
\hat{\mathrm{r}}_k\delta_{lm}
+\frac{i}{\mathrm{p}^2}\delta_{km}\mathrm{\mathrm{p}}_l
-\hat{\mathrm{r}}_k\frac{\mathrm{p}_l\mathrm{p}_m}{\mathrm{p}^2}
-i\frac{\mathrm{p}_k\mathrm{p}_l\mathrm{p}_m}{\mathrm{p}^4},
\end{equation}
which coincides with the proven form (\ref{Q|}).

We will now demonstrate the identity transversal on two sides of this operator. Right-sided identity transversality will be easier to explicitly demonstrate here:
\begin{equation}
\hat{\mathrm{R}}_{k,lm}\mathrm{p}_m\equiv 
\hat{\mathrm{r}}_k\mathrm{p}_l
+\frac{i}{\mathrm{p}^2}\mathrm{p}_k\mathrm{\mathrm{p}}_l
-\hat{\mathrm{r}}_k\mathrm{p}_l
-i\frac{\mathrm{p}_k\mathrm{p}_l}{\mathrm{p}^2}
\equiv 0.
\end{equation}
This condition means that the considered position operator vanishes on non-physical states. This follows directly from the definition, but above it has been shown {\it explicite} based on the index form. The more important left-sided identity transversal, when computing on indexes, requires transferring to the left side of the non-commutative operator  $\hat{\mathrm{r}}_k=i\partial/\partial\mathrm{p}_k$:
\begin{equation}
\mathrm{p}_l\hat{\mathrm{R}}_{k,lm}\equiv 
\mathrm{p}_m\hat{\mathrm{r}}_k
+i\delta_{km}
-\mathrm{p}_l\hat{\mathrm{r}}_k\frac{\mathrm{p}_l\mathrm{p}_m}{\mathrm{p}^2}
-i\frac{\mathrm{p}_k\mathrm{p}_m}{\mathrm{p}^2}
\equiv 0.
\end{equation}

Let us calculate the Hermitian conjugate of the purely transverse position operator using indices:
\begin{equation}
(\hat{\mathrm{R}}_{k,ml})^{\dagger}\equiv 
\hat{\mathrm{r}}_k\delta_{lm}
-\frac{i}{\mathrm{p}^2}\delta_{kl}\mathrm{\mathrm{p}}_m
-\frac{\mathrm{p}_m\mathrm{p}_l}{\mathrm{p}^2}\hat{\mathrm{r}}_k
+i\frac{\mathrm{p}_k\mathrm{p}_m\mathrm{p}_l}{\mathrm{p}^4},
\end{equation}
which, after transferring the non-commuting operator $\hat{\mathrm{r}}_k$, leads to:
\begin{equation}
(\hat{\mathrm{R}}_{k,ml})^{\dagger}\equiv 
\hat{\mathrm{R}}_{k,ml}.
\end{equation}
This means the identity Hermitian of the operator, which also results directly from the definition.

The form of the full position operator $\mathbf{Q}$ remains to be calculated, which also acts on non-physical modes, but does not mix them with physical modes. This operator, therefore, contains an additional term of longitudinal modes, which in index form can be expressed as follows:
\begin{equation}
\hat{\mathrm{Q}}_{k,ml}\equiv 
\hat{\mathrm{R}}_{k,ml}
+\frac{\mathrm{p}_l\mathrm{p}_j}{\mathrm{p}^2}
  \frac{i\partial}{\partial\mathrm{p}_k}
  \frac{\mathrm{p}_j\mathrm{p}_m}{\mathrm{p}^2}.
\end{equation}
Using the formulas (\ref{P0rP0}), (\ref{QI}) leads to:
\begin{equation}
\hat{\mathrm{Q}}_{k,lm}\equiv 
\hat{\mathrm{r}}_k\delta_{lm}
+\frac{i}{\mathrm{p}^2}\delta_{km}\mathrm{\mathrm{p}}_l
-\frac{i}{\mathrm{p}^2}\delta_{kl}\mathrm{\mathrm{p}}_m.
\end{equation}
This expression for $\hbar:=1$ coincides with (\ref{qk}), so it is also equivalent to (\ref{qq}). Q.E.D.

As we can see, the $\hat{\mathbf{Q}}$ operator operating in an extended space than the $\hat{\mathbf{R}}$ operator turned out to have a simpler form. Among other things, this property constitutes the sense of using such an extended operator, in which the extension does not affect the physical states and significantly simplifies the form of the operator. In this simplification, however, we lose the identity transversality of the operator, limiting ourselves to the transversality of the operator on physical states.

\subsection{3.2. Minimal Form of the Photon Position Operator}

An even more simplified form of the photon position operator can be considered, which will not differ in its operation on physical states. To do this, it is enough to omit the operator terms that disappear on transversal physical states (they contain the index $\mathrm{p}_m$). To obtain such an operator, in the operator $\hat{\mathrm{Q}}_{k,lm}\equiv\hat{\mathrm{q}}_{k,lm}$ just enought omit the last term, and in the operator $\hat{\mathrm{R}}_{k,lm}$ the last two terms, giving:
\begin{equation}
\hat{\mathrm{Q}}_k|_{lm}:\equiv 
\hat{\mathrm{r}}_k\delta_{lm}
+\frac{i\hbar}{\hat{\mathrm{p}}^2}\delta_{km}\mathrm{\hat{\mathrm{p}}}_l,
\end{equation}
which in the positional representation takes the form (in the SI system):
\begin{equation}
\hat{\mathrm{Q}}_k|_{lm}\equiv 
\mathrm{r}_k\delta_{lm}
-\Delta^{-1}\delta_{km}\partial_l.
\end{equation}
Such a minimal form of the photon position operator was given in the master's thesis \cite{Koczan 2002} of the author of the present work. The result was obtained by another method based on covariant quantisation using the Gupta-Bleuler method with a Loretz-type gauge condition. The method turned out to be subtly physically equivalent to the Coulomb gauge. Therefore, we can say that this method's essence lies in the transversality condition, which was encoded in the commutation conditions of the appropriate field operators. In any case, in physical states, this operator is equal to the other operators considered so far:
\begin{equation}
\hat{\mathbf{Q}}|=\hat{\mathbf{Q}}=\hat{\mathbf{q}}=\hat{\mathbf{x}}=\hat{\mathbf{R}}\ \ \ \ \text{on} \ \ \ \ \mathcal{H}_{phys}.
\end{equation}

An identity equivalent to $\hat{\mathbf{Q}}|$ was proposed already in 1977 by Kraus \cite{Kraus} -- however, the author, writing \cite{Koczan 2002}, was not familiar with this study, while the work \cite{Jadczyk} was written earlier, in 1973. In assessing the independence of the works, the widespread belief either in the non-existence of the photon position operator or in problems with this operator should be taken into account. If the works from 1973 and 1977 were better known, the above belief would probably not hold.

\section{4. Definition of the photon position operator as a projection onto helicity eigenspaces}

\subsection{4.1. Riemann--Silberstein Form of Maxwell Equations}

In fact, problems with the relativistic position operator had already arisen in the late 1920s due to the {\it Zitterbewegung} (trembling) problem. This phenomenon was that a free electron described by the Dirac equation would perform microscopic fluctuations. This theoretical effect was described by Schr\"odinger in 1930, who called it {\it Zitterbewegung} \cite{Schrodinger 1930}. Schr\"odinger focused on the electron velocity operator, while later works focused on the position operator (e.g., \cite{Barut}).

The simplest solution to the {\it Zitterbewegung} problem is the division into positive and negative frequencies (energies) described in Davydov's textbook from 1967 \cite{Davydov}. This division covers not only wave functions, but also operators, separating their even and odd parts. The even part of the operator does not mix negative and positive frequencies (electrons with positrons).  Momentum and the Dirac Hamiltonian are even operators, but the usual coordinates $x$, $y$, $z$ are not. This means that a correct definition of the position operator should be an even part of the vector $\mathbf{r}=[x,y,z]^T$.

It is difficult to say who was the first to use such a definition of the position operator and such a solution to the trembling problem. According to Bacry \cite{Bacry 1988}, citing Schr\"odinger's collection of works \cite{Schrodinger 1984}, this was already done by Schr\"odinger himself. This opinion was repeated by the author in the work \cite{Koczan 2002}, although the solution discussed is not found in Schr\"odinger's work from 1930 \cite{Schrodinger 1930}.

Similarly to the Dirac particle position operator, the photon position operator can be approached. However, here the situation becomes a bit more complicated, because instead of division into positive and negative frequencies, we have essentially three eigenvalues of helicity $-1$, 0, and +1, of which zero is an unphysical value. Nevertheless, in this way, Jadczyk and Jancewicz obtained the photon position operator in 1973 \cite{Jadczyk}. The Schrödinger-like method in relation to the photon was also used by the above-mentioned Bacry in 1988 \cite{Bacry 1988}.

Thus, the problem of {\it Zitterbewegung} for the electron has a specific bearing on the issue of describing the position of the photon. This is proven by Silenko's work on the {\it Zitterbewegung} phenomenon for massless particles \cite{Silenko}, which contains an extensive bibliography of the problem. Additionally, the work contains Foldy-Wouthuysen-type transformations for photons. Such a transformation for massless particles can be problematic and requires, for example, a special boundary transition (see \cite{Koczan 2002}).

However, fundamentally, the photon field (electromagnetic field) is real, not complex, so the positive and negative frequency parts of this field do not describe different particles, like electron and positron. Therefore, the method of calculating the even-frequency photon position operator is limited here only to the method of quantising the complex Riemann--Silberstein field $\mathbf{F}=\mathbf{E}+ic\mathbf{B}$. Maxwell's equations written for this field take a form resembling the Schr\"odinger equation with an additional "gauge" type equation (we ignore charges and currents):
\begin{equation}
i\frac{\partial \mathbf{F}}{\partial t}=c\nabla\times\mathbf{F}, \ \ \ \ \nabla\cdot\mathbf{F}=0.
\end{equation}
Instead of the three-dimensional complex Riemann--Silberstein vector, one can also write a similar equation for the six-dimensional vector \cite{Silenko}.

The Riemann-Silberstein field satisfies the wave equation, so it can be decomposed into positive and negative frequencies using auxiliary projective operators:
\begin{equation}
\hat{P}_+ \cong \frac{1}{2}+\frac{i\hbar}{2\hat{\mathrm{p}}c}\frac{\partial}{\partial t},\ \ \ \ 
\hat{P}_- \cong \frac{1}{2}-\frac{i\hbar}{2\hat{\mathrm{p}}c}\frac{\partial}{\partial t},
\end{equation}
where the approximate equality symbol means that this is not yet the final strict identity form of these operators.
Thanks to the field equations, these operators are Hermitian, for a scalar product of the type ($\ref{skalarny}$), and formally meet the criteria of projective operators:
\begin{equation}
\label{rzutowe P+}
\hat{P}^2_+=\hat{P}_+,\ \ \ \hat{P}_-^2=\hat{P}_-,\ \ \  \hat{P}_+\hat{P}_-=\hat{P}_-\hat{P}_+=0
\ \ \ \text{on} \ \ \ \mathcal{H}_{phys}.
\end{equation}
However, these equations do not have an identity character, but refer to operations on Riemann-Silberstein vectors satisfying the field equations. 

Therefore, at this level of description, without additional clarifications, nonphysical states and the superdomain cannot be considered. Consequently, it will not be possible to operate these operators on the nonphysical operator $\mathbf{r}$, which is not a transversal operator even on states. Therefore, let us transform these operators according to the field equation to the form:
\begin{equation}
\label{P+P-}
\hat{P}_{+lm} \cong \frac{1}{2}\delta_{lm}+\frac{i}{2\hat{\mathrm{p}}}\varepsilon_{ljm}\hat{\mathrm{p}}_j,\ \ \ \ 
\hat{P}_{-lm} \cong \frac{1}{2}\delta_{lm}-\frac{i}{2\hat{\mathrm{p}}}\varepsilon_{ljm}\hat{\mathrm{p}}_j.
\end{equation}
Let us square the first of these operators in this form, without taking into account the condition for states:
\begin{equation}
\hat{P}_{+ln}\hat{P}_{+nm} \cong \frac{1}{2}\Big(\delta_{lm}-\frac{\hat{\mathrm{p}}_l\hat{\mathrm{p}}_m}{\hat{\mathrm{p}}^2}\Big)+\frac{i}{2\hat{\mathrm{p}}}\varepsilon_{ljm}\hat{\mathrm{p}}_j.
\end{equation}
We can see that the square generated a correction by replacing the ordinary Kronecker delta with a transversal delta (i.e. the projective operator $\hat{P}$). Let us denote the double of the second term of the above operator by $\hat{\lambda}$ (this is the operator of the helicity sign and at the same time the frequency sign operator in this description). 

We can now provide a more precise definition of projective operators on subspaces of a given frequency sign, identity-correct over the entire space:
\begin{equation}
\label{dobre P+}
\hat{P}_+:\equiv \frac{1}{2}\hat{P}+\frac{1}{2}\hat{\lambda},\ \ \ \ 
\hat{P}_-:\equiv \frac{1}{2}\hat{P}-\frac{1}{2}\hat{\lambda}.
\end{equation}
This form of operators satisfies the identity relation of the type (\ref{rzutowe P+}) over the entire extended space. The sum of these operators is $\hat{P}$, so the projection operator $\hat{P}_0$ onto nonphysical modes has the same form as before (\ref{P0}).

We can now conveniently reinterpret the sign-based frequency projective operators into sign-based helicity projective operators (right-handed, left-handed). Let us note that in the definitions of the projection operators $\hat{P}_+$ and $\hat{P}_-$ there is a helicity operator, i.e. the operator of the projection of the spin vector on the direction of momentum:
\begin{equation}
\label{helicity}
\hat{\Lambda}:\equiv\hbar \hat{\lambda}:\equiv
\frac{1}{\hat{\mathrm{p}}}\hat{\mathbf{p}}\cdot\hat{\mathbf{S}},
\ \ \ \ 
\hat{\Lambda}_{lm}\equiv\hbar \hat{\lambda}_{lm}\equiv\frac{-i\hbar}{\hat{\mathrm{p}}}\hat{\mathrm{p}}_j\varepsilon_{jlm}.
\end{equation}

Such an operator (for a photon) has two physical eigenvalues $+\hbar$, $-\hbar$ and one non-physical eigenvalue 0. The state with eigenvalue $+\hbar$ corresponds to right-handed circular polarisation, and the state with eigenvalue $-\hbar$ is left-handed circular polarisation. We interpret the unphysical state with an eigenvalue of 0 as longitudinal polarisation that does not exist for the photon. It is known that circular polarisation can be composed of two perpendicular linear polarisations in antiphase. Depending on the sign of this composition, we can obtain right-handed or left-handed circular polarisation. Therefore, if both circular polarisations can be obtained from two corresponding linear polarisations, then any linear polarisation can also be obtained by combining the appropriate circular polarisations.

The eigenvalues of helicity result from a simple relation for the square of this operator, which in practice is the identity operator (one) on the states:
\begin{equation}
\label{lambda^2}
\hat{\lambda}^2\equiv(\hat{\Lambda}/\hbar)^2\equiv\hat{P}.
\end{equation}
This property also results in the correctness of projective operators (\ref{dobre P+}). 

At the same time, we can reinterpret these operators from projection operators for positive and negative frequencies to projection operators for right-handed and left-handed helicity:
\begin{equation}
\label{P+helicity}
\hat{P}_R :\equiv \frac{1}{2}\hat{P}+\frac{1}{2}\hat{\lambda}\equiv \hat{P}_+,
\ \ \ \ 
\hat{P}_L :\equiv \frac{1}{2}\hat{P}-\frac{1}{2}\hat{\lambda}\equiv{P}_-.
\end{equation}
This reinterpretation is also a generalisation. The interpretation of frequency signs applies only to the representation of the Riemann--Silberstein vector type, and the helicity interpretation has a general character. 

It can be checked that the operator $\hat{P}_-$ in the form (\ref{dobre P+}) does not zero the part of the vector potential (\ref{A+}) with positive frequency:
\begin{equation}
\hat{P}_-\mathbf{A}_{(+)}=\frac{\mathbf{A}_{(+)}}{2}-\frac{\nabla\times\mathbf{A}_{(+)}}{2\sqrt{-\Delta}}=\frac{\mathbf{A}_{(+)}}{2}-\frac{1}{2\sqrt{-\Delta}}\mathbf{B}_{(+)}\nequiv 0.
\end{equation}
Zeroing would be closer to implementation if, instead of the magnetic field $\mathbf{B}$, there were an electric field $\mathbf{E}$, which is equal to $-\partial\mathbf{A}/\partial t$. Even then, there would be a problem with the sign and the imaginary unit.
Let's see how the analogous zeroing works for the Riemann--Silberstein vector:
\begin{equation}
\hat{P}_-\mathbf{F}_{(+)}=\frac{\mathbf{F}_{(+)}}{2}-\frac{\nabla\times\mathbf{F}_{(+)}}{2\sqrt{-\Delta}}=\frac{\mathbf{F}_{(+)}}{2}-\frac{i/c}{2\sqrt{-\Delta}}\frac{\partial\mathbf{F}_{(+)}}{\partial t}=0.
\end{equation}

Therefore, formally identical projection operators on states with different frequency signs or on states with different helicity eigenvalues differ only in their interpretation and scope of operation. Nevertheless, even with this second interpretation, they constitute identity-correct Hermitian projective operators:
\begin{equation}
\hat{P}^2_R\equiv\hat{P}_R,\ \ \ \hat{P}_L^2\equiv\hat{P}_L,\ \ \  \hat{P}_R\hat{P}_L\equiv\hat{P}_L\hat{P}_R\equiv 0,
\end{equation}
\begin{equation}
\hat{P}_R\hat{P}_0\equiv\hat{P}_0\hat{P}_R\equiv \hat{P}_L\hat{P}_0\equiv\hat{P}_0\hat{P}_L\equiv 0.
\end{equation}

\subsection{4.2. Even Photon Position Operator}

Therefore, it is now possible to define (like Schr\"odinger for the Dirac electron) a photon position operator that does not mix different signs of the electromagnetic field frequency, including the value 0 for non-physical mode:
\begin{equation}
\label{X}
\hat{\mathbf{X}}:\equiv\hat{P}_+\ \hat{\mathbf{r}}\ \hat{P}_++\hat{P}_-\ \hat{\mathbf{r}}\ \hat{P}_-+\hat{P}_0\ \hat{\mathbf{r}}\ \hat{P}_0.
\end{equation}
This definition and its interpretation are based on the representation of the photon wave function by the Riemann--Silberstein vector (modulo special normalisation simplifying the dot product to the form (\ref{skalarny})). 

However, from this point on, it will be treated formally more generally and will be reinterpreted in the language of helicity eigenvalues instead of frequency signs. So we can define a photon position operator that does not mix the helicity eigenstates:
\begin{equation}
\label{Y}
\hat{\mathbf{Y}}:\equiv\hat{P}_R\ \hat{\mathbf{r}}\ \hat{P}_R+\hat{P}_L\ \hat{\mathbf{r}}\ \hat{P}_L+\hat{P}_0\ \hat{\mathbf{r}}\ \hat{P}_0.
\end{equation}
It is also worth considering the purely physical part of this operator:
\begin{equation}
\label{Z}
\hat{\mathbf{Z}}:\equiv\hat{P}_R\ \hat{\mathbf{r}}\ \hat{P}_R+\hat{P}_L\ \hat{\mathbf{r}}\ \hat{P}_L.
\end{equation}

It turns out that these new photon position operator formulas based on helicity or frequency signs still define the same photon position operator.
\\
\\
STATEMENT 3 (A photon position operator that does not mix helicity)

\textit{The photon position operators (\ref{X}) and (\ref{Y}) that do not mix the frequency or helicity signs are identical to the operator (\ref{QPP}) that does not mix transversal and longitudinal states (the same Pryce--Bacry and Born--Infeld operators for a photon):
\begin{equation}
\hat{\mathbf{X}}\equiv\hat{\mathbf{Y}}\equiv\hat{\mathbf{Q}}\equiv\hat{\mathbf{x}}\equiv \hat{\mathbf{r}}+
\frac{\hat{\mathbf{p}}\times\hat{\mathbf{S}}}{\hat{\mathrm{p}}^2}\equiv \hat{\mathbf{q}}.
\end{equation}
Also, the position operator projected onto the two subspaces of the physical eigenmodes of the helicity (\ref{Z}) is identity-equal to the position operator projected onto the physical subspace:
\begin{equation}
\hat{\mathbf{Z}}\equiv\hat{\mathbf{R}}.
\end{equation}
Therefore, all considered photon position operators (does not apply to ordinary coordinates $\mathbf{r}$) are equal in the subspace of physical states:}
\begin{equation}
\hat{\mathbf{X}}=\hat{\mathbf{Y}}=\hat{\mathbf{Z}}=\hat{\mathbf{R}}=\hat{\mathbf{Q}}=\hat{\mathbf{Q}}|=\hat{\mathbf{x}}=\hat{\mathbf{q}}\ \ \ \ \text{on} \ \ \ \ \mathcal{H}_{phys}.
\end{equation}
\\
PROOF. In the proof, we assume that $\hbar:=1$. Using the definition (\ref{Z}) and the formulas (\ref{P+helicity}), we transform the position operator projected into two physical eigensubspaces of helicity:
\begin{equation}
\label{Z proof}
\hat{\mathbf{Z}}\equiv\frac{1}{2}\hat{P}\ \hat{\mathbf{r}}\ \hat{P}+
\frac{1}{2}\hat{\lambda}\ \hat{\mathbf{r}}\ \hat{\lambda}\equiv
\frac{1}{2}\hat{\mathbf{R}}+
\frac{1}{2}\hat{\lambda}\ \hat{\mathbf{r}}\ \hat{\lambda}.
\end{equation}
So just focus on doubling the last expression (for $k$ component):
\begin{equation}
(\hat{\lambda}\ \hat{\mathrm{r}}_k\ \hat{\lambda})_{lm}\equiv
-\frac{\mathrm{p}_j}{\mathrm{p}}\varepsilon_{jln}\hat{\mathrm{r}}_k\frac{\mathrm{p}_s}{\mathrm{p}}\varepsilon_{snm}\equiv
\frac{\mathrm{p}_j}{\mathrm{p}}\hat{\mathrm{r}}_k\frac{\mathrm{p}_s}{\mathrm{p}}(\delta_{js}\delta_{lm}-\delta_{jm}\delta_{ls}),
\end{equation}
\begin{equation}
(\hat{\lambda}\ \hat{\mathrm{r}}_k\ \hat{\lambda})_{lm}\equiv
\frac{\mathrm{p}_j}{\mathrm{p}}\hat{\mathrm{r}}_k\frac{\mathrm{p}_j}{\mathrm{p}}\delta_{lm}
-\frac{\mathrm{p}_m}{\mathrm{p}}\hat{\mathrm{r}}_k\frac{\mathrm{p}_l}{\mathrm{p}}.
\end{equation}
We shift the $\hat{\mathrm{r}}_k=i\partial/\partial \mathrm{p}_k$ operator to the left, taking into account the commutation derivatives:
\begin{equation}
(\hat{\lambda}\ \hat{\mathrm{r}}_k\ \hat{\lambda})_{lm}\equiv
\hat{\mathrm{r}}_k\hat{P}_{lm}
-i\frac{\partial}{\partial \mathrm{p}_k}\Big(\frac{\mathrm{p}_j}{\mathrm{p}}\Big)\frac{\mathrm{p}_j}{\mathrm{p}}\delta_{lm}
+i\frac{\partial}{\partial \mathrm{p}_k}\Big(\frac{\mathrm{p}_m}{\mathrm{p}}\Big)\frac{\mathrm{p}_l}{\mathrm{p}}.
\end{equation}
It remains to calculate the derivatives (the first one reduces after narrowing):
\begin{equation}
(\hat{\lambda}\ \hat{\mathrm{r}}_k\ \hat{\lambda})_{lm}\equiv
\hat{\mathrm{r}}_k\hat{P}_{lm}
+i\frac{\mathrm{p}_l}{\mathrm{p}^2}\delta_{km}
-i\frac{\mathrm{p}_k\mathrm{p}_l\mathrm{p}_m}{\mathrm{p}^4},
\end{equation}
which coincides with (\ref{QI}), so also by (\ref{Z proof}) we have:
\begin{equation}
(\hat{\lambda}\ \hat{\mathrm{r}}_k\ \hat{\lambda})_{lm}\equiv
\hat{\mathrm{R}}_{k,lm}\equiv
\hat{\mathrm{Z}}_{k,lm}.
\end{equation}
The above operator equality entails the following:
\begin{equation}
\hat{\mathbf{Y}}\equiv
\hat{\mathbf{Z}}+\hat{P}_0\ \hat{\mathbf{r}}\ \hat{P}_0\equiv
\hat{\mathbf{R}}+\hat{P}_0\ \hat{\mathbf{r}}\ \hat{P}_0\equiv
\hat{\mathbf{Q}},
\end{equation}
which is also equal to $\hat{\mathbf{x}}$ and $\hat{\mathbf{q}}$ (the Born--Infeld operator for the photon), as well as the operator $\hat{\mathbf{X}}$, which from $\hat{\mathbf{Y }}$ differs only in interpretation and markings in the mathematically identical definition. Q.E.D. 

During the proof of the statement, another potential definition of the photon position operator, identity-equivalent in the domain of physical states, emerged:
\begin{equation}
\hat{\mathbf{\Theta}}:\equiv
\hat{\lambda}\hat{\mathbf{r}}\hat{\lambda}\equiv
\hat{\mathbf{R}}\equiv
\hat{\mathbf{Z}},
\end{equation}
which just shows the importance of the helicity operator for the concept of the photon position operator. Therefore, the operator defined above (or rather the definition form) will be called the helical position operator. 

\subsection{4.3. Convenient Properties of the Photon Position Operator}

At the same time, the following is met:
\\
\\
LEMMA 1 (Commutativity of the position and helicity operators)

\textit{The helicity operator identity commutes with the Cartesian photon position operator, both those defined in the extended space and the physical space:}
\begin{equation}
[\hat{\mathbf{Q}},\hat{\lambda}]\equiv 0,\ \ \ \ [\hat{\mathbf{R}},\hat{\lambda}]\equiv 0.
\end{equation}
\\
PROOF. The lemma basically follows from the commutativity of helicity with the generators of the Poincaré group, which can be used to define the photon position operator. Nevertheless, it is worth proving the lemma directly on the indices. In addition, an independent, shorter synthetic proof will be provided at the end.

Based on the definition of $\hat{\mathbf{R}}$ and the relation (\ref{lambda^2}), the second commutator of the lemma can be transformed:
\begin{equation}
[\hat{\mathbf{R}},\hat{\lambda}]\equiv [\hat{\lambda}^2\hat{\mathbf{r}}\hat{\lambda}^2,\hat{\lambda}]\equiv
\hat{\lambda}^2[\hat{\mathbf{r}},\hat{\lambda}]\hat{\lambda}^2.
\end{equation}
Therefore, it is enough to check the commutator of ordinary coordinates with the helicity operator (let $\hbar:=1$):
\begin{equation}
[\hat{\mathrm{r}}_k,\hat{\lambda}]_{lm}\equiv
i\frac{\partial}{\partial \mathrm{p}_k}
\Big(-i\frac{\mathrm{p}_j}{\mathrm{p}}\varepsilon_{jlm}\Big)\equiv
\frac{1}{\mathrm{p}}\varepsilon_{klm}
-\frac{\mathrm{p}_k\mathrm{p}_j}{\mathrm{p}^3}\varepsilon_{jlm}.
\end{equation}
Let's multiply this on the left by the helicity square:
\begin{equation}
\hat{\lambda}^2_{nl}[\hat{\mathrm{r}}_k,\hat{\lambda}]_{lm}\equiv
\Big(\delta_{nl}-\frac{\mathrm{p}_n\mathrm{p}_l}{\mathrm{p}^2}\Big)
\Big(\frac{1}{\mathrm{p}}\varepsilon_{klm}
-\frac{\mathrm{p}_k\mathrm{p}_j}{\mathrm{p}^3}\varepsilon_{jlm}\Big),
\end{equation}
\begin{equation}
\hat{\lambda}^2_{nl}[\hat{\mathrm{r}}_k,\hat{\lambda}]_{lm}\equiv
\frac{1}{\mathrm{p}}\varepsilon_{knm}
-\frac{\mathrm{p}_k\mathrm{p}_j}{\mathrm{p}^3}\varepsilon_{jnm}
-\frac{\mathrm{p}_n\mathrm{p}_l}{\mathrm{p}^3}\varepsilon_{klm}.
\end{equation}
Now let's multiply this on the right side by the helicity square:
\begin{equation}
\Big(\frac{1}{\mathrm{p}}\varepsilon_{knm}
-\frac{\mathrm{p}_k\mathrm{p}_j}{\mathrm{p}^3}\varepsilon_{jnm}
-\frac{\mathrm{p}_n\mathrm{p}_l}{\mathrm{p}^3}\varepsilon_{klm}\Big)
\Big(\delta_{ms}-\frac{\mathrm{p}_m\mathrm{p}_s}{\mathrm{p}^2}\Big),
\end{equation}
which, after multiplication, gives:
\begin{equation}
[\hat{\mathrm{R}}_k,\hat{\lambda}]_{ns}\equiv
\frac{1}{\mathrm{p}}\varepsilon_{kns}
-\frac{\mathrm{p}_k\mathrm{p}_j}{\mathrm{p}^3}\varepsilon_{jns}
-\frac{\mathrm{p}_n\mathrm{p}_l}{\mathrm{p}^3}\varepsilon_{kls}
-\frac{\mathrm{p}_s\mathrm{p}_m}{\mathrm{p}^3}\varepsilon_{knm}.
\end{equation}
The right side vanishes identity-wise based on some non-trivial identity for the Levi-Civita tensor (for $\mathrm{p}\neq 0$). However, this must be checked for all matrix components, e.g. for $\hat{\mathrm{R}}_1$:
\begin{equation}
[\hat{\mathrm{R}}_1,\hat{\lambda}]_{11}\equiv
0
-0
-0
-0\equiv 0,
\end{equation}
\begin{equation}
[\hat{\mathrm{R}}_1,\hat{\lambda}]_{12}\equiv
0
-\frac{\mathrm{p}_1\mathrm{p}_3}{\mathrm{p}^3}\varepsilon_{312}
-\frac{\mathrm{p}_1\mathrm{p}_3}{\mathrm{p}^3}\varepsilon_{132}
-0\equiv 0,
\end{equation}
\begin{equation}
[\hat{\mathrm{R}}_1,\hat{\lambda}]_{13}\equiv
0
-\frac{\mathrm{p}_1\mathrm{p}_2}{\mathrm{p}^3}\varepsilon_{213}
-\frac{\mathrm{p}_1\mathrm{p}_2}{\mathrm{p}^3}\varepsilon_{123}
-0\equiv 0,
\end{equation}
\begin{equation}
[\hat{\mathrm{R}}_1,\hat{\lambda}]_{21}\equiv
0
-\frac{\mathrm{p}_1\mathrm{p}_3}{\mathrm{p}^3}\varepsilon_{321}
-0
-\frac{\mathrm{p}_1\mathrm{p}_3}{\mathrm{p}^3}\varepsilon_{123}
\equiv 0,
\end{equation}
\begin{equation}
[\hat{\mathrm{R}}_1,\hat{\lambda}]_{22}\equiv
0
-0
-\frac{\mathrm{p}_2\mathrm{p}_3}{\mathrm{p}^3}\varepsilon_{132}
-\frac{\mathrm{p}_2\mathrm{p}_3}{\mathrm{p}^3}\varepsilon_{123}
\equiv 0,
\end{equation}
\begin{equation}
[\hat{\mathrm{R}}_1,\hat{\lambda}]_{23}\equiv
\frac{1}{\mathrm{p}}
-\frac{\mathrm{p}_1\mathrm{p}_1}{\mathrm{p}^3}\varepsilon_{123}
-\frac{\mathrm{p}_2\mathrm{p}_2}{\mathrm{p}^3}\varepsilon_{123}
-\frac{\mathrm{p}_3\mathrm{p}_3}{\mathrm{p}^3}\varepsilon_{123}
\equiv 0,
\end{equation}
\begin{equation}
[\hat{\mathrm{R}}_1,\hat{\lambda}]_{31}\equiv
0
-\frac{\mathrm{p}_1\mathrm{p}_2}{\mathrm{p}^3}\varepsilon_{231}
-0
-\frac{\mathrm{p}_1\mathrm{p}_2}{\mathrm{p}^3}\varepsilon_{132}
\equiv 0,
\end{equation}
\begin{equation}
[\hat{\mathrm{R}}_1,\hat{\lambda}]_{32}\equiv
\frac{-1}{\mathrm{p}}
-\frac{\mathrm{p}_1\mathrm{p}_1}{\mathrm{p}^3}\varepsilon_{132}
-\frac{\mathrm{p}_3\mathrm{p}_3}{\mathrm{p}^3}\varepsilon_{132}
-\frac{\mathrm{p}_2\mathrm{p}_2}{\mathrm{p}^3}\varepsilon_{132}
\equiv 0,
\end{equation}
\begin{equation}
[\hat{\mathrm{R}}_1,\hat{\lambda}]_{33}\equiv
0
-0
-\frac{\mathrm{p}_3\mathrm{p}_2}{\mathrm{p}^3}\varepsilon_{123}
-\frac{\mathrm{p}_3\mathrm{p}_2}{\mathrm{p}^3}\varepsilon_{132}
\equiv 0.
\end{equation}
Similarly, the commutators for the $\hat{\mathrm{R}}_2$ and $\hat{\mathrm{R}}_3$ components will be equal zero, because the $\hat{\mathrm{R}}_1$ component was not distinguished.

What remains to be calculated is the helicity commutator with the position operator $\hat{\mathbf{Q}}$ defined over the entire extended space. This operator differs from the $\hat{\mathbf{R}}$ operator only in its projection onto the subspace of zero mode, which disappears under the action of helicity:
\begin{equation}
\hat{\lambda}\hat{P}_0\equiv\hat{P}_0\hat{\lambda}\equiv
(1-\hat{\lambda}^2)\hat{\lambda}\equiv 
\hat{\lambda}-\hat{\lambda}^3\equiv 0.
\end{equation}
The above equation results from the zero eigenvalue of helicity for the zero mode. However, it is worth confirming this equality with direct calculation:
\begin{equation}
\hat{\lambda}^3_{lm}\equiv \hat{P}_{ln} \hat{\lambda}_{nm}
\equiv
\Big(\delta_{ln}-\frac{\mathrm{p}_l\mathrm{p}_n}{\mathrm{p}^2}\Big)
\Big(-i\frac{\mathrm{p}_s}{\mathrm{p}}\varepsilon_{snm}\Big),
\end{equation}
\begin{equation}
\label{lambda^3}
\hat{\lambda}^3_{lm}\equiv 
-i\frac{\mathrm{p}_s}{\mathrm{p}}\varepsilon_{slm}\equiv
\hat{\lambda}_{lm}.
\end{equation}
Therefore:
\begin{equation}
[\hat{P}_0\hat{\mathbf{r}}\hat{P}_0,\hat{\lambda}]\equiv 
\hat{P}_0\hat{\mathbf{r}}\hat{P}_0\hat{\lambda}
-\hat{\lambda}\hat{P}_0\hat{\mathbf{r}}\hat{P}_0\equiv 0-0
\equiv 0.
\end{equation}
This means that:
\begin{equation}
[\hat{\mathbf{Q}},\hat{\lambda}]\equiv
[\hat{\mathbf{R}}+\hat{P}_0\hat{\mathbf{r}}\hat{P}_0,\hat{\lambda}]\equiv 
0+0\equiv 0,
\end{equation}
according to the main thesis of the lemma. 

The proof of the lemma can be performed much more simply on the symbols of projective operators without using indices. To do this, just use a simple relation resulting directly from (\ref{P+helicity}):
\begin{equation}
\hat{\lambda}\equiv
\hat{P}_R-\hat{P}_L.
\end{equation}
Using (\ref{Y}) we calculate the first expression of the commutator with the position operator:
\begin{equation}
\hat{\mathbf{Q}}\hat{P}_R\equiv
\hat{\mathbf{Y}}\hat{P}_R\equiv
\big(\hat{P}_R\ \hat{\mathbf{r}}\ \hat{P}_R+\hat{P}_L\ \hat{\mathbf{r}}\ \hat{P}_L+\hat{P}_0\ \hat{\mathbf{r}}\ \hat{P}_0\big)\hat{P}_R,
\end{equation}
\begin{equation}
\hat{\mathbf{Q}}\hat{P}_R\equiv
\hat{P}_R\ \hat{\mathbf{r}}\ \hat{P}_R,
\end{equation}
where it was enough to use the basic properties of projection operators onto orthogonal eigensubspaces. The expression in reverse order looks analogous:
\begin{equation}
\hat{P}_R\hat{\mathbf{Q}}\equiv
\hat{P}_R\big(\hat{P}_R\ \hat{\mathbf{r}}\ \hat{P}_R+\hat{P}_L\ \hat{\mathbf{r}}\ \hat{P}_L+\hat{P}_0\ \hat{\mathbf{r}}\ \hat{P}_0\big)\equiv
\hat{P}_R\ \hat{\mathbf{r}}\ \hat{P}_R.
\end{equation}
Therefore, we have that:
\begin{equation}
[\hat{\mathbf{Q}},\hat{P}_R]\equiv
\hat{\mathbf{Q}}\hat{P}_R-\hat{P}_R\hat{\mathbf{Q}}\equiv
\hat{P}_R\ \hat{\mathbf{r}}\ \hat{P}_R-\hat{P}_R\ \hat{\mathbf{r}}\ \hat{P}_R\equiv 0.
\end{equation}
In the full analogy, the commutator with the projection operator onto left-handed photons also disappears:
\begin{equation}
[\hat{\mathbf{Q}},\hat{P}_L]\equiv 0.
\end{equation}
Combining the above results, we finally have:
\begin{equation}
[\hat{\mathbf{Q}},\hat{\lambda}]\equiv 
[\hat{\mathbf{Q}},\hat{P}_R-\hat{P}_L]\equiv 0-0\equiv 0,
\end{equation}
which is the thesis of the lemma demonstrated in the second way. Q.E.D.

The lemma leads to an essential and unexpected conclusion that helicity not only does not depreciate the position operator, but is consistent with it. This even suggests the existence of common eigenstates of the helicity and the position operator component. Bacry even provided such states \cite{Bacry 1988, Bacry 2}. However, in the opinion of some scholars, such a situation was impossible. For example, even Kraus claimed that the photon position operator does not commute with helicity \cite{Kraus}.
\\
\\
THEOREM 1 (Matrix element of photon position operator)

\textit{Although the photon position operator is quite a complicated operator, when applied to the nominal scalar product (\ref{skalarny}) of transversal wave functions, it takes the form of multiplication by ordinary coordinates $\mathrm{r}_k$:
\begin{equation}
\label{matrix element}
(\Phi |\hat{\mathrm{Q}}_k\Psi)
=(\Phi |\hat{\mathrm{R}}_k\Psi)
=
\int{\Phi_l^*(\mathbf{r})\  \mathrm{r}_k\Psi_l(\mathbf{r})d^3\mathrm{r}}.
\end{equation}
This applies even more to the minimal form of the photon position operator:}
\begin{equation}
\label{matrix element Q|}
(\Phi |\hat{\mathrm{Q}}_{k|}\Psi)
=
\int{\Phi_l^*(\mathbf{r})\  \mathrm{r}_k\Psi_l(\mathbf{r})d^3\mathrm{r}}.
\end{equation}
\\
PROOF. According to the definition of the photon position operator, its matrix element can be written as follows:
\begin{equation}
(\Phi |\hat{\mathrm{Q}}_k\Psi)
=
\int{\Phi_l^*\cdot\big((\hat{P}\  \mathrm{r}_k \hat{P})_{lm}+(\hat{P}_0\  \mathrm{r}_k \hat{P}_0)_{lm}\big)\Psi_m d^3\mathrm{r}}.
\end{equation}
The conditions for state transversality have the form:
\begin{equation}
\hat{P}_0\mathbf{\Psi}=0,\ \ \ \ 
\hat{P}\mathbf{\Psi}=\mathbf{\Psi}, \ \ \ \
\hat{P}^{\dagger}\mathbf{\Phi}=\mathbf{\Phi},
\end{equation}
from which follows the form (\ref{matrix element}) of the matrix element for $\hat{\mathbf{Q}}$, and even more so for $\hat{\mathbf{R}}$, which does not contain zero mode.

The case of the minimal form of the position operator requires direct proof. The matrix element of this operator contains only one additional element of the form:
\begin{equation}
\int{\Phi_l^*(\mathbf{r})\frac{\partial_l}{-\Delta}\Psi_k(\mathbf{r})d^3\mathrm{r}}=
\int{\partial_l\big(\Phi_l^*(\mathbf{r})\big)\frac{1}{\Delta}\Psi_k(\mathbf{r})d^3\mathrm{r}}=0,
\end{equation}
where the Hermitian nature of the $i\partial_l$ operator and the transversality condition were used. This last condition is provided by the form (\ref{matrix element Q|}). Q.E.D.

This, in fact, is a simple theorem, resulting from a good definition, and is intended to show that, contrary to appearances, the photon position operator is a natural observable closely related to the used geometric coordinates $\mathrm{r}_k$. In other words, the complications with the concept of the photon position operator result from the properties of the photon and its wave function, and not from the problems of the existence or non-existence of the photon position operator.

However, it is worth knowing that quantum mechanics has more difficulties than classical mechanics. For example, the components of the photon position operator are not commutative -- just like the components of the angular momentum operator. Moreover, Theorem 1 does not transfer to quadratic quantities $\hat{\mathrm{Q}}_1^2$ or $\hat{\mathrm{Q}}_1\hat{\mathrm{Q}}_2$, which cannot be converted into $x^2$ or $xy$ in the dot product.

%******
%Użycie splitu
%\begin{equation}
%\label{lambda}
%\begin{split}
%\lambda=\frac{\mathrm{u}-\mathrm{v}}{(1-\frac{\mathrm{uv}}{c^2})(\gamma_{\mathrm{u}}%\mathrm{u}-\gamma_{\mathrm{v}}\mathrm{v})}
%=\frac{(\mathrm{u}-\mathrm{v})(\gamma_{\mathrm{u}}^{-1}+\gamma_{\mathrm{v}}^{-1})}{(1-\frac{\mathrm{uv}}{c^2})(\mathrm{u}-\mathrm{v}+\frac{\gamma_{\mathrm{u}}}{\gamma_{\mathrm{v}}}\mathrm{u}-\frac{\gamma_{\mathrm{v}}}{\gamma_{\mathrm{u}}}\mathrm{v})}=\\
%=\frac{\gamma_{\mathrm{u}}^{-1}+\gamma_{\mathrm{v}}^{-1}}{(1-\frac{\mathrm{uv}}{c^2})(1+\gamma_{\mathrm{u}}\gamma_{\mathrm{v}}(1+\frac{\mathrm{uv}}{c^2}))}=
%\frac{\gamma_{\mathrm{u}}^{-1}+\gamma_{\mathrm{v}}^{-1}}{1-\frac{\mathrm{uv}}{c^2}+\gamma_{\mathrm{u}}\gamma_{\mathrm{v}}(1-\frac{\mathrm{u^2v^2}}{c^2})}.
%\end{split}
%\end{equation}
%***** 

\section{5. Radial component of the photon position operator}

\subsection{5.1. Some Definitions of the Radial Position Operators}

The issue of the radial component of the photon position operator is more problematic than that of the Cartesian components. From a formal point of view, only the Cartesian components operators $\hat{\mathrm{Q}}_1$, $\hat{\mathrm{Q}}_2$, $\hat{\mathrm{Q}}_3$ have been defined so far, while the radial component operator has not been defined yet here. 

In quantum mechanics, the definition of the curvilinear component does not have to be as unambiguous as in classical mechanics. For the sake of clarity, we will focus on the radial axis component $\rho=\sqrt{x^2+y^2}$ in the sense of a cylindrical coordinate, rather than the radial component $r=\sqrt{x^2+y^2+z^2}$ as a spherical coordinate. 

Considering the previous definitions, the following transversal type operator seems to be the simplest (transversal radial operator):
\begin{equation}
\label{Qrho}
    \hat{\mathrm{Q}}_{\rho}:\equiv
    \hat{P}\rho\hat{P}+\hat{P}_0\rho\hat{P}_0,
\end{equation}
the use of zero term here is, as before, of a formal nature. A similar operator, but for $\rho^2$ (see below), was defined in \cite{Jadczyk}, but these operators do not have to be equivalent up to the square or square root. 

The use of the operators $\hat{P}_R$ and $\hat{P}_L$ would give analogous formula (even radial operator):
\begin{equation}
    \hat{\mathrm{Y}}_{\rho}:\equiv
    \hat{P}_R\rho\hat{P}_R+\hat{P}_L\rho\hat{P}_L+\hat{P}_0\rho\hat{P}_0.
\end{equation}
A radial operator can also be defined, like a Cartesian operator, using helicity (helical radial operator):
\begin{equation}
    \hat{\mathrm{\Theta}}_{\rho}:\equiv
    \hat{\lambda}\rho\hat{\lambda}.
\end{equation}
For simplicity, the unphysical zero mode is omitted here. Up to the omitted zero modes, the second radial operator introduced is equal to the arithmetic mean of the first and third radial operators:
\begin{equation}
\label{radial relation}
\hat{\mathrm{Y}}_{\rho}\equiv
\frac{1}{2}
(\hat{\mathrm{Q}}_{\rho}+\hat{\mathrm{\Theta}}_{\rho})+
    \frac{1}{2}\hat{P}_0\rho\hat{P}_0.
\end{equation}

According to the first concept of this paper, one can also use the (radial) component of the boost generator:
\begin{equation}
\hat{\mathrm{q}}_{\rho}:\equiv
\frac{1}{\sqrt{\hat{\mathrm{p}}c}}\hat{\mathrm{N}}_{[\rho]}
\frac{1}{\sqrt{\hat{\mathrm{p}}c}},
\end{equation}
where the following notation is used:
\begin{equation}
\label{Nrho}
    \hat{\mathrm{N}}_{[\rho]}:\equiv
    \sqrt{\hat{\mathrm{N}}_1^2+\hat{\mathrm{N}}_2^2}.
\end{equation}
In a similar way, one could easily define the radial component of the position operator based on its Cartesian components:
\begin{equation}
\label{Q[rho]}
    \hat{\mathrm{Q}}_{[\rho]}:\equiv
    \sqrt{\hat{\mathrm{Q}}_1^2+\hat{\mathrm{Q}}_2^2}\equiv:
     \hat{\mathrm{Y}}_{[\rho]}\equiv:
      \hat{\mathrm{q}}_{[\rho]}.
\end{equation}
Although this method gives a consistent result for all three consistent ways of defining the Cartesian components, the method is difficult, and it is not certain that it is valid in quantum mechanics.

At the end of defining radial operators, we can give the previously mentioned definition with the square (square radial operator):
\begin{equation}
    \hat{\mathrm{Q}}^2_{\rho^2}:\equiv
    \hat{P}\rho^2\hat{P},
\end{equation}
but it is unknown whether such an operator is actually the square of another operator or, in other words, whether it has a root. In particular, the square radial operator is not equal to the square of the transversal radial operator, nor the square of the rooter radial operator:
\begin{equation}
    \hat{\mathrm{Q}}^2_{\rho^2}\neq
    (\hat{\mathrm{Q}}_{\rho})^2, \ \ \ \
    \hat{\mathrm{Q}}^2_{\rho^2}\neq
    (\hat{\mathrm{Q}}_{[\rho]})^2.
\end{equation}
These properties follow from writing out the definitions,
but it remains to be shown that the differently looking expressions are not equal (not directly proven here, but the second relation will emerge further). The square radial operator does not literally define the radial component operator.

\subsection{5.2. Simple Properties of Radial Position Operators}

Therefore, we have as many as five definitions of the radial position operator ($\hat{\mathrm{Q}}_{\rho}$, $\hat{\mathrm{Y}}_{\rho}$, $\hat{\mathrm{\Theta}}_{\rho}$, $\hat{\mathrm{q}}_{\rho}$, $\hat{\mathrm{Q}}_{[\rho]}$) and we do not know the general relations between them.
Unfortunately, this time the definitions introduced are not the same, neither in terms of identity, nor even in terms of physical states (this applies, for example, to the first and last operators -- what will be deduced next):
\begin{equation}
\label{neq1}
\hat{\mathrm{Q}}_{\rho}\nequiv
    \hat{\mathrm{Q}}_{[\rho]}, \: \: \: \:
    \hat{\mathrm{Q}}_{\rho}\neq
    \hat{\mathrm{Q}}_{[\rho]}.
\end{equation}

In the rest of this paper we will mainly use operators $\hat{\mathrm{Q}}_{\rho}$ and $\hat{\mathrm{\Theta}}_{\rho}$, but now we will also examine operator $\hat{\mathrm{Y}}_{\rho}$.
\\
\\
LEMMA 2 (Commutativity of some radial positions and helicity)

\textit{The helicity operator identity commutes with the three radial position operators (excluding the noncommutative operators $\hat{\mathrm{Q}}_{\rho}$ and $\hat{\mathrm{\Theta}}_{\rho}$ -- what will be deduced next):}
\begin{equation}
[\hat{\mathrm{Y}}_{\rho},\hat{\lambda}]\equiv 0,\ \ \ \ [\hat{\mathrm{q}}_{\rho},\hat{\lambda}]\equiv 0,\ \ \ \ [\hat{\mathrm{Q}}_{[\rho]},\hat{\lambda}] \equiv 0.
\end{equation}
\\
PROOF. In the case of the first operator, it is enough to use the equality $\hat{\lambda}\equiv\hat{P}_R-\hat{P}_L$ and check the commutativity with the terms of the difference separately:
\begin{equation}
\hat{\mathrm{Y}}_{\rho}\hat{P}_R\equiv
\hat{P}_R\ \rho \hat{P}^2_R+0\equiv
\hat{P}_R\ \rho \hat{P}_R,
\end{equation}
\begin{equation}
\hat{P}_R\hat{\mathrm{Y}}_{\rho}\equiv
\hat{P}^2_R\ \rho \hat{P}_R+0\equiv
\hat{P}_R\ \rho \hat{P}_R,
\end{equation}
than:
\begin{equation}
[\hat{\mathrm{Y}}_{\rho},\hat{P}_R]\equiv 
\hat{\mathrm{Y}}_{\rho}\hat{P}_R-\hat{P}_R\hat{\mathrm{Y}}_{\rho}\equiv
0.
\end{equation}
Similarly, for $\hat{P}_L$, which completes the proof of the commutativity of the first radial position operator.

The second radial position operator is defined in terms of the components of the boost generator and the energy operator, which is also a Poincaré generator. The helicity operator commutes with the Poincaré generators, so it also commutes with the second radial operator.

The third radial position operator is defined in terms of the components of the Cartesian position operator. The components of the Cartesian position operator commute with helicity by Lemma 1. Therefore, the third radial operator must also commute with helicity. Q.E.D.

The above lemma gives some hints about the photon position operator and its eigenstates. First, it shows that the three defined radial operators have a common property. Second, the lemma suggests that there can be common eigenstates of the radial operator and the helicity operator. Indeed, such states on the circle were given in the work of Jadczyk and Jancewicz \cite{Jadczyk}. However, one of the authors of this work (A. Jadczyk) claims in private correspondence (and on the blog -- see further) that there is some error in this work. Namely, an eigenstate of a position on a circle with a defined helicity would contradict Galindo--Amrein theorem about the non-existence of such localised states \cite{Galindo, Amrein}.

In this work, the helicity eigenstates are not studied in a particular way, but a second mode of position eigenstates (magnetic string) is introduced. It is possible that a helicity eigenstate can be constructed from the first mode (electric string) and the second mode (magnetic string). Moreover, this work suggests an alternative probability density representation suitable for a magnetic string. Thus, it casts doubt on the assumptions of Galindo--Amrein theorem.

From the point of view of physics, more important than commutation with helicity is commutation with an operator projective onto physical states. This condition is satisfied by all introduced radial operators, according to:
\\
\\
STATEMENT 4 (On the physicality of radial operators)

\textit{All five defined radial operators do not mix physical and non-physical states (and vice versa), i.e. they commute with the projective operator onto physical states:}
\begin{equation}
[\hat{\mathrm{Q}}_{\rho},\hat{P}]\equiv 0,\ \  
[\hat{\mathrm{Y}}_{\rho},\hat{P}]\equiv 0,\ \   [\hat{\mathrm{\Theta}}_{\rho},\hat{P}]\equiv 0,
\end{equation}
\begin{equation}
   [\hat{\mathrm{q}}_{\rho},\hat{P}]\equiv 0,\ \   [\hat{\mathrm{Q}}_{[\rho]},\hat{P}] \equiv 0.
\end{equation}
\\
PROOF. The proof for the first two radial operators proceeds similarly to the proof of Lemma 2:
\begin{equation}
\hat{\mathrm{Q}}_{\rho}\hat{P}\equiv
\hat{P}\ \rho \hat{P}^2+0\equiv
\hat{P}\ \rho \hat{P},
\end{equation}
\begin{equation}
\hat{P}\hat{\mathrm{Q}}_{\rho}\equiv
\hat{P}^2\ \rho \hat{P}+0\equiv
\hat{P}\ \rho \hat{P},
\end{equation}
than:
\begin{equation}
[\hat{\mathrm{Q}}_{\rho},\hat{P}]\equiv 
\hat{\mathrm{Q}}_{\rho}\hat{P}-\hat{P}\hat{\mathrm{Q}}_{\rho}\equiv
0;
\end{equation}
\begin{equation}
\hat{\mathrm{Y}}_{\rho}\hat{P}\equiv
\hat{P}_R\ \rho \hat{P}^2_R+\hat{P}_L\ \rho \hat{P}^2_L+0\equiv
\hat{P}_R\ \rho \hat{P}_R+\hat{P}_L\ \rho \hat{P}_L,
\end{equation}
\begin{equation}
\hat{P}\hat{\mathrm{Y}}_{\rho}\equiv
\hat{P}^2_R\ \rho \hat{P}_R+\hat{P}^2_L\ \rho \hat{P}_L+0\equiv
\hat{P}_R\ \rho \hat{P}_R+\hat{P}_L\ \rho \hat{P}_L,
\end{equation}
than:
\begin{equation}
[\hat{\mathrm{Y}}_{\rho},\hat{P}]\equiv 
\hat{\mathrm{Y}}_{\rho}\hat{P}-\hat{P}\hat{\mathrm{Y}}_{\rho}\equiv
0.
\end{equation}
The zeroing of the third commutator is due to the following helicity property (see (\ref{lambda^2}) and (\ref{lambda^3})):
\begin{equation}
\label{PL}
\hat{P}\hat{\lambda}\equiv
\hat{\lambda}\hat{P}\equiv
\hat{\lambda},
\end{equation}
therefore:
\begin{equation}
[\hat{\mathrm{\Theta}}_{\rho},\hat{P}]\equiv 
\hat{\lambda} \rho \hat{\lambda} \hat{P}-\hat{P}\hat{\lambda} \rho \hat{\lambda}\equiv
0.
\end{equation}
The last two radial operators can be constructed from Cartesian components based on formulas (\ref{Nrho}) and (\ref{Q[rho]}). Finally, with (\ref{pq+qp}) or (\ref{pqp}) and (\ref{Q=q}) both operators are expressed using $\hat{\mathrm{Q}}_1$ and $\hat{\mathrm{Q}}_2$. And these last operators commute with $\hat{P}$ by virtue of (\ref{QPP}):
\begin{equation}
[\hat{\mathrm{Q}}_1,\hat{P}]\equiv 0,\ \ \ \
[\hat{\mathrm{Q}}_2,\hat{P}]\equiv 0.
\end{equation}
Therefore the radial operators $\hat{\mathrm{q}}_{\rho}$ and $\hat{\mathrm{Q}}_{[\rho]}$ also commute with $\hat{P}$.
Q.E.D.

Since $\hat{P}_0=1-\hat{P}$, all radial operators also commute with the projective operator $\hat{P}_0$ onto non-physical states (see Appendix D).

\section{6. Photon states on a straight line -- open photon strings}

We will now find, write and prove the eigenstates of two position components $\hat{\mathrm{Q}}_1$ and $\hat{\mathrm{Q}}_2$ simultaneously. Even though such components do not commute identity-wise, such a commutator will equal zero on eigenstates. Furthermore, there are two types of such states: electric photon string and magnetic photon string.

\subsection{6.1. Photon Electric Open String}

The positional eigenstates should be as close as possible to Dirac deltas in the positional representation. The transversality condition limits the localizability of photons to unbounded or closed lines (applies to electric strings, but not magnetic ones). This suggests the following simple formula for the wave function of a photon located on the infinite line $\Gamma$:
\begin{equation}
\label{Gamma}
    \mathbf{\Psi}_{\Gamma}(\mathbf{r}):=\int_{-\infty}^{+\infty}{\delta^{(3)}\big(\mathbf{r}-\mathbf{r}'(s)\big)d\mathbf{r}'(s)}=:\ket{\Gamma},
\end{equation}
or on the closed line $\Omega$:
\begin{equation}
\label{Omega}
    \mathbf{\Psi}_{\Omega}(\mathbf{r}):=\ointctrclockwise_{\Omega}{\delta^{(3)}\big(\mathbf{r}-\mathbf{r}'(s)\big)d\mathbf{r}'(s)}=:\ket{\Omega}.
\end{equation}
In both integrals, the primes denote the parametric equation of the line ($s \rightarrow \mathbf{r}'(s) \rightarrow \mathbf{r}$), not the derivatives.
The above states in the momentum representation (their Fourier transformation) were proposed in the conclusion of the publication \cite{Jadczyk}. However, these eigenstates were not identical to the specific examples of states given in \cite{Jadczyk}.

We will start by demonstrating the transversality of these states:
\begin{equation}
    \begin{split}
    \nabla\cdot\mathbf{\Psi}(\mathbf{r})=-\int{\frac{\partial}{\partial\mathbf{r}'}\delta^{(3)}\big(\mathbf{r}-\mathbf{r}'(s)\big)d\mathbf{r}'(s)}=\\
   = \delta^{(3)}\big(\mathbf{r}-\mathbf{r}'(s_1)\big)-\delta^{(3)}\big(\mathbf{r}-\mathbf{r}'(s_2)\big)=0,
    \end{split}
\end{equation}
where zeroing is trivial for a closed curve in which the values $s_1$, $s_2$ describe the same starting and ending points. In the case of infinite limits, one must refer to the distribution definition on test functions with compact support. Then each component of the calculated difference equals zero in the limit up to infinity.

Let's apply our formula to the line parallel to the $Oz$ axis, for $x'=x_0$, $y'=y_0$, $z'=s$:
\begin{equation}
\label{electric line}
  \ket{x_0, y_0}(\mathbf{r}):=\int_{-\infty}^{+\infty}{\delta^{(3)}\big(\mathbf{r}-\mathbf{r}'(s)\big)d\mathbf{r}'(s)},
\end{equation}
\begin{equation}
    \ket{x_0,y_0}(\mathbf{r})=
    \left(\begin{array}{c}
0\\
0\\
\delta(x-x_0)\delta(y-y_0)\int_{-\infty}^{+\infty}{\delta(z-z')dz'}
\end{array} \right),
\end{equation}
\begin{equation}
\label{electric}
    \ket{x_0,y_0}(\mathbf{r})=
    \left(\begin{array}{c}
0\\
0\\
\delta(x-x_0)\delta(y-y_0)
\end{array} \right).
\end{equation}
These states are delta-normalized in $x$ and $y$, but are not normalized with respect to $z$:
\begin{equation}
\braket{x_1,y_1|x_2,y_2}=\delta(x_1-x_2)\delta(y_1-y_2)\int{dz}.
\end{equation}

Because the transversality condition is met automatically, it is quite easy to prove the eigenvalues of the position operator for this state. It is most convenient to use the minimal form of the position operator in matrix notation:
\begin{equation}
    \hat{\mathrm{Q}}|_1\ket{x_0,y_0}=
    \left(\begin{array}{ccc}
x-\Delta^{-1}\partial_x & 0 & 0\\
-\Delta^{-1}\partial_y & x & 0\\
-\Delta^{-1}\partial_z & 0 & x
\end{array} \right)\ket{x_0,y_0}=x_0\ket{x_0,y_0}.
\end{equation}
The calculation of the eigenvalue for the $y$ component is analogous:
\begin{equation}
    \hat{\mathrm{Q}}|_2\ket{x_0,y_0}=
    \left(\begin{array}{ccc}
y & -\Delta^{-1}\partial_x & 0\\
0 & y-\Delta^{-1}\partial_y & 0\\
0 & -\Delta^{-1}\partial_z & y
\end{array} \right)\ket{x_0,y_0}=y_0\ket{x_0,y_0}.
\end{equation}
Of course, the above eigenequations apply equivalently to the general form $\hat{\mathbf{Q}}$ operator, and not only to the position operator in minimal form:
\begin{equation}
\hat{\mathrm{Q}}_1\ket{x_0,y_0}=\hat{P}x\ket{x_0,y_0}=x_0\ket{x_0,y_0},
\end{equation}
\begin{equation}
\hat{\mathrm{Q}}_2\ket{x_0,y_0}=\hat{P}y\ket{x_0,y_0}=y_0\ket{x_0,y_0}.
\end{equation}

\begin{figure}[h!]
\centering
\includegraphics[width=8.5cm]{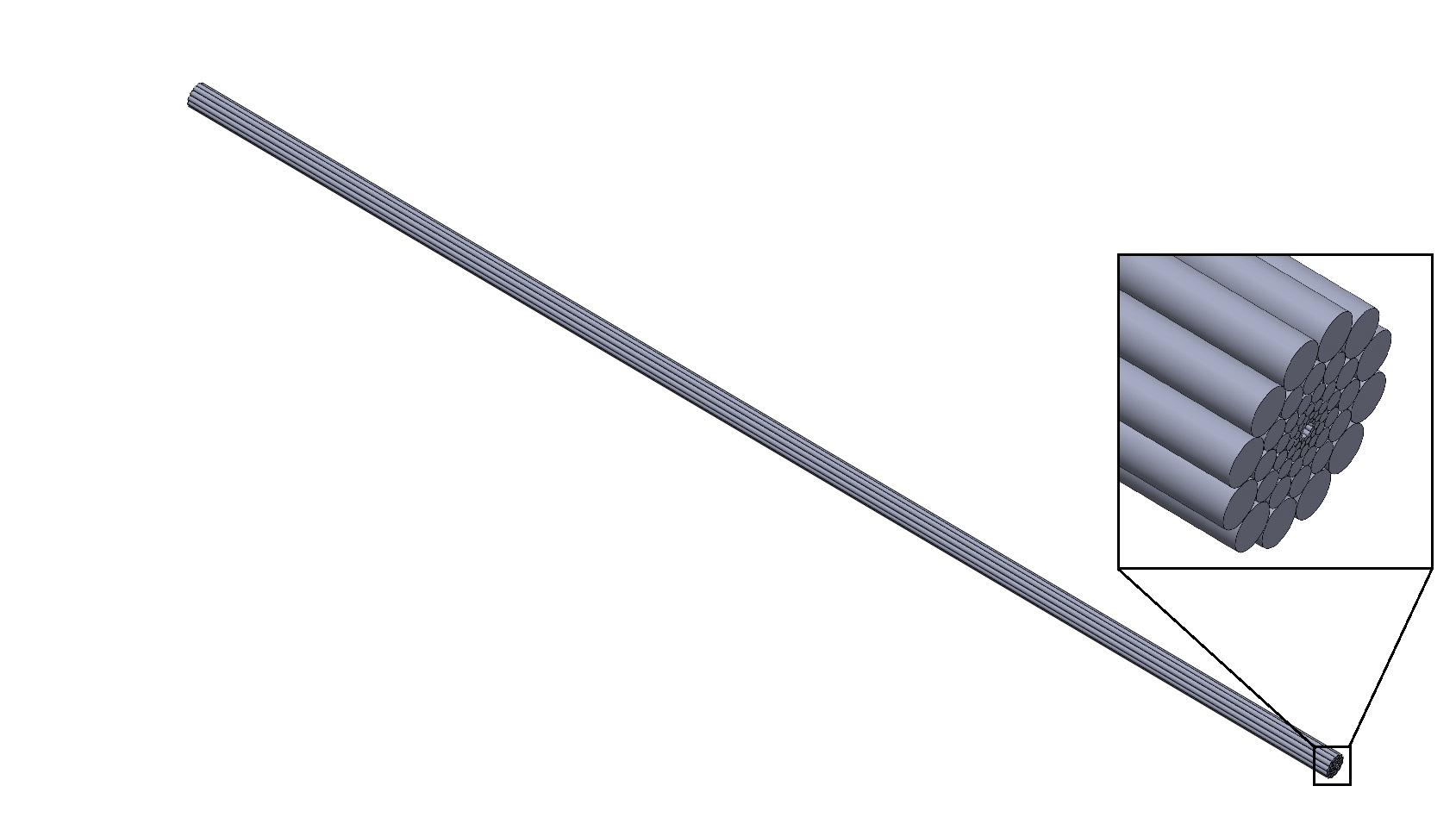}
\caption{Graphical visualisation of the $\ket{x_0,y_0}$ photon state as a linear electric string, which is an infinitely thin and infinitely dense bundle (vector Dirac delta). The string is open, but it is infinitely long -- its cross-section is for illustrative purposes only}
\label{state1}
\end{figure}

The state describing a photon lying on an axis parallel to the $z$ axis can be called a linear string. The wave vector of this state is also parallel to the $z$ axis and is infinitely large as the Dirac delta function. Thus, this straight-fibered state can be called an electric linear string (see Fig.\ref{state1}) in contrast to the spiral magnetic string considered later. The adjectives electric and magnetic are explained in more detail in section 9.

 %rysunek
 %(see Fig.\ref{diagram}):
It turns out that a linear electric string is an eigenstate of the two radial operators:
\begin{equation}
\hat{\mathrm{Q}}_{[\rho]}\ket{x_0,y_0}=\sqrt{x_0^2+y_0^2}\ket{x_0,y_0},
\end{equation}
\begin{equation}
\hat{\mathrm{Q}}_\rho\ket{x_0,y_0}=\hat{P}\rho\ket{x_0,y_0}=\sqrt{x_0^2+y_0^2}\ket{x_0,y_0}.
\end{equation}

It turns out that the above axial eigenstates of the two components of the photon position operator are not helicity eigenstates. However, such common eigenstates also exist and were found and published \cite{Jadczyk}. However, they are more mathematically complex and cause interpretation problems -- for this reason, they were not supposed to be analysed in this work, but they were (see below). Instead, other types of photon eigenstates on a straight line are given -- magnetic photon strings.

\subsection{6.2. Photon Magnetic Open String}

In addition to the usual (\ref{Gamma}) or (\ref{Omega}) line integrals, there are integrals with rotation:
\begin{equation}
  \int_{\Gamma}{\nabla \delta^{(3)}\big(\mathbf{r}-\mathbf{r}'(s)\big)\times d\mathbf{r}'(s)}.
\end{equation}
The use of a vector product requires the introduction of the vector nabla operator. Rotation in the integral expression automatically ensures zeroing of the divergence of the wave function vector, i.e. the condition of transversality of the physical state:
\begin{equation}
  \nabla\cdot \int_{\Gamma}{\nabla \delta^{(3)}\big(\mathbf{r}-\mathbf{r}'(s)\big)\times d\mathbf{r}'(s)}\equiv 0.
\end{equation}
The condition is met for finite and infinite, closed and unclosed curves $\Gamma:s \rightarrow \mathbf{r}'(s) \rightarrow \mathbf{r}$. Prime here does not mean derivative.

States defined by rotation integrals have different units from the states of electric strings, which are position eigenstates. Therefore, it is worth changing the normalisation of these states as follows:
\begin{equation}
  \ket{\Gamma^{\star}}:=\frac{1}{\sqrt{-\Delta}} \int_{\Gamma}{\nabla \delta^{(3)}\big(\mathbf{r}-\mathbf{r}'(s)\big)\times d\mathbf{r}'(s)},
\end{equation}
where the star denotes a spiral (rotational) state, i.e. a magnetic string along the $\Gamma$ curve.

So let's consider a magnetic string along a line parallel to the $z$ axis ($x'=x_0,y'=y_0,z'=s$):
\begin{equation}
\label{magnetic line}
  \ket{x_0^{\star}, y_0^{\star}}:=\frac{1}{\sqrt{-\Delta}} \int_{-\infty}^{+\infty}{\nabla \delta^{(3)}\big(\mathbf{r}-\mathbf{r}'(s)\big)\times d\mathbf{r}'(s)}.
\end{equation}
Calculating the line integral with rotation leads to:
\begin{equation}
    \ket{x_0^{\star},y_0^{\star}}=\frac{1}{\sqrt{-\Delta}}
    \left(\begin{array}{c}
\delta(x-x_0)\delta'(y-y_0)\\
-\delta'(x-x_0)\delta(y-y_0)\\
0
\end{array} \right).
\end{equation}
Since the wave vector does not depend on $z$, we can use (\ref{lap 2D}) to obtain:
\begin{equation}
\label{magnetic}
    \ket{x_0^{\star},y_0^{\star}}=\frac{1}{2\pi}
    \left(\begin{array}{c}
\partial_y\frac{1}{\sqrt{(x-x_0)^2+(y-y_0)^2}}\\
-\partial_x\frac{1}{\sqrt{(x-x_0)^2+(y-y_0)^2}}\\
0
\end{array} \right).
\end{equation}
For this state, the following occurs:
\\
\\
\\
STATEMENT 5 (Eigenstates of a linear magnetic string)

\textit{The states $\ket{x_0^{\star},y_0^{\star}}$ are simultaneously the eigenstates of the components of the position operator $\hat{\mathrm{Q}}_1$ and $\hat{\mathrm{Q}}_2$:}
\begin{equation}
\hat{\mathrm{Q}}_1\ket{x_0^{\star}, y_0^{\star}}=x_0\ket{x_0^{\star}, y_0^{\star}}, 
\end{equation}
\begin{equation}
 \hat{\mathrm{Q}}_2\ket{x_0^{\star}, y_0^{\star}}=y_0\ket{x_0^{\star}, y_0^{\star}}.
\end{equation}
\\
PROOF. The proof will be done twice using two different methods in: the position representation and the momentum representation. The calculations in the position representation are a bit non-standard, so it is worth confirming them in the momentum representation. In order not to repeat the same methods for the same components, the methods will be applied to two different components of the position operator.

The proof for the operator $\hat{\mathrm{Q}}_1$ will be done, rather without loss of generality, for the case $x_0=0$ and $y_0=0$:
\begin{equation}
    \ket{0^{\star},0^{\star}}=\frac{1}{2\pi}
    \left(\begin{array}{c}
\partial_y\ \rho^{-1}\\
-\partial_x\ \rho^{-1}\\
0
\end{array} \right)=:\mathbf{\Psi}.
\end{equation}
The partial derivatives occurring here are distributive in nature, so it is better not to calculate them directly. In the previous subsection 6.1, a simplified version of the $\hat{\mathrm{Q}}|_1$ operator was used, and now the full version $\hat{\mathrm{Q}}_1$ will be used equivalently (the matrices of all components of both operator forms  can be seen in poster \cite{Poster}):
\begin{equation}
    \hat{\mathrm{Q}}_1\mathbf{\Psi}=
    \frac{1}{2\pi}
    \left(\begin{array}{ccc}
x & \Delta^{-1}\partial_y & \Delta^{-1}\partial_z\\
-\Delta^{-1}\partial_y & x & 0\\
-\Delta^{-1}\partial_z & 0 & x
\end{array} \right)
\left(\begin{array}{c}
\partial_y\ \rho^{-1}\\
-\partial_x\ \rho^{-1}\\
0
\end{array} \right),
\end{equation}
\begin{equation}
    \hat{\mathrm{Q}}_1\mathbf{\Psi}=
    \frac{1}{2\pi}
    \left(\begin{array}{c}
x\partial_y\ \rho^{-1} - \partial_y\partial_x\Delta_2^{-1} \rho^{-1} \\
-\partial^2_y\Delta_2^{-1}\rho^{-1} - x \partial_x\rho^{-1}\\
0
\end{array} \right).
\end{equation}
Where the commutativity of the inverse operator of the Laplacian with partial derivatives is assumed (which is obvious in the momentum representation).
Thanks to the following property of the two-dimensional Laplacian:
\begin{equation}
    \Delta_2\ \rho=\rho^{-1} \ \ \ \longrightarrow \ \ \ \Delta_2^{-1}\rho^{-1}=\rho,
\end{equation}
we can simplify the calculations:
\begin{equation}
    \hat{\mathrm{Q}}_1\mathbf{\Psi}=
    \frac{1}{2\pi}
    \left(\begin{array}{c}
x\partial_y\ \rho^{-1} - \partial_y\partial_x\rho \\
-\partial^2_y\rho - x \partial_x\rho^{-1}\\
0
\end{array} \right).
\end{equation}
Now all we need to do is calculate the first or second order partial derivatives:
\begin{equation}
    \hat{\mathrm{Q}}_1\mathbf{\Psi}=
    \frac{1}{2\pi}
    \left(\begin{array}{c}
x\partial_y\ \rho^{-1} - \partial_y x\rho^{-1} \\
-\rho^{-1}+y^2 \rho^{-3} + x^2 \rho^{-3}\\
0
\end{array} \right)=\left(\begin{array}{c}
0 \\
0\\
0
\end{array} \right)=0,
\end{equation}
with the condition that the distributional part of the second-order derivatives disappears under the multiplier $x$ or $y$.
Thus, the zero eigenvalue of the first component of the position operator for the magnetic string has just been proven:
\begin{equation}
\hat{\mathrm{Q}}_1\ket{0^{\star}, 0^{\star}}=\hat{\mathrm{Q}}_1\mathbf{\Psi}=0=0\ket{0^{\star}, 0^{\star}}.
\end{equation}

The proof for the second position component will be done in the momentum representation, which is the Fourier transform of the positional representation. In manipulating the Fourier transform or its inverse, the integral representation of the Dirac delta function, proven in detail in the work of \cite{Dirac}, can be helpful. The general form of the state under consideration, in the momentum representation, and the form of the second component of the position operator (with simplification $\hbar:=1$) are:
\begin{equation}
    \widetilde{\ket{x_0^{\star},y_0^{\star}}}=
    \left(\begin{array}{c}
\frac{i\mathrm{p}_2}{\mathrm{p}} \ e^{-i(\mathrm{p}_1x_0+\mathrm{p}_2y_0)}\ \delta(\mathrm{p}_3) \\
\frac{-i\mathrm{p}_1}{\mathrm{p}} \ e^{-i(\mathrm{p}_1x_0+\mathrm{p}_2y_0)}\ \delta(\mathrm{p}_3)\\
0
\end{array} \right)=:\widetilde{\mathbf{\Psi}},
\end{equation}
\begin{equation}
    \widetilde{\hat{\mathrm{Q}}}_2=
    \left(\begin{array}{ccc}
i\frac{\partial}{\partial \mathrm{p}_2} & i\mathrm{p}_1/\mathrm{p}^2 & 0\\
-i\mathrm{p}_1/\mathrm{p}^2 & i\frac{\partial}{\partial \mathrm{p}_2} & -i\mathrm{p}_3/\mathrm{p}^2\\
0 & i\mathrm{p}_3/\mathrm{p}^2 & i\frac{\partial}{\partial \mathrm{p}_2}
\end{array} \right).
\end{equation}
Therefore:
\begin{equation}
    \widetilde{\hat{\mathrm{Q}}}_2\widetilde{\mathbf{\Psi}}=
    \left(\begin{array}{c}
\frac{-1}{\mathrm{p}}+\frac{\mathrm{p}_2^2}{\mathrm{p}^3}+\frac{i\mathrm{p}_2}{\mathrm{p}}y_0 + \frac{\mathrm{p}_1^2}{\mathrm{p}^3} \\
\frac{\mathrm{p}_1\mathrm{p}_2}{\mathrm{p}^3} - \frac{\mathrm{p}_1\mathrm{p}_2}{\mathrm{p}^3}-\frac{i\mathrm{p}_1}{\mathrm{p}}y_0 \\
\frac{\mathrm{p}_1\mathrm{p}_3}{\mathrm{p}^3}
\end{array} \right) \ e^{-i(\mathrm{p}_1x_0+\mathrm{p}_2y_0)}\ \delta(\mathrm{p}_3),
\end{equation}
\begin{equation}
    \widetilde{\hat{\mathrm{Q}}}_2\widetilde{\mathbf{\Psi}}=
    \left(\begin{array}{c}
\frac{i\mathrm{p}_2}{\mathrm{p}}y_0  \\
-\frac{i\mathrm{p}_1}{\mathrm{p}}y_0 \\
0
\end{array} \right) \ e^{-i(\mathrm{p}_1x_0+\mathrm{p}_2y_0)}\ \delta(\mathrm{p}_3),
\end{equation}
because $\mathrm{p}_3=0$ due to the Dirac delta. So finally:
\begin{equation}
    \widetilde{\hat{\mathrm{Q}}}_2 \widetilde{\ket{x_0^{\star},y_0^{\star}}}=\widetilde{\hat{\mathrm{Q}}}_2\widetilde{\mathbf{\Psi}}=y_0\widetilde{\mathbf{\Psi}}=
    y_0\widetilde{\ket{x_0^{\star},y_0^{\star}}},
\end{equation}
which means the eigenstate of the second component. Q.E.D.

\begin{figure}[h!]
\centering
\includegraphics[width=8.5cm]{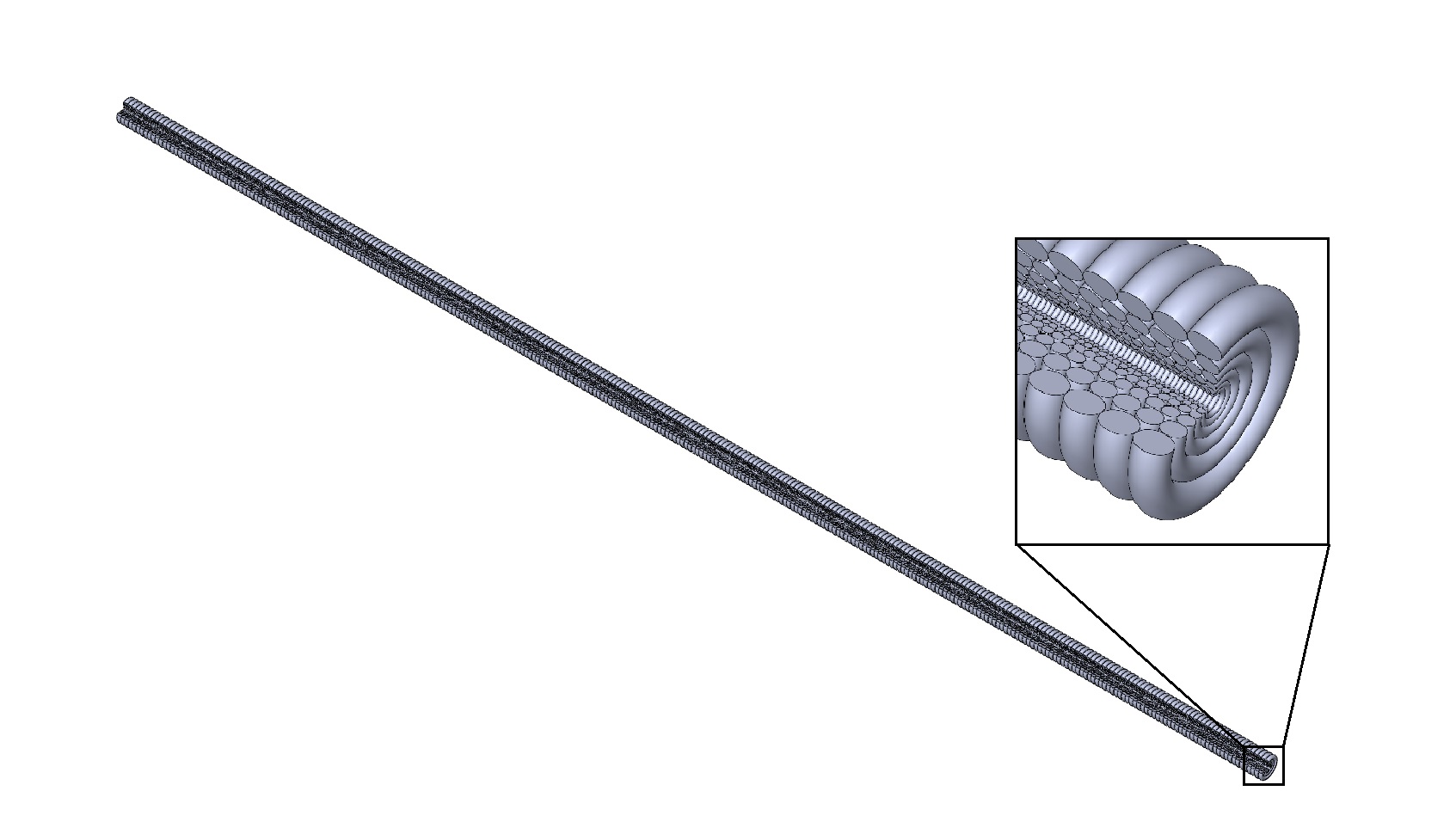}
\caption{Graphical visualization of the $\ket{x_0^{\star},y_0^{\star}}$ photon state as a linear magnetic string, which in the $\hat{\mathrm{p}}\ket{x_0^{\star},y_0^{\star}}(\mathbf{r})$ "energetic" representation is an infinitely tight and dense spiral vector field (a rotation of the vector Dirac delta). The rings are complete but have been cut for illustrative purposes. Basically, this magnetic string is infinitely long, but it can be shortened by separating but not cutting the rings -- without violating the transversality condition of the state}
\label{state2}
\end{figure}

Since the state under consideration is described by a photon lying on an axis parallel to the $z$ axis, like a linear electric string, then it is also a linear string. This time, however, the state vector is not parallel to the photon line (string), but winds around it infinitely tightly and densely (see Fig. \ref{state2}). This is why this photon state is called a linear magnetic string (i.e. rotational or spiral).

The existence of eigenstates of a magnetic string next to an electric string leads to the question of the relations of these states. It turns out that these states can be related by the helicity operator, but they are linearly independent:
\\
\\
THEOREM 2 (About magnetic and electric strings)

\textit{A magnetic string (linear) is created from an electric string (linear) by the action of the helicity operator (and vice versa):
\begin{equation}
\ket{x_0^{\star}, y_0^{\star}}=\hat{\lambda}\ket{x_0, y_0}, 
\end{equation}
\begin{equation}
\ket{x_0, y_0}=\hat{\lambda}\ket{x_0^{\star}, y_0^{\star}}, 
\end{equation}
nevertheless, these states are orthogonal for any eigenvalues:}
\begin{equation}
\braket{x_1^{\star}, y_1^{\star}| x_2, y_2}=0. 
\end{equation}
\textit{Additionally, the helicity operator has a simple formula for acting on the wave function vector:}
\begin{equation}
  \hat{\lambda}\ \mathbf{\Psi}=\frac{1}{\sqrt{-\Delta}} \nabla \times \mathbf{\Psi}.
\end{equation}
PROOF. Let's start with the last formula. According to the definition of (\ref{helicity}), helicity has the form:
\begin{equation}
 \hat{\lambda}\equiv
\frac{1}{\hbar \hat{\mathrm{p}}}\hat{\mathbf{p}}\cdot\hat{\mathbf{S}},
\ \ \ \ 
 \hat{\lambda}_{lm}\equiv\frac{-i}{\hat{\mathrm{p}}}\hat{\mathrm{p}}_j\varepsilon_{jlm}.
\end{equation}
Using the index form in the action on the state, we get:
\begin{equation}
 \hat{\lambda}_{lm}\Psi_m=\frac{-i}{\hat{\mathrm{p}}}\hat{\mathrm{p}}_j\varepsilon_{jlm}\Psi_m=
 \frac{1}{\sqrt{-\Delta}}\varepsilon_{ljm}\partial_j\Psi_m.
\end{equation}
At the end, we see the index definition of rotation, so:
\begin{equation}
 \hat{\lambda}_{lm}\Psi_m=
 \frac{1}{\sqrt{-\Delta}}(\nabla \times \mathbf{\Psi})_l.
\end{equation}
The formula for normalised rotation appears in the definition (\ref{magnetic line}) of a magnetic string, and we can relate this formula to the definition (\ref{electric line}) of an electric string.
This proves, at the level of definition, that magnetic and electric strings differ only in their helicity action. The proof in the other direction follows from the square property of the helicity operator:
\begin{equation}
\hat{\lambda}\ket{x_0^{\star}, y_0^{\star}}=\hat{\lambda}\big(\hat{\lambda}\ket{x_0, y_0}\big)=
\hat{P}\ket{x_0, y_0}=\ket{x_0, y_0}. 
\end{equation}

However, the orthogonality of the states of a magnetic string and an electric string results algebraically from the forms (\ref{magnetic}) and (\ref{electric}) of these states. Q.E.D.

The above theorem was proved by synthetic methods, as opposed to the detailed computational proof of the statement which preceded this theorem. Thanks to this, the formula of the magnetic string state is better verified, although its eigenstate follows from Theorem 2 and Lemma 1:
\begin{equation}
\hat{\mathrm{Q}}_1\ket{x_0^{\star}, y_0^{\star}}=\hat{\mathrm{Q}}_1\hat{\lambda}\ket{x_0, y_0}=
\hat{\lambda}\ x_0\ket{x_0, y_0}=x_0 \ket{x_0^{\star}, y_0^{\star}},
\end{equation}
\begin{equation}
\hat{\mathrm{Q}}_2\ket{x_0^{\star}, y_0^{\star}}=\hat{\mathrm{Q}}_2\hat{\lambda}\ket{x_0, y_0}=
\hat{\lambda}\ y_0\ket{x_0, y_0}=y_0 \ket{x_0^{\star}, y_0^{\star}}.
\end{equation}

Theorem 2 also directly implies the normalisation of the states of magnetic strings:
\begin{equation}
\braket{x_1^{\star},y_1^{\star}|x_2^{\star},y_2^{\star}}=\braket{x_1,y_1|\hat{\lambda}^{\dagger}\hat{\lambda}|x_2,y_2}=\braket{x_1,y_1|x_2,y_2},
\end{equation}
\begin{equation}
\braket{x_1^{\star},y_1^{\star}|x_2^{\star},y_2^{\star}}=\delta(x_1-x_2)\delta(y_1-y_2)\int{dz}.
\end{equation}

The linear magnetic string is also an eigenstate of the rooter radial operator, which is a function of Cartesian components:
\begin{equation}
\hat{\mathrm{Q}}_{[\rho]}\ket{x_0^{\star}, y_0^{\star}}=\sqrt{x_0^2+y_0^2} \ket{x_0^{\star}, y_0^{\star}}.
\end{equation}

\subsection{6.3. Right- and Left-handed Photon Linear Helical Strings}

Theorem 2 has yet another important consequence: it shows how one can construct common eigenstates of position and helicity. Using the definition of projective operators into right-handed (positive) and left-handed (negative) states, let us consider the states:
\begin{equation}
\ket{x_0, y_0,\pm }=\sqrt{2}\hat{P}_{\pm}\ket{x_0, y_0}=
\frac{1}{\sqrt{2}}\ket{x_0, y_0}\pm \frac{1}{\sqrt{2}}\ket{x_0^{\star}, y_0^{\star}}. 
\end{equation}
These states, by definition and by Theorem 2, are eigenstates of helicity:
\begin{equation}
\hat{\lambda}\ket{x_0, y_0,\pm }=\pm \ket{x_0, y_0,\pm }. 
\end{equation}
Equally obvious are the position eigenvalues for these states:
\begin{equation}
\hat{\mathrm{Q}}_1\ket{x_0, y_0,\pm }=x_0\ket{x_0, y_0,\pm }, 
\end{equation}
\begin{equation}
\hat{\mathrm{Q}}_2\ket{x_0, y_0,\pm }=y_0\ket{x_0, y_0,\pm }.
\end{equation}
The normalisation condition has the same form as for linear electric and magnetic strings:
\begin{equation}
\braket{x_1,y_1,\pm|x_2,y_2,\pm}=\delta(x_1-x_2)\delta(y_1-y_2)\int{dz}.
\end{equation}
However, states with different helicity are obviously orthogonal:
\begin{equation}
\braket{x_1,y_1,\pm|x_2,y_2,\mp}=0.
\end{equation}

Additionally, the helical strings considered here are eigenstates of the radial operator (rooter version):
\begin{equation}
\hat{\mathrm{Q}}_{[\rho]}\ket{x_0, y_0,\pm }=\rho_0\ket{x_0, y_0,\pm }.
\end{equation}
It is possible that this also applies to the $\hat{\mathrm{q}}_{\rho}$ operator based on the boost generator. However, this work does not study the radial boost operator in detail. In any case, it is known that the above eigenequation is not satisfied by the remaining radial operators (transversal, helical and even).

\subsection{6.4. Study of Radial Operators for Linear Strings}

The linear electric and magnetic strings are eigenstates of the rooter radial operator with eigenvalue $\rho_0$ (or zero in the special, simplest case). The same is true for the electric string and the transversal radial operator.
However, for a magnetic string, the situation is no longer trivial:
\\
\\
STATEMENT 6 (Radial operators for open magnetic string)

\textit{The state $\ket{x_0^{\star},y_0^{\star}}$ is the eigenstate of the helical radial position operator $\hat{\mathrm{\Theta}}_{\rho}$, but is not an eigenstate of transversal radial operator $\hat{\mathrm{Q}}_{\rho}$ and square radial operator $\hat{\mathrm{Q}}^2_{\rho^2}$:}
\begin{equation}
    \hat{\mathrm{\Theta}}_{\rho}\ket{x_0^{\star},y_0^{\star}}=
  \sqrt{x_0^2+y_0^2}\ket{x_0^{\star},y_0^{\star}},
\end{equation}
\begin{equation}
    \hat{\mathrm{Q}}_{\rho}\ket{x_0^{\star},y_0^{\star}}\neq
   \sqrt{x_0^2+y_0^2}\ket{x_0^{\star},y_0^{\star}},
\end{equation}
\begin{equation}
    \hat{\mathrm{Q}}^2_{\rho^2}\ket{x_0^{\star},y_0^{\star}}\neq
   (x_0^2+y_0^2)\ket{x_0^{\star},y_0^{\star}}.
\end{equation}

PROOF. The first relation follows from the definition of the given radial operator and Theorem 2:
\begin{equation}
\hat{\mathrm{\Theta}}_{\rho}\ket{x_0^{\star}, y_0^{\star}}=\hat{\lambda}\rho\hat{\lambda} \ket{x_0^{\star}, y_0^{\star}}=\hat{\lambda}\rho \ket{x_0, y_0},
\end{equation}
and further with the property of an electric string:
\begin{equation}
\hat{\lambda}\rho \ket{x_0, y_0}=\sqrt{x_0^2+y_0^2}\hat{\lambda} \ket{x_0, y_0}=
\sqrt{x_0^2+y_0^2}\ket{x_0^{\star}, y_0^{\star}},
\end{equation}
where again is used at the end of Theorem 2.

It remains to prove the strange property of the absence of an eigenstate for the transversal radial operator. From the definition of the operator and the condition of the physicality of the state, we have:
\begin{equation}
    \hat{\mathrm{Q}}_{\rho}\ket{x_0^{\star},y_0^{\star}}=
    \hat{P}\rho\hat{P}\ket{x_0^{\star},y_0^{\star}}=
    \hat{P}\rho\ket{x_0^{\star},y_0^{\star}}.
\end{equation}
Let's start calculating this in the simplest case of zero values:
\begin{equation}
    \hat{P}\rho\ket{0^{\star},0^{\star}}=\frac{\hat{P}}{2\pi}
    \left(\begin{array}{c}
\rho\ \partial_y \rho^{-1}\\
-\rho\ \partial_x \rho^{-1}\\
0
\end{array} \right)=
\frac{\hat{P}}{2\pi}
    \left(\begin{array}{c}
-y \rho^{-2}\\
x \rho^{-2}\\
0
\end{array} \right),
\end{equation}
where thanks to the $\rho$ multiplier, it was possible to calculate partial derivatives without worrying about distributional singularities in the neighbourhood of zero. Note that the state on the right is a physical transversal state:
\begin{equation}
\nabla\cdot
    \left(\begin{array}{c}
-y \rho^{-2}\\
x \rho^{-2}\\
0
\end{array} \right)=-y\partial_x(\rho^{-2})+x\partial_y(\rho^{-2})=0,
\end{equation}
where again, we do not have to worry about distributions due to the multipliers $x$, $y$ and the symmetry of the subtracted terms. We can therefore leave the projection operator to such a transversal state:
\begin{equation}
\label{not comute}
    \hat{\mathrm{Q}}_{\rho}\ket{0^{\star},0^{\star}}=
     \hat{P}\rho\ket{0^{\star},0^{\star}}=
\frac{1}{2\pi}
    \left(\begin{array}{c}
-y \rho^{-2}\\
x \rho^{-2}\\
0
\end{array} \right) \neq 0.
\end{equation}
Therefore, the radial transversal operator is not an eigenstate for zero eigenvalues $x_0=0$ and $x_0=0$.

Unfortunately, the zero radial value can be singular for a radial operator, and this must be proven separately for nonzero values. So let's consider simple Pythagorean values $x_0=3$, $y_0=4$, $\rho_0=5$ (default in meters). Let's also use a simplified notation of the state vector (beyond the singularity at point (3,4)):
\begin{equation}
   \mathbf{f}:=
     2\pi\ket{3^{\star},4^{\star}}=
    \left(\begin{array}{c}
\frac{4-y}{\sqrt{(x-3)^2+(y-4)^2}^3}\\
\frac{x-3}{\sqrt{(x-3)^2+(y-4)^2}^3}\\
0
\end{array} \right) .
\end{equation}
Proving by contradiction, let us assume that this is an eigenstate of the transversal radial operator:
\begin{equation}
    \hat{\mathrm{Q}}_{\rho}\mathbf{f}=\hat{P}\rho\ \mathbf{f}
   \overset{?}{=}
    5\ \mathbf{f},
\end{equation}
\begin{equation}
    \rho\ \mathbf{f}-\Delta^{-1}\nabla (\nabla\cdot \rho\mathbf{f})
   \overset{?}{=}
    5\ \mathbf{f}.
\end{equation}
Therefore, the following equality should hold:
\begin{equation}
    \Delta(\rho\mathbf{f})-\nabla (\nabla\cdot \rho\mathbf{f})
   \overset{?}{=}
    5\Delta \mathbf{f}.
\end{equation}
In particular, for the first component:
\begin{equation}
    \Delta(\rho\mathrm{f}_x)-\partial_x (\nabla\cdot \rho\mathbf{f})
   \overset{?}{=}
    5\Delta \mathrm{f}_x.
\end{equation}
The divergence is easy to calculate:
\begin{equation}
    \nabla\cdot \rho\mathbf{f}=\frac{4x-3y}{\sqrt{x^2+y^2}\sqrt{(x-3)^2+(y-4)^2}^3}\neq 0.
\end{equation}
Laplacians are more complex to calculate, but can be tested, e.g. on the Wolfram Alpha platform:
\begin{equation}
    \Delta (\rho\mathrm{f}_x)=\frac{100+48x-8x^2-11y-8y^2}{\sqrt{x^2+y^2}\sqrt{(x-3)^2+(y-4)^2}^5},
\end{equation}
\begin{equation}
    5\Delta \mathrm{f}_x=\frac{15(4-y)}{\sqrt{(x-3)^2+(y-4)^2}^5}.
\end{equation}
Similarly, one can calculate the partial derivative of the divergence:
\begin{displaymath}
    \partial_x(\nabla\cdot \rho\mathbf{f})=
    \frac{-12x^4+12x^3(y+3)-x^2y(8y+45)}
    {\sqrt{x^2+y^2}^3\sqrt{(x-3)^2+(y-4)^2}^5}+
\end{displaymath}
\begin{equation}
    +\frac{3xy(4y^2-4y+25)+y^2(4y^2-59y+100)}
    {\sqrt{x^2+y^2}^3\sqrt{(x-3)^2+(y-4)^2}^5}.
\end{equation}
The last three expressions cannot be reduced because one of them does not contain an irrational term $\rho=\sqrt{x^2+y^2}$:
\begin{equation}
    \Delta(\rho\mathrm{f}_x)-\partial_x (\nabla\cdot \rho\mathbf{f})
   \neq
    5\Delta \mathrm{f}_x.
\end{equation}
Which proves by contradiction that it is not an eigenstate of the transversal radial operator, also for nonzero values. 

The situation is analogous for the square radial operator. Again, the simplest calculation is for zero values:
\begin{equation}
    \hat{\mathrm{Q}}^2_{\rho^2}\ket{0^{\star},0^{\star}}=
\frac{\hat{P}}{2\pi}
    \left(\begin{array}{c}
-y \rho^{-1}\\
x \rho^{-1}\\
0
\end{array} \right)=
\frac{1}{2\pi}
    \left(\begin{array}{c}
-y \rho^{-1}\\
x \rho^{-1}\\
0
\end{array} \right)\neq 0.
\end{equation}
The case of zero values has also been excluded (by a calculation analogous to the one above), so there is no reason to mention it here. Q.E.D.

The lack of eigenstates of a magnetic string with respect to the radial transversal operator implies other deficiencies.
\\
\\
STATEMENT 7 (Some radial operators for a straight electric string)

\textit{The state of electric string $\ket{x_0,y_0}$ is not the eigenstate of the helical radial position operator $\hat{\mathrm{\Theta}}_{\rho}$:
\begin{equation}
    \hat{\mathrm{\Theta}}_{\rho}\ket{x_0,y_0}\neq
  \sqrt{x_0^2+y_0^2}\ket{x_0,y_0},
\end{equation}
similarly as for the radial even operator $\hat{\mathrm{Y}}_{\rho}$:
\begin{equation}
    \hat{\mathrm{Y}}_{\rho}\ket{x_0,y_0}\neq
   \sqrt{x_0^2+y_0^2}\ket{x_0,y_0},
\end{equation}
which does not satisfy the eigenequation even for a magnetic string:}
\begin{equation}
    \hat{\mathrm{Y}}_{\rho}\ket{x_0^{\star},y_0^{\star}}\neq
   \sqrt{x_0^2+y_0^2}\ket{x_0^{\star},y_0^{\star}}.
\end{equation}

PROOF. Let's start with the last relation. By (\ref{radial relation}), the even operator is practically the arithmetic mean of the transversal and helical operators. If a magnetic string is not an eigenstate of the transversal operator, then it is also not an eigenstate of the radial even operator:
\begin{equation}
    \hat{\mathrm{Y}}_{\rho}\ket{x_0^{\star},y_0^{\star}}=
    \frac{1}{2}(\hat{\mathrm{Q}}_{\rho}+\hat{\mathrm{\Theta}}_{\rho})\ket{x_0^{\star},y_0^{\star}},
\end{equation}
\begin{equation}
    \hat{\mathrm{Y}}_{\rho}\ket{x_0^{\star},y_0^{\star}}=
    \frac{1}{2}(\hat{\mathrm{Q}}_{\rho}+\rho_0)\ket{x_0^{\star},y_0^{\star}}\neq
    \rho_0\ket{x_0^{\star},y_0^{\star}}.
\end{equation}
Further, using the lack of an eigenstate of the transversal operator for the magnetic string:
\begin{equation}
    \hat{\mathrm{Q}}_{\rho}\ket{x_0^{\star},y_0^{\star}}=
    \hat{P}\rho\hat{P}\ket{x_0^{\star},y_0^{\star}}
 \neq \rho_0\ket{x_0^{\star},y_0^{\star}},
\end{equation}
we can, by Theorem 2, write using an electric string:
\begin{equation}
    \hat{P}\rho\hat{P}\hat{\lambda}\ket{x_0,y_0}
 \neq \rho_0\hat{\lambda}\ket{x_0,y_0}.
\end{equation}
Acting on both sides of the equation with the helicity operator based on (\ref{PL}), we get:
\begin{equation}
    \hat{\lambda}\hat{P}\rho\hat{P}\hat{\lambda}\ket{x_0,y_0}
 \neq \rho_0\hat{\lambda}^2\ket{x_0,y_0},
\end{equation}
\begin{equation}
    \hat{\lambda}\rho\hat{\lambda}\ket{x_0,y_0}
 \neq \rho_0\ket{x_0,y_0},
\end{equation}
the action is reversible and by using helicity again we could go back to the previous version. This means that an electric string cannot be an eigenstate of a helical radial operator:
\begin{equation}
    \hat{\mathrm{\Theta}}_{\rho}\ket{x_0,y_0}
 \neq \rho_0\ket{x_0,y_0}.
\end{equation}

Therefore, the electric string is an eigenstate of the radial transversal operator, but not an eigenstate of the radial helical operator. Further, as before, based on (\ref{radial relation}) it follows that:
\begin{equation}
    \hat{\mathrm{Y}}_{\rho}\ket{x_0,y_0}
 \neq \rho_0\ket{x_0,y_0},
\end{equation}
which means that the radial even operator does not achieve eigenvalues also on the electric string. Q.E.D.

Radial operators are important for the next Section 7 on circular strings. However, the study of a straight magnetic string has enabled us to learn about the properties of radial operators:
\\
\\
THEOREM 3 (On the inequivalence of radial operators)

\textit{The four main radial position operators considered in this work: transversal $\hat{\mathrm{Q}}_{\rho}$, helical $\hat{\mathrm{\Theta}}_{\rho}$, even $\hat{\mathrm{Y}}_{\rho}$ and rooter $\hat{\mathrm{Q}}_{[\rho]}$ are pairwise inequivalent on the physical states:}
\begin{equation}
    \hat{\mathrm{Q}}_{\rho}\neq\hat{\mathrm{\Theta}}_{\rho}\neq\hat{\mathrm{Y}}_{\rho}\neq
    \hat{\mathrm{Q}}_{[\rho]}\neq\hat{\mathrm{Q}}_{\rho}\neq\hat{\mathrm{Y}}_{\rho}, \ \ \
    \hat{\mathrm{\Theta}}_{\rho}\neq\hat{\mathrm{Q}}_{[\rho]}.
\end{equation}

PROOF. The theorem is a corollary of Statements 6 and 7, so the proof relies on them and on the states of the electric string and the magnetic string. The radial rooter operator is different from the others because it only has eigenvalues on the electric and magnetic string states. Similarly, the even operator is different from the others because it is the only one that does not have eigenvalues on these string states. The remaining two radial operators (transversal and helical) are also different, because the first has eigenvalues only on the electric string state, and the second only on the magnetic string state. Q.E.D.

Based on the above, we can draw conclusions about radial operators. Namely, the "even" radial operator $\hat{\mathrm{Y}}_{\rho}$ did not work at all. However, the $\hat{\mathrm{Q}}_{\rho}$ operator only worked for an electric string, and the $\hat{\mathrm{\Theta}}_{\rho}$ operator only for a magnetic string. For a rectilinear string, the radial operator $\hat{\mathrm{Q}}_{[\rho]}$ has shown universal working. However, the latter may be misleading, since in general the complicated operator $\hat{\mathrm{Q}}_{[\rho]}$ simply fits the eigenstates of the Cartesian components.

The above difficulties with the radial operator do not in any way concern the electric and magnetic eigenstates of the Cartesian position operators.

\section{7. Photon on a circle -- closed string}

The simplest example of a closed line is a circle. Let us consider a circle with radius $R$ lying on the plane $z=0$ (or more generally $z=z_0$) with its center at the point $x=0$, $y=0$. As with the photon open string, there are two types of such elementary strings: ordinary electric (fibrous) and magnetic (rotational-spiral). And from these elementary strings on the circle, (hybrid) helical strings can be built: right-handed and left-handed.

\subsection{7.1. Photon Electric Circle String}

Consider the line integral (\ref{Omega}) of Dirac deltas over the above described circle:
\begin{equation}
\ket{\mathring{R},z_0=0}=\delta(z)\ointctrclockwise_{x^2+y^2=R^2}{\delta^{(2)}\big(\mathbf{r}-\mathbf{r}'(s)\big)d\mathbf{r}'(s)}.
\end{equation}
Integration can be performed in polar (cylindrical) coordinates:
\begin{equation}
 \ket{\mathring{R},0}=\delta(z)\int_0^{2\pi}{\frac{1}{\rho}\delta(\rho-R)\ \delta(\varphi-\varphi')}
\left(\begin{array}{c}
-R\ \sin \varphi'\\
R\ \cos \varphi'\\
0
 \end{array} \right)d\varphi',
\end{equation}
where $\rho$ in the denominator results from the Jacobian for polar coordinates (on the circle $\rho=R$). Thanks to the Dirac delta property, calculating the integral in the polar system is trivial:
\begin{equation}
 \ket{\mathring{R},0}=\delta(z)\delta(\rho-R)
\left(\begin{array}{c}
-\sin \varphi\\
\cos \varphi\\
0
 \end{array} \right),
\end{equation}
which in the Cartesian system takes the form:
\begin{equation}
\label{circle}
 \ket{\mathring{R},0}=\frac{1}{\sqrt{x^2+y^2}}
\left(\begin{array}{c}
-y\ \delta(\sqrt{x^2+y^2}-R)\ \delta(z)\\
x\ \delta(\sqrt{x^2+y^2}-R)\ \delta(z)\\
0
 \end{array} \right).
\end{equation}
Due to the radial Dirac delta, the root in the denominator can be replaced by $R$, and the second Dirac delta can be generalised to $z_0$:
\begin{equation}
\label{circle}
 \ket{\mathring{R},z_0}=\frac{1}{R}
\left(\begin{array}{c}
-y\ \delta(\sqrt{x^2+y^2}-R)\ \delta(z-z_0)\\
x\ \delta(\sqrt{x^2+y^2}-R)\ \delta(z-z_0)\\
0
 \end{array} \right).
\end{equation}

The momentum equivalent of a similar state (accurate to the normalisation factor and projection onto the helicity eigenstates) was given in the publication from 1973 \cite{Jadczyk}. However, the momentum representation required the use of a special function -- the Bessel function of the second kind.

The condition for normalisation of photon states on circles perpendicular to the $Oz$ axis, the centers of which lie on the $Oz$ axis, is as follows:
\begin{equation}
\braket{\mathring{R}_1,z_1|\mathring{R}_2,z_2}=2\pi R_1\ \delta(R_1-R_2)\ \delta(z_1-z_2).
\end{equation}
It would be possible to get rid of the normalisation factor $2\pi R_1$ (or $2\pi R_2$) here, but this would require modifying the simple definition (\ref{Omega}) relative to the definition (\ref{Gamma}). 

However, the normalisation factor does not influence the statement about the eigenvalues of this state:
\\
\\
\\ 
\\
STATEMENT 8 (Eigenstates of a circular electric string)

\textit{The state $\ket{\mathring{R},z_0}$ are simultaneously the eigenstate of the components of the position operator $\hat{\mathrm{Q}}_{\rho}$ and $\hat{\mathrm{Q}}_3$:}
\begin{equation}
    \hat{\mathrm{Q}}_{\rho}\ket{\mathring{R},z_0}=
   R\ket{\mathring{R},z_0},
\end{equation}
\begin{equation}
    \hat{\mathrm{Q}}_3\ket{\mathring{R},z_0}=
   z_0\ket{\mathring{R},z_0}.
\end{equation}

PROOF. Thanks to the radial Dirac delta function, the state of an electric string on a circle is even an eigenstate of the nonphysical ordinary operator $\rho$: 
\begin{equation}
    \rho\ket{\mathring{R},z_0}=
   R\ket{\mathring{R},z_0}.
\end{equation}
However, the state itself is physical and satisfies the condition of transversality:
\begin{equation}
\hat{P}\ket{\mathring{R}, z_0}=\ket{\mathring{R}, z_0}, \ \ \ \ \hat{P}_0\ket{\mathring{R}, z_0}=0.
\end{equation}
Therefore, the following occurs:
\begin{equation}
    \hat{\mathrm{Q}}_{\rho}\ket{\mathring{R},z_0}=
   \hat{P}\rho\hat{P}\ket{\mathring{R},z_0}=
   \hat{P}R\ket{\mathring{R},z_0}=
   R\ket{\mathring{R},z_0}.
\end{equation}

The eigenstate of the third position component can be checked by the matrix representation (see \cite{Poster}):
\begin{equation}
    \hat{\mathrm{Q}}|_3\ket{\mathring{R},z_0}=
    \left(\begin{array}{ccc}
z & 0 & -\Delta^{-1}\partial_x\\
0 & z & -\Delta^{-1}\partial_y\\
0 & 0 & z-\Delta^{-1}\partial_z
\end{array} \right)
\left(\begin{array}{c}
\frac{-y}{R}\ \delta(\rho-R)\ \delta(z-z_0)\\
\frac{x}{R}\ \delta(\rho-R)\ \delta(z-z_0)\\
0
 \end{array} \right),
\end{equation}
where a simplified form of the position operator is used, which for transversal states is equivalent to using the general form.
Thanks to the zero in the third component of the state vector, the matrix operation is trivial and reduces to multiplying the Dirac delta term by $z$, so:
\begin{equation}
    \hat{\mathrm{Q}}|_3\ket{\mathring{R},z_0}=
   z\ket{\mathring{R},z_0}=
   z_0\ket{\mathring{R},z_0}.
\end{equation}

It gets a bit more difficult if we use the general, rather than simplified, form of the matrix operator (see \cite{Poster}):
\begin{equation}
    \hat{\mathrm{Q}}_3\ket{\mathring{R},z_0}=
    \left(\begin{array}{ccc}
z & 0 & -\Delta^{-1}\partial_x\\
0 & z & -\Delta^{-1}\partial_y\\
\Delta^{-1}\partial_x & \Delta^{-1}\partial_y & z
\end{array} \right)\ket{\mathring{R},z_0},
\end{equation}
\begin{equation}
    \hat{\mathrm{Q}}_3\ket{\mathring{R},z_0}=
\left(\begin{array}{c}
\frac{-y}{R}\ \delta(\rho-R)z_0\delta(z-z_0)\\
\frac{x}{R}\ \delta(\rho-R)z_0 \delta(z-z_0)\\
\frac{1}{\Delta}\frac{-yx+xy}{R^2}\ \delta'(\rho-R)\delta(z-z_0)
 \end{array} \right)=z_0\ket{\mathring{R},z_0}.
\end{equation}
Therefore, it is an eigenstate of both position components. Q.E.D.

\begin{figure}[h!]
\centering
\includegraphics[width=8.5cm]{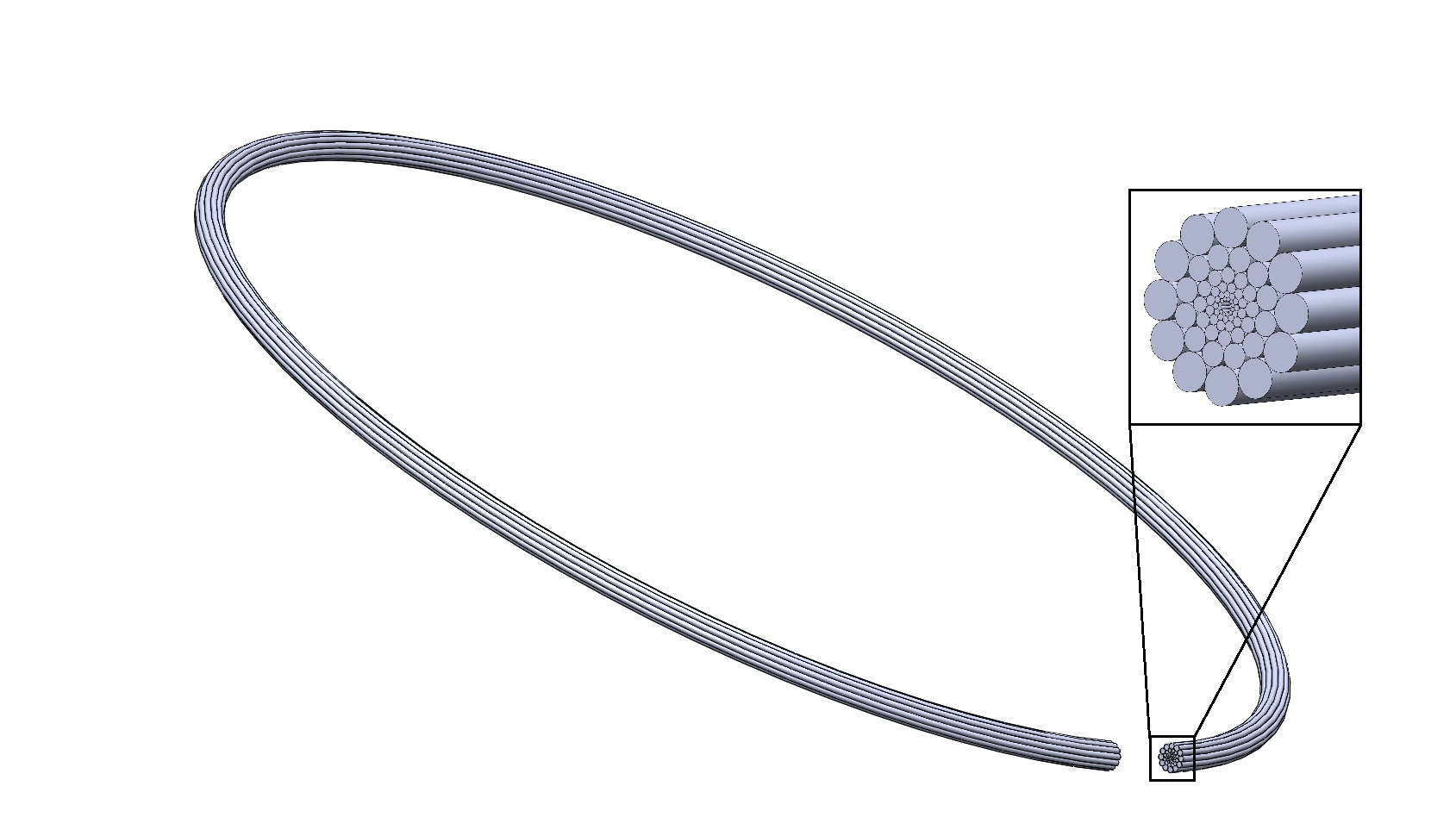}
\caption{Graphical visualisation of the $\ket{\mathring{R},z_0}$ photon state as an electric circular string, which is an infinitely thin and infinitely dense bundle on the circle (vector tangent Dirac delta). The string is closed, but its cross-section is for illustrative purposes only}
\label{state3}
\end{figure}

The above eigenequations and the Dirac deltas in the $\ket{\mathring{R},z_0}$ state vector notation justify the name electric circular string (see Fig.\ref{state3}).

The issue of the eigenvalue of the radial position operator is generally more problematic. It has been proven that the states on a circle are eigenstates of the radial position operator of the first kind (transversal). However, it looks like the states on the circle are not eigenstates of the radial position operator of the second kind (rooter):
\begin{equation}
\label{neq2}
    \hat{\mathrm{Q}}_{[\rho]}\ket{\mathring{R},0}\neq R\ket{\mathring{R},0}\ \ \ \longleftrightarrow \ \ \
    \hat{\mathrm{Q}}^2_{[\rho]}\ket{\mathring{R},0}\neq R^2\ket{\mathring{R},0}.
\end{equation}
This thesis is not proven in this work, but is stated in publication \cite{Jadczyk 2024} for the square of this radial operator.

At the end of this section, it is worth asking whether there are common eigenstates of the position operators $\hat{\mathrm{Q}}_{\rho}$, $\hat{\mathrm{Q}}_3$ and the helicity operator on the circle $\hat{\Lambda}$? It is easy to check that the state (\ref{circle}) is not an eigenstate of helicity. If the operator $\hat{\mathrm{Q}}_{\rho}$ commutes with helicity (commutes if the transversal radial operator is equivalent to an even radial operator), then the states on the circle (\ref{circle}) can be projected and divided into eigenspaces of helicity. However, we already know that the radial transversal and even operators are not equivalent, so the radial transversal operator probably does not commute with helicity. Probably, because for example, the rooter radial operator commutes with helicity, and is not equal to the radial even operator. Therefore, the problem of finding any eigenstates of position and helicity simultaneously is quite complex in the context of a circle. 

Attempts to find such states were already made in a publication from 1973 \cite{Jadczyk}. However, some mistake must have been made there, because the existence of such a state contradicts the Galindo--Amrein theorem (based on correspondence with A. Jadczyk and his post on the blog: https://www.salon24.pl/u/arkadiusz-jadczyk/66463,blad-profesora). This theorem states that there are no photon states with a definite helicity, which are localised in a finite region of space.

It is worth recalling that Bacry in 1988 gave in \cite{Bacry 1988, Bacry 2} a certain class of common eigenstates $\hat{\mathrm{Q}}_3$ and $\hat{\lambda}$, but it has not been examined whether this class includes states on the circle. In any case, this issue will be discussed later in the work and will be partially resolved --  to the greatest extent possible.

\subsection{7.2. Photon circular magnetic string}

In analogy to a linear string, let us consider the magnetic version of a string on a circle:
\begin{equation}
\ket{\mathring{R}, z_0^{\star}}:=\hat{\lambda}\ket{\mathring{R}, z_0}= \frac{1}{\sqrt{-\Delta}}\nabla \times \ket{\mathring{R}, z_0}. 
\end{equation}
Before examining the properties of this state, it is worth computing it explicitly in the positional representation (for $z_0=0$). 

The main difficulty is based on the calculation of the integral operator (\ref{lap-1/2}) from appendix A:
\begin{equation}
    I_1:=\frac{1}{\sqrt{-\Delta}}\Big(\frac{x}{R}\delta(\rho-R)\delta(z)\Big),
\end{equation}
\begin{equation}
    I_1=\frac{1}{2\pi^2R}\int_0^{2\pi}\frac{R \cos(\phi')R d\phi'}{R^2-2R\rho\cos{(\phi'-\phi)}+\rho^2+z^2},
\end{equation}
where polar coordinates are used. By making a simple substitution of the variable $\beta:=\phi'-\phi$ within the full range of the cosine function, we have:
\begin{equation}
    I_1=\frac{R}{2\pi^2}\int_0^{2\pi}\frac{\cos(\phi)\cos{\beta}-\sin{\phi}\sin{\beta} }{R^2-2R\rho\cos{\beta}+\rho^2+z^2}d\beta.
\end{equation}
The integral of the sine disappears, and for the integral of the cosine
we can use doubling the (\ref{Int2}) integral formula from Appendix A:
\begin{equation}
    I_1=\frac{x}{2\pi\rho^2}\Bigg(\frac{R^2+\rho^2+z^2}{\sqrt{(R^2+\rho^2+z^2)^2-4R^2\rho^2}}-1\Bigg).
\end{equation}
An analogous integral is also needed:
\begin{equation}
    I_2:=\frac{1}{\sqrt{-\Delta}}\Big(\frac{y}{R}\delta(\rho-R)\delta(z)\Big),
\end{equation}
\begin{equation}
    I_2=\frac{y}{2\pi\rho^2}\Bigg(\frac{R^2+\rho^2+z^2}{\sqrt{(R^2+\rho^2+z^2)^2-4R^2\rho^2}}-1\Bigg).
\end{equation}
We still need to calculate the rotations:
\begin{equation}
 \ket{\mathring{R},0^{\star}}=\nabla \times
\left(\begin{array}{c}
-I_2\\
I_1\\
0
 \end{array} \right)=
 \left(\begin{array}{c}
-\partial_z I_1\\
-\partial_z I_2\\
\partial_x I_1+\partial_y I_2
 \end{array} \right).
\end{equation}
The first derivative has the value:
\begin{equation}
   \partial_z I_1=\frac{x}{2\pi\rho^2} \cdot \frac{-8 R^2 \rho^2 z}{\sqrt{(R^2+\rho^2+z^2)^2-4R^2\rho^2}^3}.
\end{equation}
The second one is analogous:
\begin{equation}
   \partial_z I_2=\frac{2R^2}{\pi} \cdot \frac{-2y z}{\sqrt{(R^2+\rho^2+z^2)^2-4R^2\rho^2}^3}.
\end{equation}
To calculate the last component, let us use simplifying notations:
\begin{equation}
    F:=\frac{R^2+r^2}{\sqrt{(R^2+r^2)^2-4R^2\rho^2}}.
\end{equation}
It can be calculated that:
\begin{equation}
    \partial_xF:=4R^2x\cdot\frac{R^2-\rho^2+z^2}{\sqrt{(R^2+r^2)^2-4R^2\rho^2}^3},
\end{equation}
\begin{equation}
    \partial_yF:=4R^2y\cdot\frac{R^2-\rho^2+z^2}{\sqrt{(R^2+r^2)^2-4R^2\rho^2}^3}.
\end{equation}
If we notice that:
\begin{equation}
   \partial_x \Big(\frac{x}{2\pi\rho^2}\Big)+\partial_y \Big(\frac{y}{2\pi\rho^2}\Big)=\delta(x)\delta(y),
\end{equation}
we can perform the calculation:
\begin{equation}
   \partial_x I_1+\partial_y I_2=\frac{x}{2\pi\rho^2}\partial_x F+\frac{y}{2\pi\rho^2} \partial_y F.
\end{equation}
The expression $(F-1)$ for $\rho=0$ becomes zero thanks to Dirac deltas.
Inserting partial results, we get:
\begin{equation}
    \partial_x I_1+\partial_y I_2=\frac{2R^2}{\pi}\cdot\frac{R^2-\rho^2+z^2}{\sqrt{(R^2+r^2)^2-4R^2\rho^2}^3}.
\end{equation}
Therefore, the sought state vector of the circular magnetic string takes the final form:
\begin{equation}
 \ket{\mathring{R},0^{\star}}(\mathbf{r})=\frac{2R^2}{\pi}
\left(\begin{array}{c}
\frac{2xz}{\sqrt{(R^2+x^2+y^2+z^2)^2-4R^2(x^2+y^2)}^3}\\
\frac{2yz}{\sqrt{(R^2+x^2+y^2+z^2)^2-4R^2(x^2+y^2)}^3}\\
\frac{R^2-x^2-y^2+z^2}{\sqrt{(R^2+x^2+y^2+z^2)^2-4R^2(x^2+y^2)}^3}
 \end{array} \right).
\end{equation}
The calculated state can now be analysed:
\\
\\
STATEMENT 9 (Magnetic string on a circle)

\textit{The state $\ket{\mathring{R},z_0^{\star}}$ is simultaneously the eigenstate of the helical radial position operator $\hat{\mathrm{\Theta}}_{\rho}$ and the third Cartesian component of position operator $\hat{\mathrm{Q}}_3$:}
\begin{equation}
    \hat{\mathrm{\Theta}}_{\rho}\ket{\mathring{R},z_0^{\star}}=
   R\ket{\mathring{R},z_0^{\star}},
\end{equation}
\begin{equation}
    \hat{\mathrm{Q}}_3\ket{\mathring{R},z_0^{\star}}=
   z_0\ket{\mathring{R},z_0^{\star}}.
\end{equation}

PROOF. The first eigenequation is a simple consequence of the definition of the state, the definition of the radial operator and the properties of the electric string from which the magnetic string arises:
\begin{equation}
    \hat{\mathrm{\Theta}}_{\rho}\ket{\mathring{R},z_0^{\star}}=
\hat{\lambda}\rho\hat{\lambda}\hat{\lambda}\ket{\mathring{R},z_0}=
\hat{\lambda}\rho\ket{\mathring{R},z_0},
\end{equation}
\begin{equation}
\hat{\lambda}\rho\ket{\mathring{R},z_0}=
R\hat{\lambda}\ket{\mathring{R},z_0}=
R\ket{\mathring{R},z_0^{\star}}.
\end{equation}

The proof of the second eigenequation proceeds similarly, where we use Lemma 1 of the commutativity of the position operator with helicity:
\begin{equation}
    \hat{\mathrm{Q}}_3\ket{\mathring{R},z_0^{\star}}=
\hat{\mathrm{Q}}_3\hat{\lambda}\ket{\mathring{R},z_0}=
\hat{\lambda}\hat{\mathrm{Q}}_3\ket{\mathring{R},z_0},
\end{equation}
\begin{equation}
\hat{\lambda}\hat{\mathrm{Q}}_3\ket{\mathring{R},z_0}=
\hat{\lambda}z_0\ket{\mathring{R},z_0}=
z_0\ket{\mathring{R},z_0^{\star}}.
\end{equation}

However, the above proof assumes that the state has been correctly calculated. It is therefore worthwhile, at least in one case, to perform a direct calculating check. Let us hence check computationally how it is possible that a function that does not vanish outside the plane $z_0=0$ is an eigenfunction of a unique third-component position operator.

Let's start simplifying the state notation:
\begin{equation}
\mathbf{g}:=\frac{\pi}{2R^2}
 \ket{\mathring{R},0^{\star}}=
\left(\begin{array}{c}
\frac{2xz}{\sqrt{(R^2+r^2)^2-4R^2\rho^2}^3}\\
\frac{2yz}{\sqrt{(R^2+r^2)^2-4R^2\rho^2}^3}\\
\frac{R^2-\rho^2+z^2}{\sqrt{(R^2+r^2)^2-4R^2\rho^2}^3}
 \end{array} \right).
\end{equation}
The eigenequation of the third component of the position should have the form (according to the transversal definition):
\begin{equation}
    z\ \mathbf{g}-\Delta^{-1}\nabla (\nabla\cdot z\mathbf{g})
   \overset{?}{=}
    0\ \mathbf{g}.
\end{equation}
Since $\nabla\cdot\mathbf{g}=0$, the equation can be transformed equivalently:
\begin{equation}
    z\ \mathbf{g}\overset{?}{=}
    \Delta^{-1}\nabla (\mathrm{g}_z).
\end{equation}
For the further equivalent transformation, we need to weaken this equation by a harmonic functions vector $\mathbf{h}_f$ ($\Delta\mathbf{h}_f=0$) defined over the entire space:
\begin{equation}
    z\ \mathbf{g}+\mathbf{h}_f\overset{?}{=}
    \Delta^{-1}\nabla (\mathrm{g}_z) \ \
    \longleftrightarrow \ \ 
    \Delta(z\ \mathbf{g})\overset{?}{=}
    \nabla (\mathrm{g}_z).
\end{equation}
Checking the last equality will be a sufficient computational check of the thesis, due to the unlimitedness of the harmonic function (e.g. linear) and the previous formal proof. We can calculate using Wolfram Alpha that for the first component we have:
\begin{equation}
\Delta(z\mathrm{g}_x)=
     4x\frac{(R^2-\rho^2)^2-R^2z^2-\rho^2z^2}
     {\sqrt{(R^2+r^2)^2-4R^2\rho^2}^5}=
     \partial_x\mathrm{g}_z.
\end{equation}
Due to axial symmetry, the second component holds similarly. The third component remains:
\begin{equation}
\Delta(z\mathrm{g}_z)=
     4z\frac{(2\rho^2+z^2-R^2)\rho^2-(R^2+z^2)^2}
     {\sqrt{(R^2+r^2)^2-4R^2\rho^2}^5}=
     \partial_z\mathrm{g}_z.
\end{equation}
So the equality was confirmed by calculation. Q.E.D.

\begin{figure}[h!]
\centering
\includegraphics[width=8.5cm]{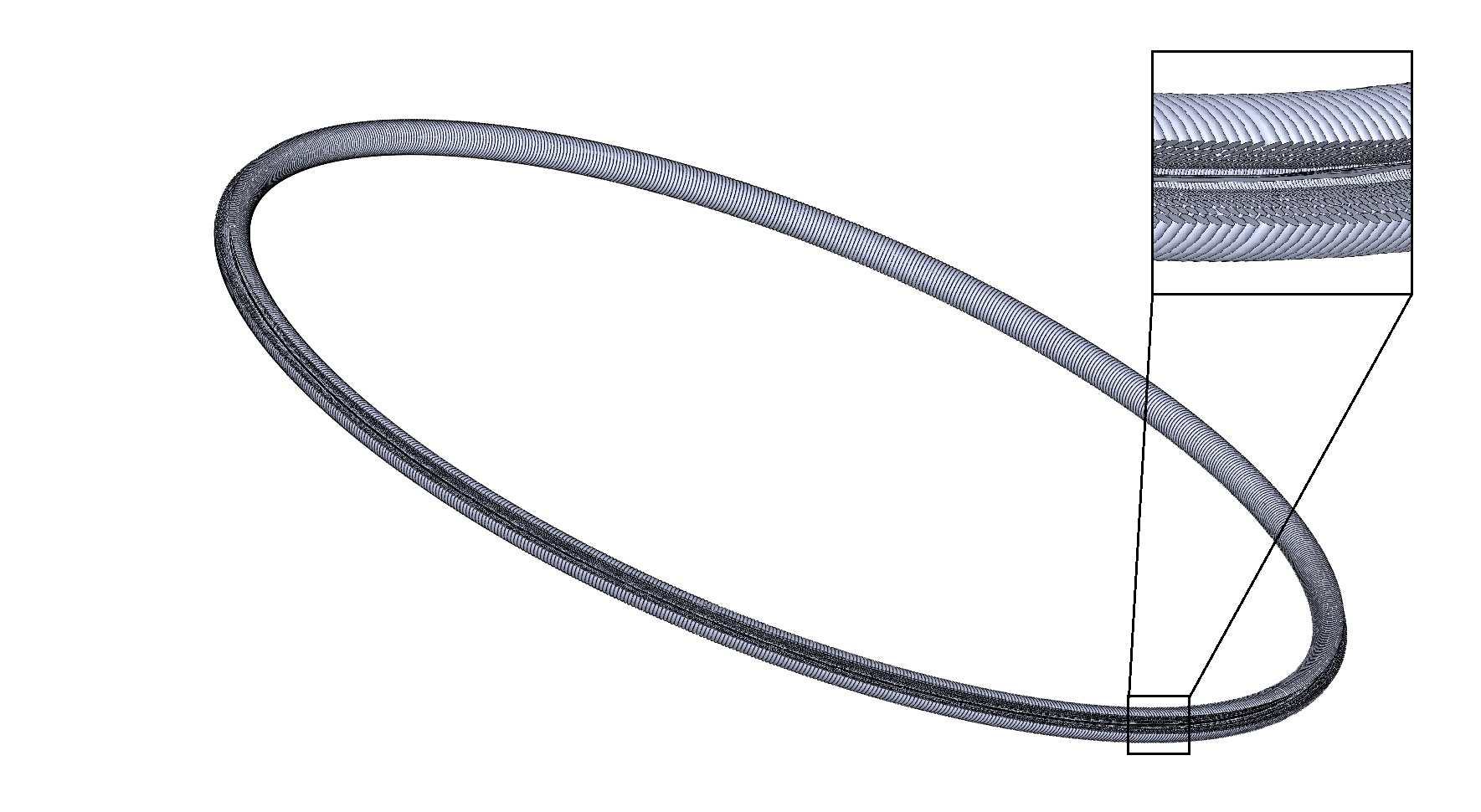}
\caption{Graphical visualization of the $\ket{\mathring{R},z_0^{\star}}$ photon state as a circular magnetic string, which in the $\hat{\mathrm{p}}\ket{x_0^{\star},y_0^{\star}}(\mathbf{r})$ "energetic" representation is an infinitely tight and dense spiral vector field on the circle (a rotation of the vector tangent Dirac delta). The small rings are complete but have been cut for illustrative purposes. Basically, this magnetic string is closed, but it can be opened by separating but not cutting the small rings -- without violating the transversality condition of the state}
\label{state4}
\end{figure}

The above statement constitutes the shape of the circle of the considered state called a circular magnetic string. This is also presented in the visualisation on Fig.\ref{state4}.

Unfortunately, we only know that the radial eigenequation is satisfied only by the helical radial operator. It can be checked that the two simplest radial operators do not satisfy this eigenequation:
\\
\\
STATEMENT 10 (Broken some radiality)

\textit{The state $\ket{\mathring{R},z_0^{\star}}$ is not the eigenstate of the tranversal radial operator $\hat{\mathrm{Q}}_{\rho}$ and square radial operator $\hat{\mathrm{Q}}^2_{\rho^2}$:}
\begin{equation}
    \hat{\mathrm{Q}}_{\rho}\ket{\mathring{R},z_0^{\star}}\neq
   R\ket{\mathring{R},z_0^{\star}},
\end{equation}
\begin{equation}
    \hat{\mathrm{Q}}^2_{\rho^2}\ket{\mathring{R},z_0^{\star}}\neq
   R^2\ket{\mathring{R},z_0^{\star}}.
\end{equation}

PROOF. Since the statement is negative, it is enough to show that equality does not hold for the example parameters $R=1$, $z_0=0$ (default in meters):
\begin{equation}
\mathbf{g}=
\left(\begin{array}{c}
\frac{2xz}{\sqrt{(1+r^2)^2-4\rho^2}^3}\\
\frac{2yz}{\sqrt{(1+r^2)^2-4\rho^2}^3}\\
\frac{1-\rho^2+z^2}{\sqrt{(1+r^2)^2-4\rho^2}^3}
 \end{array} \right).
\end{equation}

In the proof by contradiction for the transversal operator, we should assume:
\begin{equation}
    \rho \mathbf{g}-\Delta^{-1}\nabla (\nabla\cdot \rho\mathbf{g})
   \overset{?}{=}
    \mathbf{g}.
\end{equation}
The divergence can be easily calculated:
\begin{equation}
   \nabla\cdot \rho\mathbf{g}=\mathbf{g}\cdot \nabla\rho=
   \frac{2\rho z}{\sqrt{(1+r^2)^2-4\rho^2}^3}=:\mathrm{g}_{\rho}.
\end{equation}
Therefore, after the transformations, the equality should hold:
\begin{equation}
    \Delta(\rho\mathbf{g}-\mathbf{g})
    \overset{?}{=}
    \nabla \mathrm{g}_{\rho}.
\end{equation}
Let's consider the first component ($x$):
\begin{equation}
    \Delta(\rho\mathrm{g}_x-\mathrm{g}_x)
    \overset{?}{=}
    \partial_x \mathrm{g}_{\rho}.
\end{equation}
The calculations can be done in Wolfram Alpha and substitute point $(1,0,1)$:
\begin{equation}
    \Delta(\rho\mathrm{g}_x-\mathrm{g}_x)\big\rvert_{x=1,y=0,z=1}=\frac{6}{\sqrt{5}^5},
\end{equation}
\begin{equation}
    \partial_x \mathrm{g}_{\rho}\big\rvert_{x=1,y=0,z=1}=\frac{-2}{\sqrt{5}^5}.
\end{equation}
So we see that equality does not hold:
\begin{equation}
    \Delta(\rho\mathrm{g}_x-\mathrm{g}_x)
    \neq
    \partial_x \mathrm{g}_{\rho}.
\end{equation}
Similarly, we check for the square radial operator:
\begin{equation}
    \Delta(\rho^2\mathrm{g}_x-\mathrm{g}_x)
    \overset{?}{=}
    \partial_x (2\rho\mathrm{g}_{\rho}),
\end{equation}
\begin{equation}
    \Delta(\rho^2\mathrm{g}_x-\mathrm{g}_x)\big\rvert_{x=1,y=0,z=1}=\frac{32}{\sqrt{5}^5},
\end{equation}
\begin{equation}
    \partial_x(2\rho\mathrm{g}_{\rho})\big\rvert_{x=1,y=0,z=1}=\frac{16}{\sqrt{5}^5},
\end{equation}
\begin{equation}
    \Delta(\rho^2\mathrm{g}_x-\mathrm{g}_x)
    \neq
    \partial_x (2\rho\mathrm{g}_{\rho}).
\end{equation}
So this equality also does not hold. Q.E.D.

Despite the breaking of radiality in the transversal sense, helical radiality still holds. Additionally, the eigenvalue of the third component of position is satisfied. Therefore, a magnetic string on a circle should be treated on an equal level with an electric string on a circle.

\subsection{7.3. Helical strings on a circle}

Helical states with a defined helicity can easily be constructed from electric and magnetic strings. Previously, this was done for strings on a straight line, and now it will be done for strings on a circle. 

The definition is analogous and differs only in the appropriate notations:
\begin{equation}
\ket{\mathring{\tilde{R}},z_0,\pm }:=
\frac{1}{\sqrt{2}}\ket{\mathring{R},z_0}\pm \frac{1}{\sqrt{2}}\ket{\mathring{R},z_0^{\star}}. 
\end{equation}
The action of the helicity operator on this state turns the magnetic string into an electric string and vice versa (changes the position of the star), thus not changing the right-handed (positive) state, but changing to the opposite left-handed (negative) state:
\begin{equation}
\hat{\lambda}\ket{\mathring{\tilde{R}},z_0,\pm }=\pm
\ket{\mathring{\tilde{R}},z_0,\pm }. 
\end{equation}
The helical states are built from the eigenstates of the third position component, so they are also eigenstates of this component:
\begin{equation}
\hat{\mathrm{Q}}_3\ket{\mathring{\tilde{R}},z_0,\pm }=z_0
\ket{\mathring{\tilde{R}},z_0,\pm }. 
\end{equation}

Unfortunately, these states are not eigenstates of either the radial transversal or the helical operator:
\begin{equation}
\hat{\mathrm{Q}}_{\rho}\ket{\mathring{\tilde{R}},z_0,\pm }\neq
R
\ket{\mathring{\tilde{R}},z_0,\pm }
\neq
\hat{\mathrm{\Theta}}_{\rho}\ket{\mathring{\tilde{R}},z_0,\pm }. 
\end{equation}
The following formulas, which actually refer to electric and magnetic strings, can be a certain substitute for the radial nature of helical states:
\begin{equation}
\hat{\mathrm{Q}}_{\rho}\big(\ket{\mathring{\tilde{R}},z_0,+}+\ket{\mathring{\tilde{R}},z_0,-}\big)=R
\big(\ket{\mathring{\tilde{R}},z_0,+}+\ket{\mathring{\tilde{R}},z_0,-}\big), 
\end{equation}
\begin{equation}
\hat{\mathrm{\Theta}}_{\rho}\big(\ket{\mathring{\tilde{R}},z_0,+}-\ket{\mathring{\tilde{R}},z_0,-}\big)=R
\big(\ket{\mathring{\tilde{R}},z_0,+}-\ket{\mathring{\tilde{R}},z_0,-}\big). 
\end{equation}
Due to its non-trivial nature, the radial symbol $R$ is marked with a tilde (with a circle, as before). One can say that the radiativity problems of helicity eigenstates reflect the Galindo--Amrein theorem, which denies the existence of localised helicity eigenstates.

\section{8. Photon state in the vicinity of a point on a plane -- flat photon vortex}

We will now look for a photon state localised on the plane -- the eigenstate of $\hat{\mathrm{Q}}_3$ -- but localised as close as possible to the origin of the coordinate system. We assume that the location in the small circle is not the location in the center of this circle. 

From the minimal form of the position operator \cite{Poster}:
\begin{equation}
    \hat{\mathrm{Q}}|_3=
    \left(\begin{array}{ccc}
z & 0 & -\Delta^{-1}\partial_x\\
0 & z & -\Delta^{-1}\partial_y\\
0 & 0 & z-\Delta^{-1}\partial_z
\end{array} \right),
\end{equation}
it follows that it will be easiest if the desired state has a zero third component. So first, let's consider a non-physical but simple state of the above type:
\begin{equation}
    \mathbf{\Psi}_0=
    \left(\begin{array}{c}
\delta(x)\delta(y)\delta(z)\\
\delta(x)\delta(y)\delta(z)\\
0
\end{array} \right).
\end{equation}
This state does not meet the condition of transversality:
\begin{equation}   
    \nabla\cdot\mathbf{\Psi}_0=
   \delta'(x)\delta(y)\delta(z)+\delta(x)\delta'(y)\delta(z)\neq 0,
\end{equation}
which can be easily corrected by the state with antisymmetric derivatives:
\begin{equation}
    \mathbf{\Psi}'=
    \left(\begin{array}{c}
\delta(x)\delta'(y)\delta(z)\\
-\delta'(x)\delta(y)\delta(z)\\
0
\end{array} \right),
\end{equation}
\begin{equation}   
    \nabla\cdot\mathbf{\Psi}'=
   \delta'(x)\delta'(y)\delta(z)-\delta'(x)\delta'(y)\delta(z)=0.
\end{equation}
We can obtain such a transversal state vector as a rotation of the vector with the third component being the Dirac delta:
\begin{equation}
    \mathbf{\Psi}'=\nabla\times
    \left(\begin{array}{c}
0\\
0\\
\delta(x)\delta(y)\delta(z)
\end{array} \right).
\end{equation}

These types of physical, although generalised (distributions), states are not normalised to Dirac deltas -- except for the $z$ variable. This can be easily corrected by dividing the state by the square root of the minus two-dimensional Laplacian (similar to the earlier use of helicity on the states of an open electric string):
\begin{equation}
    \ket{\tilde{x}_0=0,\tilde{y}_0=0,z_0=0}=
    \left(\begin{array}{c}
\delta(z)\frac{1}{\sqrt{-\Delta}_2}\big(\delta(x)\delta'(y)\big)\\
-\delta(z)\frac{1}{\sqrt{-\Delta}_2}\big(\delta'(x)\delta(y)\big)\\
0
\end{array} \right).
\end{equation}

Before we calculate this state explicitly and check whether it is the eigenstate of $\hat{\mathrm{Q}}_3$, let's check the normalisation condition:
\begin{displaymath}
\braket{\tilde{x}_1,\tilde{y}_1,z_1|\tilde{x}_2,\tilde{y}_2,z_2}=
\end{displaymath}
\begin{displaymath}
\delta_{(z_1-z_2)}\iint{\delta_{(x-\tilde{x}_1)}\delta_{(y-\tilde{y}_1)} 
    \frac{\delta_{(x-\tilde{x}_2)}\delta''_{(y-\tilde{y}_2)}+\delta''_{(x-\tilde{x}_2)}\delta_{(y-\tilde{y}_2)}}{\Delta_2}dxdy}=
\end{displaymath}
\begin{equation}
    \delta(\tilde{x}_1-\tilde{x}_2) \delta(\tilde{y}_1-\tilde{y}_2)\delta(z_1-z_2).
\end{equation}
The Hermitian properties of the Laplacian and the anti-Hermitian properties of partial derivatives were used in the calculations.

Thanks to the formula (\ref{lap 2D}) from Appendix A, we can more clearly express the state located around the origin:
\begin{equation}
    \ket{\tilde{0},\tilde{0},0}=\frac{1}{2\pi}
    \left(\begin{array}{c}
\partial_y \big(\frac{1}{\sqrt{x^2+y^2}}\big)\delta(z)\\
-\partial_x \big(\frac{1}{\sqrt{x^2+y^2}}\big)\delta(z)\\
0
\end{array} \right).
\end{equation}
The apparently simple derivatives are not explicitly calculated here, since they have distributional values.
The values of zero for the coordinates of a point in Dirac brackets should not be confused with a similar notation for a vacuum. More generally, we can write:
\begin{equation}
\label{flat vortex}
    \ket{\tilde{x}_0,\tilde{y}_0,z_0}=\frac{1}{2\pi}
    \left(\begin{array}{c}
\partial_y \Big(\frac{1}{\sqrt{(x-x_0)^2+(y-y_0)^2}}\Big)\delta(z-z_0)\\
-\partial_x \Big(\frac{1}{\sqrt{(x-x_0)^2+(y-y_0)^2}}\Big)\delta(z-z_0)\\
0
\end{array} \right).
\end{equation}
The following can be stated about the state calculated in this way:
\\
\\
STATEMENT 11 (About photon flat vortex)

\textit{The state $\ket{\tilde{x}_0,\tilde{y}_0,z_0}$ is an infinitely short magnetic open string:
\begin{equation}
\label{short string}
    \ket{\tilde{x}_0,\tilde{y}_0,z_0}=
   \ket{x_0^{\star},y_0^{\star}}\delta(z-z_0),
\end{equation}
so that it is an eigenstate of the third component of the position operator $\hat{\mathrm{Q}}_3$:
\begin{equation}
    \hat{\mathrm{Q}}_3\ket{\tilde{x}_0,\tilde{y}_0,z_0}=
   z_0\ket{\tilde{x}_0,\tilde{y}_0,z_0},
\end{equation}
but loses the eigenstates property for $\hat{\mathrm{Q}}_1$ and $\hat{\mathrm{Q}}_2$:}
\begin{equation}
    \hat{\mathrm{Q}}_1\ket{\tilde{x}_0,\tilde{y}_0,z_0}\neq
   x_0\ket{\tilde{x}_0,\tilde{y}_0,z_0},
\end{equation}
\begin{equation}
    \hat{\mathrm{Q}}_2\ket{\tilde{x}_0,\tilde{y}_0,z_0}\neq
   y_0\ket{\tilde{x}_0,\tilde{y}_0,z_0}.
\end{equation}

PROOF. The relation (\ref{short string}) between the states of the flat photon vortex (\ref{flat vortex}) and the open magnetic string (\ref{magnetic}) follows clearly from the form of these states. It is worth noting that multiplication by the $z$-dependent Dirac delta does not violate the transversality condition of the wave vector, which has zero third component.

The pure $z$-dependent Dirac delta in the transverse state vector also provides an eigenstate of this component:
\begin{equation}
z\ket{\tilde{x}_0,\tilde{y}_0,z_0}=z_0\ket{\tilde{x}_0,\tilde{y}_0,z_0},
\end{equation}
\begin{equation}
\hat{\mathrm{Q}}_3\ket{\tilde{x}_0,\tilde{y}_0,z_0}=\hat{P}z\ket{\tilde{x}_0,\tilde{y}_0,z_0}=z_0\ket{\tilde{x}_0,\tilde{y}_0,z_0}.
\end{equation}

The absence of an eigenstate of the first and second position components will be proven in the position representation, without loss of generality, for the case $x_0=y_0=z_0=0$. Due to symmetry, limiting ourselves to the first position component is sufficient. Consider the auxiliary state vector:
\begin{equation}
   \mathbf{k}:=2\pi \ket{\tilde{0},\tilde{0},0}=
    \left(\begin{array}{c}
\partial_y \big(\rho^{-1}\big)\delta(z)\\
-\partial_x \big(\rho^{-1}\big)\delta(z)\\
0
\end{array} \right).
\end{equation}
In the proof by contradiction, the assumption of an eigenstate of the first position component means:
\begin{equation}
    x\mathbf{k}-\Delta^{-1}\nabla (\nabla\cdot x\mathbf{k})
   \overset{?}{=}
    0\ \mathbf{k}.
\end{equation}
This implies the following equation:
\begin{equation}
    \Delta(x\mathbf{k})\overset{?}{=}
    \nabla \mathrm{k}_x,
\end{equation}
in particular, for the first component of the state vector:
\begin{equation}
    \Delta(x\mathrm{k}_x)\overset{?}{=}
    \partial_x \mathrm{k}_x.
\end{equation}
It can be calculated directly:
\begin{equation}
    \Delta_2\big(x\partial_y \rho^{-1}\big)\delta(z)+
    \big(x\partial_y \rho^{-1}\big)\delta''(z)
    \overset{?}{=}
    \big(\partial_x \partial_y \rho^{-1}\big)\delta(z).
\end{equation}
The above equality is only satisfied for the two-dimensional part (at the distribution and ordinary level):
\begin{equation}
    \Delta_2\big(x\partial_y \rho^{-1}\big)\equiv
    \partial_x \partial_y \rho^{-1},
\end{equation}
\begin{equation}
    \Delta_2\big(x\partial_y \rho^{-1}\big)=
    \frac{3xy}{\rho^5}=
    \partial_x \partial_y \rho^{-1}, \ \ \ \rho\neq 0.
\end{equation}
This part of the equality confirms the proof for the infinite magnetic string, which was previously carried out using a different method. However, equality at the distributional level is not proven here, because the discussed equality in three dimensions does not hold:
\begin{equation}
    0\neq\big(x\partial_y \rho^{-1}\big)\delta''(z)
    \overset{?}{=}
    0.
\end{equation}
Lack of equality ends the proof by contradiction. Q.E.D.

The considered flat vortex state $\ket{\tilde{x}_0,\tilde{y}_0,z_0}$ is not eigenstate of the $\hat{\mathrm{Q}}_1$, $\hat{\mathrm{Q}}_2$ components of the photon position operator. Therefore, the coordinates $\tilde{x}_0$, $\tilde{y}_0$ in the designation of these state are marked with a tilde. 

 Based on the above form of the state of a photon located on the $z=0$ plane, we can conclude that the amplitude of the probability density of finding a photon is inversely proportional to the square of the distance from the highlighted point. Therefore, the probability density should be inversely proportional to the fourth power of the distance from the singular point. Therefore, the point singularity in the plane is significant, although it is not a Dirac delta.

\section{9. Two types of probability densities and two different representations}

In this work, the simplest form of the dot product of states vectors is preferred (\ref{skalarny}):
\begin{equation}
\label{scalar}
(\Phi |\Psi)
=
\int{\Phi_l^*(\mathbf{r}) \Psi_l(\mathbf{r})d^3\mathrm{r}}.
\end{equation}
Contrary to appearances, behind this simple choice there is a non-trivial representation of the wave function vector (\ref{Psi definition}) or equivalently (\ref{Psi from A}), (\ref{Psi from E}) and (\ref{Psi from B and E}). Let's choose the last form, in which we do not have to consider time dependencies (derivatives and decompositions into positive frequencies):
\begin{equation}
\label{Psi nominal definition}
\mathbf{\Psi}(\mathbf{r})
:=
\frac{\nabla\times\mathbf{B}}{\sqrt[4]{-\Delta}^3}-\frac{i}{\sqrt[4]{-\Delta}}\frac{1}{c}\mathbf{E}.
\end{equation}

Apart from the time dependence, the multiplicative constant in the form of the square root of two (compare with \cite{Koczan 2002}) has been omitted here for convenience. The $\mathbf{B}$ and $\mathbf{E}$ fields can be interpreted as magnetic and electric fields in the sense of first quantisation and close connection with $\mathbf{\Psi}$. More precisely, in the Coulomb gauge, the magnetic and electric fields are expressed by the vector potential $\mathbf{A}$, and the state vector $\mathbf{\Psi}$ is an appropriately normalised part of this potential with positive frequencies.

The simple form of the scalar product appears to provide a straightforward interpretation of the probability density (as the square of the state vector module). However, the question arises whether this approach is unique and whether there is not some transformation that changes the representation but does not change the form and the scalar product. From a mathematical point of view, the existence of such transformations seems obvious, but it does not have to be obvious from the point of view of physical interpretation. The collected set of eigenstates of the photon position operator shows that the probability density of the photon must be understood in at least two ways.

This is shown by analysing the state vectors of magnetic strings (rectilinear and circular), which in the adopted representation do not disappear outside the regions of the eigenvalues of the position coordinates (even Cartesian ones). 

Let's define a new representation of the state vector (but not a new different state vector), which we will denote by a star (not to be confused with the complex conjugate asterisk):
\begin{equation}
\{\mathbf{\Phi}\}^{\star}:=\hat{\lambda}\mathbf{\Phi}=:\mathbf{\Phi}^{\star}, \ \ \ \ 
\{\mathbf{\Psi}\}^{\star}:=\hat{\lambda}\mathbf{\Psi}=:\mathbf{\Psi}^{\star}.
\end{equation}
Due to the use of the helicity operator, this representation will be called the helical representation. The helicity operator is invertible on physical states, so it could be used to change the basis of the state vector. The helical representation with respect to the nominal one (called heterovariant in work \cite{Koczan 2002}) has very interesting properties:
\\
\\
\\
\\
THEOREM 4 (About helical representation)

\textit{The helical representation is an invariant of the standard form of the ordinary scalar product (in particular, in the positional representation):
\begin{equation}
\int{\Phi_l^{\star*}(\mathbf{r}) \Psi_l^{\star}(\mathbf{r})d^3\mathrm{r}}=
\int{\Phi_l^*(\mathbf{r}) \Psi_l(\mathbf{r})d^3\mathrm{r}}.
\end{equation}
Additionally, in the helical representation of the state vector, the magnetic field becomes an imaginary electric field, but the electric field becomes an imaginary magnetic field:
\begin{equation}
    \mathbf{B}\longrightarrow \frac{1}{ic}\mathbf{E}, \ \ \ \ \ \
    \mathbf{E}\longrightarrow ic\mathbf{B}
    \end{equation}
which based on (\ref{Psi nominal definition}) means that:}
\begin{equation}
\label{Psi star}
\mathbf{\Psi}^{\star}(\mathbf{r})
=
-\frac{i}{c}\frac{\nabla\times\mathbf{E}}{\sqrt[4]{-\Delta}^3}+\frac{1}{\sqrt[4]{-\Delta}}\mathbf{B}.
\end{equation}

PROOF. Since $\hat{\lambda}^2=\hat{P}$, the nominal representation is expressed symmetrically using the helical representation:
\begin{equation}
\label{inverse helical}
\mathbf{\Phi}=\hat{\lambda}\mathbf{\Phi}^{\star}, \ \ \ \ 
\mathbf{\Psi}=\hat{\lambda}\mathbf{\Psi}^{\star}.
\end{equation}
Therefore, the scalar product (\ref{scalar}) of states can be written as follows:
\begin{equation}
(\Phi |\Psi)
=
(\hat{\lambda}\Phi^{\star} |\hat{\lambda}\Psi^{\star})
=
(\Phi^{\star} |\hat{\lambda}^{\dagger}\hat{\lambda}\Psi^{\star}).
\end{equation}
which by virtue of the Hermitianity $\hat{\lambda}^{\dagger}=\hat{\lambda}$ leads to the thesis:
\begin{equation}
(\Phi |\Psi)
=
(\Phi^{\star} |\hat{\lambda}^2\Psi^{\star})=
(\Phi^{\star} |\hat{P}\Psi^{\star})=
(\Phi^{\star} |\Psi^{\star}).
\end{equation}

Let us compute the helical representation explicitly:
\begin{equation}
\mathbf{\Psi}^{\star}=
\hat{\lambda}\mathbf{\Psi}=
\frac{1}{\sqrt{-\Delta}}\nabla\times\mathbf{\Psi},
\end{equation}
\begin{equation}
\mathbf{\Psi}^{\star}
=
\frac{\nabla\times(\nabla\times\mathbf{B})}{\sqrt[4]{-\Delta}^5}-\frac{i}{\sqrt[4]{-\Delta}^3}\frac{1}{c}\nabla\times\mathbf{E}.
\end{equation}
Using the identities $\nabla\times(\nabla\times\mathbf{B})=\nabla(\nabla\cdot\mathbf{B})-\Delta\mathbf{B}$ and $\nabla\cdot\mathbf{B}=0$ we get:
\begin{equation}
\mathbf{\Psi}^{\star}
=
\frac{1}{\sqrt[4]{-\Delta}}\mathbf{B}-\frac{i}{\sqrt[4]{-\Delta}^3}\frac{1}{c}\nabla\times\mathbf{E},
\end{equation}
which, up to the order, coincides with the proven form of the helical representation. It is also visible that the $\mathbf{E}\rightleftarrows ic\mathbf{B}$ fields have been mutually swapped. Q.E.D.

The above theorem justifies some previous results and leads to further important conclusions. Since the nominal representation (\ref{Psi nominal definition}) favours the electric field over the magnetic field, the name electric strings is justified. In contrast, the helical representation (\ref{Psi star}) favours the magnetic field and justifies the name magnetic strings. The helical representation leads to further synthetic observations:
\\
\\
STATEMENT 12 (The importance of magnetic states and radial operators)

\textit{In the positional helical representation, the magnetic string states (linear and circular) look like electric strings, so they are Dirac deltas on these strings:
\begin{equation}
\ket{x_0^{\star}, y_0^{\star}}^{\star}(\mathbf{r})=\ket{x_0, y_0}(\mathbf{r}), 
\end{equation}
\begin{equation}
    \ket{\mathring{R},z_0^{\star}}^{\star}(\mathbf{r})=
   \ket{\mathring{R},z_0}(\mathbf{r}).
\end{equation}
Moreover, in the helical representation, a radial helical operator looks like a radial transversal operator, and a transversal operator looks like a helical operator:}
\begin{equation}
\hat{\mathrm{\Theta}}_{\rho}^{\star}=
\hat{\mathrm{Q}}_{\rho}, 
\end{equation}
\begin{equation}
\hat{\mathrm{Q}}_{\rho}^{\star}=
\hat{\mathrm{\Theta}}_{\rho}.
\end{equation}
PROOF. According to the definition of the helical representation applied to a magnetic string on a straight line, we have:
\begin{equation}
\ket{x_0^{\star}, y_0^{\star}}^{\star}(\mathbf{r})=\hat{\lambda}\ket{x_0^{\star}, y_0^{\star}}(\mathbf{r}). 
\end{equation}
Which, by Theorem 2, leads to the right-hand side of the first thesis:
\begin{equation}
\hat{\lambda}\ket{x_0^{\star}, y_0^{\star}}(\mathbf{r})=
\ket{x_0, y_0}(\mathbf{r}). 
\end{equation}
Similarly, for a string on a circle:
\begin{equation}
\ket{\mathring{R}, z_0^{\star}}^{\star}(\mathbf{r})=\hat{\lambda}\ket{\mathring{R}, z_0^{\star}}(\mathbf{r}). 
\end{equation}
Transforming the right-hand side using the definition of a magnetic string on a circle, we arrive at the right-hand side of the second thesis:
\begin{equation}
\hat{\lambda}\ket{\mathring{R}, z_0^{\star}}(\mathbf{r})=
\hat{\lambda}^2\ket{\mathring{R}, z_0}(\mathbf{r})=
\ket{\mathring{R}, z_0}(\mathbf{r}). 
\end{equation}

From (\ref{inverse helical}) and the definitions of the helical and transversal radial operators, we have:
\begin{equation}
\hat{\mathrm{\Theta}}_{\rho}^{\star}=
\hat{\lambda}^{\dagger}
\hat{\mathrm{\Theta}}_{\rho}\hat{\lambda}=
\hat{\lambda}^2\rho\hat{\lambda}^2=
\hat{P}\rho\hat{P}=
\hat{\mathrm{Q}}_{\rho}.
\end{equation}
Similarly:
\begin{equation}
\hat{\mathrm{Q}}_{\rho}^{\star}=
\hat{\lambda}^{\dagger}
\hat{\mathrm{Q}}_{\rho}\hat{\lambda}=
\hat{\lambda}\hat{P}\rho\hat{P}\hat{\lambda}=
\hat{\lambda}\rho\hat{\lambda}=
\hat{\mathrm{\Theta}}_{\rho},
\end{equation}
which is the final and fourth thesis. Q.E.D.

Thus, the two most important radial operators in this work have swapped places in the helical representation, which indicates their close relationship. On the other hand, the Cartesian photon position operator is physically invariant under the transformation to the helical representation:
\begin{equation}
\hat{\mathbf{Q}}^{\star}=
\hat{\mathbf{Q}}.
\end{equation}
The physical equivalence "=" implies equality on physical transverse states, since the helical representation cancels the longitudinal (zero) mode and prevents the consideration of the identity equality "$\equiv$" of operators. The above equality is a consequence of the physical equivalence of all Cartesian photon position operator definitions.

A proven statement shows that string magnetic states are as important as string electric states. The dual nature of the probability density in the nominal and helical representations seems to reflect the natural duality of the electro-magnetic field. However, this duality is difficult to interpret mathematically and physically. For example, the geometric interpretation of magnetic strings in the helical representation looks the same as the previously considered electric strings (in the nominal representation). 

Therefore, in order to capture the significant difference of the magnetic string states in Fig. \ref{state2} and Fig. \ref{state4}, a slightly simpler representation (compared to the helical one) is used:
\begin{equation}
\{\mathbf{\Psi}\}^E:=\frac{1}{\hbar}\hat{\mathrm{p}}\ \mathbf{\Psi}=\sqrt{-\Delta}\mathbf{\Psi}=\nabla\times\mathbf{\Psi}^{\star}.
\end{equation}
Due to the use of the energy operator, we can call it an "energetic" representation. The "energetic" representation is, in that sense, simpler than the helical representation, just as the energy operator is simpler than the helicity operator.

\section{10. Conclusions}

\subsection{10.1. Detailed summary of the conclusions obtained}
The first goal of providing at least three different but consistent and equivalent definitions of the photon position operator has been achieved (see Tab. \ref{Cartesian definitions} in Appendix C). The general definitions based on the boost generator (energy-mass center), the transversality condition (with added zero mode), and the helicity projective operators turned out to be identity-equivalent even on the Hilbert space extended with zero mode (for domain and codomain). Additionally, operators defined only in the codomain in the physical space (does not apply to the minimal form of position operator) also turned out to be, in a certain sense, identity equivalent (in the sense of extended domain). Finally, all definitions of the photon position operator (including the minimal form of the position operator) turned out to be equivalent on physical states (for domain and codomain). The way to achieve the goal was, among other things, to distinguish between identity equality and equality on physical states (transversal). It turned out that this distinction can also be correctly applied to operators with values only in physical states (codomain) and arguments that are not necessarily physical (extended domain). The greatest ambiguity of equivalent forms occurs for the unextended domain of physical states. The greatest uniqueness and the greatest simplicity of the identity forms occur for the extended domain and extended codomain of states. The use of extended spaces is not unphysical because the position operator does not mix physical and unphysical states (see Tab. \ref{transformations} in Appendix C).

The above concerns the Cartesian components of the photon position operator. In the case of radial operators (for the radial component), the situation becomes more complicated. The radial operators are hard in themselves, regardless of whether it is unique or not. Nevertheless, in complete analogy to the Cartesian components, it was possible to define the most useful radial operators: transversal and helical. Unlike the Cartesian components, the radial operator versions were not found to be entirely equivalent (see Tab. \ref{radial operators} in Appendix D). However, it has been noticed that these two versions of the radial operator transform into each other under the transformation to the helical representation. This means that a radial transversal operator and a radial helical operator can be treated as helically conjugate twins.

The second main goal of finding the common eigenstates of the two components of the photon position operator was also achieved (see Tab. \ref{eigenvalues} in Appendix F).
Despite the lack of commuting of these components, this was possible because the commutator proportional to the spin component zeroed on the eigenstates. If we talk about photon states located along a straight line, we call them open infinite photon strings. States are also given for closed strings -- circles that should be located on these circles. In fact, the states on a straight line are the eigenstates of the two components perpendicular to that line (the first and second components: $x$ and $y$), and the states on a circle are the eigenstates of the component perpendicular to the plane of the circle (the third component: $z$). The states on a circle are also eigenstates of a radial component $\rho$, which means radial operator: transversal or helical. 

Regarding vector field modes, there are generally two types of photon string states: electric and magnetic strings (see Tab. \ref{eigenstates} in Appendix F). Electric photon strings are more typical fiber bundles, while magnetic photon strings have a rotational-spiral vector field structure. Magnetic photon strings are, in a sense, more elementary, since they can be made open and finite without violating the transversality condition (as is the case for electric strings). Therefore, the state of an infinitely short magnetic string exists, but it loses its string character in favour of a flat photon vortex. 
This type of state, similarly to states on a circle, is lying strictly on the plane but located in the vicinity of a point of this plane in the sense of a relationship inversely proportional to the second (for wave function) or fourth (for module square) power of the distance from this point. Such states are normalised to three-dimensional Dirac deltas.

Helical photon strings have also been defined, i.e. those that are right-handed or left-handed. Helical strings on a straight line, apart from their infinite length, do not pose any significant problems in definition and properties. The situation gets a bit more complicated for (hybrid) helical strings on a circle. Unfortunately, such strings are not eigenstates of the radial operators, neither transversal nor helical -- but they are for the components (electric and magnetic, these helical strings). Therefore, it is consistent with the Galindo--Amrein theorem, which prohibits the existence of helical states fully localised in a finite region. 

In this context, an important result of this work is Lemma  1, which proves that the photon position operator commutes with the helicity operator. This lemma suggests that there are common eigenstates of the position operator and the helicity operator. This has been partially confirmed -- only for helical straight line strings -- and for helical strings on a circle, it would contradict the Galindo-Amrein theorem. Therefore, helical strings on a circle are called hybrid strings, and their radial component $R=\rho$ does not have the status of a strict radial eigenvalue.
Nevertheless, the problem still seems to be incompletely investigated in view of the existence of the new helical positional representation.

Indeed, Theorem 1 and Theorem 2 trivialise the form of the photon position operator on matrix elements (in scalar product) in both the nominal and the helical representations. However, the existence of a positional helical representation that is invariant with respect to the Cartesian photon position operator seems surprising.

\subsection{10.2. A concise summary of the work}

This work introduced the photon position operator using many independent and mutually consistent methods. The photon position operator reflects Maxwell's equations, special relativity, quantum mechanics, and quantum field theory to the extent permitted by the scope of the subject matter. 

However, the photon position operator does not satisfy the sometimes \textit{a priori} assumed \textit{ad hoc} postulate of commutating components. Works respecting this \textit{ad hoc} postulate lead to violations of invariance under three-dimensional rotations. The thesis denying the existence of the photon position operator is undermined by the simultaneous eigenstate of two non-commuting position components.

The such eigenstates of the photon position operator are one-dimensional curves in three-dimensional space, here called photon strings.
There are essentially two main types of photon strings: electric and magnetic. Line integral formulas for electric and magnetic strings provide an alternative to the formalisms of position measures and positional projective operators used for commuting position operators.

\section{Acknowledgments}

I would like to thank Arkadiusz Jadczyk -- co-author of an important source publication \cite{Jadczyk} and author of a newer, competing publication \cite{Jadczyk 2024} -- for his active help. 
Arkadiusz Jadczyk was also the second author of the conference poster \cite{Poster}, which is an introduction to this work.
Extensive consultations concerned both the definitions of the photon position operator (the use of zero mode and definitions equivalence) and the eigenstates (on a straight line and a circle). Moreover, the consultations resulted in Lemma 1 on the commutation of the photon position operator with the helicity operator, as well as the Galindo-Amrein theorem.

I would also like to thank Piotr Stachura for mathematical consultations on the distributional character of the first and second derivatives of the $\rho^{-1}$ function. These non-trivial distributions have not yet been published, but a preliminary paper on the integral representation of the Dirac delta has been written together \cite{Dirac}.

\section{Declarations}
The author declares no conflict of interest. The author declares 100\% contribution to the publication. The author declares all standard consents associated with this scientific publication. The author declares that no specific research data was produced for this work. All necessary data can be found within the text of the work.

\section{Appendix A. Integral operators}

The various negative powers of the minus Laplacian are the following integral operators:
\begin{equation}
    (-\Delta)^{-1}f(\mathbf{r})=\frac{1}{4\pi}\int{ \frac{f(\mathbf{r}')}{|\mathbf{r}'-\mathbf{r}|}d^3\mathrm{r}'},
\end{equation}
\begin{equation}
    (-\Delta)^{-3/4}f(\mathbf{r})=\int{ \frac{f(\mathbf{r}')}{(2\pi\ |\mathbf{r}'-\mathbf{r}|\ )^{3/2}}d^3\mathrm{r}'},
\end{equation}
\begin{equation}
\label{lap-1/2}
    (-\Delta)^{-1/2}f(\mathbf{r})=\frac{1}{2\pi^2}\int{ \frac{f(\mathbf{r}')}{|\mathbf{r}'-\mathbf{r}|^2}d^3\mathrm{r}'},
\end{equation}
\begin{equation}
    (-\Delta)^{-1/4}f(\mathbf{r})=\pi\int{ \frac{f(\mathbf{r}')}{(2\pi\ |\mathbf{r}'-\mathbf{r}|\ )^{5/2}}d^3\mathrm{r}'}.
\end{equation}

It is similar in two dimensions:
\begin{equation}
\label{lap 2D}
    (-\Delta_2)^{-1/2}f(\mathbf{r})=\frac{1}{2\pi}\int{ \frac{f(\mathbf{r}')}{|\mathbf{r}'-\mathbf{r}|}d^2\mathrm{r}'}.
\end{equation}
Ordinary definite integrals:
\begin{equation}
\label{Int1}
\int_0^{\pi}\frac{d\phi}{a-b\cos{\phi}}=\frac{\pi}{\sqrt{a^2-b^2}},\ \ \
a>b>0,
\end{equation}
\begin{equation}
\label{Int2}
\int_0^{\pi}\frac{b \cos{\phi}\ d\phi}{a-b\cos{\phi}}=\frac{\pi a}{\sqrt{a^2-b^2}}-\pi,\ \ \
a>b>0.
\end{equation}

\section{Appendix B. The photon position operator in the Cartesian position representation}
Minimal form on physical (transversal) states:
\begin{equation}
\hat{\mathrm{Q}}_k|_{lm}=
\mathrm{r}_k\delta_{lm}
-\Delta^{-1}\delta_{km}\partial_l,
\end{equation}
\begin{equation}
\hat{\mathrm{Q}}_k|\mathbf{\Psi}=
\mathrm{r}_k\mathbf{\Psi}
-\Delta^{-1}\nabla \Psi_k.
\end{equation}

General form on an extended space:
\begin{equation}
\hat{\mathrm{Q}}_{k,lm}\equiv 
\mathrm{r}_k\delta_{lm}
-\Delta^{-1}\delta_{km}\partial_l
+\Delta^{-1}\delta_{kl}\partial_m,
\end{equation}
\begin{equation}
\hat{\mathbf{Q}}\equiv \mathbf{r}+
\frac{\hat{\mathbf{p}}\times\hat{\mathbf{S}}}{\hat{\mathrm{p}}^2}.
\end{equation}

Identity transversal and hermitian form:
\begin{equation}
\hat{\mathrm{R}}_{k,lm}\equiv 
\mathrm{r}_k\delta_{lm}
-\Delta^{-1}\delta_{km}\partial_l
+\mathrm{r}_k\frac{\partial_l\partial_m}{-\Delta}
+\frac{\partial_k\partial_l\partial_m}{\Delta^2}.
\end{equation}

The matrix form of the basic versions of the photon position operator can be found in the poster \cite{Poster}.

\section{Appendix C. Definitions of the Cartesian photon position operator}

Below are 9 specific ways to define the Cartesian photon position operator. There were 3 primary definition methods based on the boost generator, the transversality condition, and the helicity operator. However, the approach to the unphysical longitudinal mode (mode zero) or the way of notation generated more formal definitions. Their names and designations are given in Tab. \ref{Cartesian definitions}. All 9 operators are physically equivalent on the space of physical states, and many of them are identity equivalent on the extended space.

\begin{table}[h!]
%\label{Cartesian definitions}
\centering
\caption{List of physically equivalent definitions of the Cartesian photon position operator. The content of a longitudinal mode (zero mode) refers to the extended codomain, assuming the domain is extended by such a mode. This does not apply to the "lack" case, in which the domain is not extended (and consequently the codomain is not either)}
%\begin{ruledtabular}
\begin{tabular}{l c c}
\hline
\hline
% The codes above determine the horizontal alignment in each column.
% Options are l (left), r (right), c (centered), and p (paragraph).
% The p option allows an entry to be broken into multiple lines, and
% therefore requires a width specification, in this case 5 centimeters.
Definition name & Designation & Zero mode \\
\hline	% horizontal line to separate headings from data
Born--Infeld (mass-energy center) & $\hat{\mathbf{q}}$ &  contains \\
Pryce--Bacry (by spin) & $\hat{\mathbf{x}}$ &  contains \\
Transversal (not mixing) & $\hat{\mathbf{Q}}$ &  added \\
Even (in frequency) & $\hat{\mathbf{X}}$ &  added\\
Even (in helicity) & $\hat{\mathbf{Y}}$ &  added\\
Even pure (in helicity) & $\hat{\mathbf{Z}}$ &  not added\\
Transversal pure& $\hat{\mathbf{R}}$ &  not added \\
Helical & $\hat{\mathbf{\Theta}}$ &  not added \\
Minimal form & $\hat{\mathbf{Q}}|$ &  lack \\
\hline
\hline
%$^*$ Original authorial velocities
\end{tabular}
%\end{ruledtabular}
\label{Cartesian definitions}
\end{table}
Born--Infeld definition (by boost generator):
\begin{equation}
\hat{\mathbf{q}}:\equiv\frac{1}{2c}(\hat{\mathrm{p}}^{-1}\ \hat{\mathbf{N}}+\hat{\mathbf{N}}\ \hat{\mathrm{p}}^{-1})
\equiv
\frac{1}{\sqrt{\hat{\mathrm{p}}c}}\hat{\mathbf{N}}
\frac{1}{\sqrt{\hat{\mathrm{p}}c}}.
\end{equation}
Pryce-Bacry definition (by spin operator):
\begin{equation}
\hat{\mathbf{x}}:\equiv \mathbf{r}+
\frac{\hat{\mathbf{p}}\times\hat{\mathbf{S}}}{\hat{\mathrm{p}}^2}.
\end{equation}
Transversal definition (with zero mode):
\begin{equation}
\hat{\mathbf{Q}}:\equiv\hat{P}\ \hat{\mathbf{r}}\ \hat{P}+\hat{P}_0\ \hat{\mathbf{r}}\ \hat{P}_0.
\end{equation}
Even (in frequency) definition (with zero mode):
\begin{equation}
\hat{\mathbf{X}}:\equiv\hat{P}_+\ \hat{\mathbf{r}}\ \hat{P}_++\hat{P}_-\ \hat{\mathbf{r}}\ \hat{P}_-+\hat{P}_0\ \hat{\mathbf{r}}\ \hat{P}_0.
\end{equation}
Even (in helicity) definition (with zero mode):
\begin{equation}
\hat{\mathbf{Y}}:\equiv\hat{P}_R\ \hat{\mathbf{r}}\ \hat{P}_R+\hat{P}_L\ \hat{\mathbf{r}}\ \hat{P}_L+\hat{P}_0\ \hat{\mathbf{r}}\ \hat{P}_0.
\end{equation}
Even (in helicity) pure definition (without zero mode):
\begin{equation}
\hat{\mathbf{Z}}:\equiv\hat{P}_R\ \hat{\mathbf{r}}\ \hat{P}_R+\hat{P}_L\ \hat{\mathbf{r}}\ \hat{P}_L.
\end{equation}
Transversal pure definition (without zero mode):
\begin{equation}
\hat{\mathbf{R}}:\equiv\hat{P}\ \hat{\mathbf{r}}\ \hat{P}.
\end{equation}
Helical definition:
\begin{equation}
\hat{\mathbf{\Theta}}:\equiv\hat{\lambda}\ \hat{\mathbf{r}}\ \hat{\lambda}.
\end{equation}
Minimal form (without divergence derivatives $\partial_m$):
\begin{equation}
\hat{\mathrm{Q}}_k|_{lm}:\equiv\hat{\mathrm{Q}}_{k,lm}(\xcancel{\partial_m})
:\equiv\hat{\mathrm{R}}_{k,lm}(\xcancel{\partial_m}).
\end{equation}

Definitions equivalence (in the sense of Cartesian components):
\begin{equation}
\hat{\mathbf{q}}\equiv\hat{\mathbf{x}}\equiv\hat{\mathbf{Q}}\equiv\hat{\mathbf{X}}\equiv\hat{\mathbf{Y}},
\end{equation}
\begin{equation}
\hat{\mathbf{R}}\equiv\hat{\mathbf{Z}}\equiv \hat{\mathbf{\Theta}},
\end{equation}
\begin{equation}
\hat{\mathbf{Q}}=\hat{\mathbf{Q}}|=\hat{\mathbf{q}}=\hat{\mathbf{x}}=\hat{\mathbf{R}}=\hat{\mathbf{X}}=\hat{\mathbf{Y}}=\hat{\mathbf{Z}}=\hat{\mathbf{\Theta}}\ \ \ \ \text{on} \ \ \ \ \mathcal{H}_{phys}.
\end{equation}

To better illustrate how identity equality differs from ordinary equality and what the inclusion of a longitudinal mode involves, a Tab. \ref{transformations} of transformations of transverse and longitudinal modes for operators that are not identity equivalent was prepared. The non-equivalent parts of the operators additionally concern the non-physical states under consideration.

\begin{table}[h!]
\centering
\caption{A summary of the transformations of transverse ($\perp$) and longitudinal ($\parallel$) modes by the position operator in three forms that are not identity equivalent, but are physically equivalent (and equal)}
%\begin{ruledtabular}
\begin{tabular}{c c c c}
\hline
\hline
% The codes above determine the horizontal alignment in each column.
% Options are l (left), r (right), c (centered), and p (paragraph).
% The p option allows an entry to be broken into multiple lines, and
% therefore requires a width specification, in this case 5 centimeters.
Operator & State & & Image \\
\hline	% horizontal line to separate headings from data
$\hat{\mathbf{Q}}$ & $\perp$ & $\longrightarrow$ &  $\perp$\\
$\hat{\mathbf{R}}$ & $\perp$ & $\longrightarrow$ &  $\perp$\\
$\hat{\mathbf{Q}}|$ & $\perp$ & $\longrightarrow$ &  $\perp$\\

$\hat{\mathbf{Q}}$ & $\parallel$ & $\longrightarrow$ &  $\parallel$\\
$\hat{\mathbf{R}}$ & $\parallel$ & $\longrightarrow$ &  $0$\\
$\hat{\mathbf{Q}}|$ & $\parallel \ \notin D$ & $\overset{?}{\longrightarrow}$ &  $\parallel \ , \perp$\\

\hline
\hline
%$^*$ Original authorial velocities
\end{tabular}
%\end{ruledtabular}
\label{transformations}
\end{table}

\section{Appendix D. Radial position operators}

The idea is to give the quantum equivalent of the radial polar variable $\rho=\sqrt{x^2+y^2}$ for the photon. This problem is not fundamentally unambiguous, because $\rho$ is not a Cartesian coordinate. Therefore, the definitions of the 6 radial operators listed in the Tab. \ref{radial operators} are given below.

\begin{table}[h!]
\centering
\caption{List of non-equivalent radial photon position operators}
%\begin{ruledtabular}
\begin{tabular}{l c c}
\hline
\hline
% The codes above determine the horizontal alignment in each column.
% Options are l (left), r (right), c (centered), and p (paragraph).
% The p option allows an entry to be broken into multiple lines, and
% therefore requires a width specification, in this case 5 centimeters.
Operator name & Designation & Zero mode \\
\hline	% horizontal line to separate headings from data
Transversal radial& $\hat{\mathrm{Q}}_{\rho}$ &  added \\
Even radial& $\hat{\mathrm{Y}}_{\rho}$ &  added \\
Helical radial& $\hat{\mathrm{\Theta}}_{\rho}$ &  not added \\
Boost radial& $\hat{\mathrm{q}}_{\rho}$ &  contains\\
Rooter radial& $\hat{\mathrm{Q}}_{[\rho]}$ &  contains\\
Square radial & $\hat{\mathrm{Q}}_{\rho^2}^{2}$ &  not added\\
\hline
\hline
%$^*$ Original authorial velocities
\end{tabular}
%\end{ruledtabular}
\label{radial operators}
\end{table}

Transversal radial operator (with zero mode):
\begin{equation}
    \hat{\mathrm{Q}}_{\rho}:\equiv
    \hat{P}\rho\hat{P}+\hat{P}_0\rho\hat{P}_0.
\end{equation}
Even radial operator (with zero mode):
\begin{equation}
    \hat{\mathrm{Y}}_{\rho}:\equiv
    \hat{P}_R\rho\hat{P}_R+\hat{P}_L\rho\hat{P}_L+\hat{P}_0\rho\hat{P}_0.
\end{equation}
Helical radial operator:
\begin{equation}
    \hat{\mathrm{\Theta}}_{\rho}:\equiv
    \hat{\lambda}\rho\hat{\lambda}.
\end{equation}
Boost radial position operator:
\begin{equation}
\hat{\mathrm{q}}_{\rho}:\equiv
\frac{1}{\sqrt{\hat{\mathrm{p}}c}}\hat{\mathrm{N}}_{[\rho]}
\frac{1}{\sqrt{\hat{\mathrm{p}}c}}.
\end{equation}
Rooter radial operator (square root):
\begin{equation}
    \hat{\mathrm{Q}}_{[\rho]}:\equiv
    \sqrt{\hat{\mathrm{Q}}_1^2+\hat{\mathrm{Q}}_2^2}\equiv:
     \hat{\mathrm{Y}}_{[\rho]}\equiv:
      \hat{\mathrm{q}}_{[\rho]}.
\end{equation}
Square radial operator:
\begin{equation}
    \hat{\mathrm{Q}}^2_{\rho^2}:\equiv
    \hat{P}\rho^2\hat{P}.
\end{equation}

The action of radial operators is too complex mathematically to give a direct representation (explicit mathematical form -- except for the square radial operator).

\section{Appendix E. Commutators}
Commutators equal to zero:
\begin{equation}
[\hat{\mathbf{Q}},\hat{\Lambda}]\equiv
\hbar[\hat{\mathbf{Q}},\hat{\lambda}]\equiv 0,
\end{equation}
\begin{equation}
[\hat{\mathbf{Q}},\hat{P}]\equiv
[\hat{\mathbf{Q}},\hat{P}_L]\equiv
[\hat{\mathbf{Q}},\hat{P}_R]\equiv 
[\hat{\mathbf{Q}},\hat{P}_0]\equiv 0,
\end{equation}
\begin{equation}
[\hat{\mathrm{Y}}_{\rho},\hat{\lambda}]\equiv
[\hat{\mathrm{q}}_{\rho},\hat{\lambda}]\equiv 
[\hat{\mathrm{Q}}_{[\rho]},\hat{\lambda}]\equiv 0,
\end{equation}
\begin{equation}
[\hat{\mathrm{Q}}_{\rho},\hat{P}]\equiv
[\hat{\mathrm{Y}}_{\rho},\hat{P}]\equiv
[\hat{\mathrm{\Theta}}_{\rho},\hat{P}]\equiv
[\hat{\mathrm{q}}_{\rho},\hat{P}]\equiv 
[\hat{\mathrm{Q}}_{[\rho]},\hat{P}]\equiv 0,
\end{equation}
\begin{equation}
[\hat{\mathrm{Q}}_{\rho},\hat{P}_0]\equiv
[\hat{\mathrm{Y}}_{\rho},\hat{P}_0]\equiv
[\hat{\mathrm{\Theta}}_{\rho},\hat{P}_0]\equiv
[\hat{\mathrm{q}}_{\rho},\hat{P}_0]\equiv 
[\hat{\mathrm{Q}}_{[\rho]},\hat{P}_0]\equiv 0.
\end{equation}
Position operator commutators with momentum and energy:
\begin{equation}
[\hat{\mathrm{Q}}_k,\hat{\mathrm{p}}_l]\equiv i\hbar\ \delta_{kl},
\end{equation}
\begin{equation}
[\hat{\mathbf{Q}},\hat{\mathrm{p}}]\equiv i\hbar\frac{\hat{\mathbf{p}}}{\hat{\mathrm{p}}}.
\end{equation}
Commutator of components of the photon position operator \cite{Koczan 2002}:
\begin{equation}
[\hat{\mathrm{Q}}_k,\hat{\mathrm{Q}}_l]_{mn}= \Delta^{-1}(\hat{\delta}_{mk}^{\perp}\delta_{ln}-\hat{\delta}_{ml}^{\perp}\delta_{kn}),
\end{equation}
where: $\hat{\delta}^{\perp}=\hat{P}$.

\section{Appendix F. Eigenstates of the photon position operator}

The most characteristic eigenstates of the photon position operator are the string states, which are eigenstates of two orthogonal position components. The eigenstates of three orthogonal position components do not exist because the position operator has non-commuting components. The simplest to describe are straight line strings and circular strings. Since a photon has two polarisation states, strings also have two main modes of vector function structures. Fibrous strings are called electric strings, while spiral strings are called magnetic strings. The (hybrid) combination of an electric and magnetic string produces a helical string. Therefore, due to the shape and type of structure, we have 6 types of strings, which with an additional flat vortex state are listed in Tab. \ref{eigenstates}.

First, general definitions of the photon states are given: electric string and magnetic string in the positional representation. Then all considered string states of the photon are listed, along with their basic properties. Tab. \ref{eigenvalues} summarises the eigenvalues of the individual eigenstates of the position operator.

Electric photon strings (infinite, closed):
\begin{equation}
    \ket{\Gamma}:=\mathbf{\Psi}_{\Gamma}(\mathbf{r}):=\int_{-\infty}^{+\infty}{\delta^{(3)}\big(\mathbf{r}-\mathbf{r}'(s)\big)d\mathbf{r}'(s)},
\end{equation}
\begin{equation}
\label{Omega}
    \ket{\Omega}:=\mathbf{\Psi}_{\Omega}(\mathbf{r}):=\ointctrclockwise_{\Omega}{\delta^{(3)}\big(\mathbf{r}-\mathbf{r}'(s)\big)d\mathbf{r}'(s)}.
\end{equation}
In both integrals, the primes denote the parametric equation of the line ($s \rightarrow \mathbf{r}'(s) \rightarrow \mathbf{r}$), not the derivatives.

Magnetic photon strings (infinite, closed):
\begin{equation}
  \ket{\Gamma^{\star}}:=\frac{1}{\sqrt{-\Delta}} \int_{\Gamma}{\nabla \delta^{(3)}\big(\mathbf{r}-\mathbf{r}'(s)\big)\times d\mathbf{r}'(s)},
\end{equation}
\begin{equation}
  \ket{\Gamma^{\star}}= \hat{\lambda}\ket{\Gamma},
\end{equation}
\begin{equation}
  \ket{\Omega^{\star}}:=\frac{1}{\sqrt{-\Delta}} \int_{\Omega}{\nabla \delta^{(3)}\big(\mathbf{r}-\mathbf{r}'(s)\big)\times d\mathbf{r}'(s)},
\end{equation}
\begin{equation}
  \ket{\Omega^{\star}}= \hat{\lambda}\ket{\Omega}.
\end{equation}
\begin{table}[h!]
\centering
\caption{List of specific generalised eigenstates (strings) of the photon position operator}
%\begin{ruledtabular}
\begin{tabular}{l c c}
\hline
\hline
% The codes above determine the horizontal alignment in each column.
% Options are l (left), r (right), c (centered), and p (paragraph).
% The p option allows an entry to be broken into multiple lines, and
% therefore requires a width specification, in this case 5 centimeters.
Eigenstate name & Designation & Character \\
\hline	% horizontal line to separate headings from data
Electric linear string & $\ket{x_0, y_0}$ &  infinit-open \\
Magnetic linear string & $\ket{x^{\star}_0,y^{\star}_0}$ &  infinit-open \\
Helical linear string & $\ket{x_0, y_0,\pm}$ &  infinit-open\\
Electric circular string & $\ket{\mathring{R}, z_0}$ &   close \\
Magnetic circular string & $\ket{\mathring{R}, z_0^{\star}}$ &  close \\
Helical circular string & $\ket{\mathring{\tilde{R}}, z_0, \pm}$ &  close \\
Flat vortex (short magnetic string) & $\ket{\tilde{x}_0, \tilde{y}_0, z_0}$ &  open \\
\hline
\hline
%$^*$ Original authorial velocities
\end{tabular}
%\end{ruledtabular}
\label{eigenstates}
\end{table}

\begin{table}[h!]
\centering
\caption{Eigenvalues of the components of the photon position operator for the eigenstates}
%\begin{ruledtabular}
\begin{tabular}{l c c c c c c}
\hline
\hline
% The codes above determine the horizontal alignment in each column.
% Options are l (left), r (right), c (centered), and p (paragraph).
% The p option allows an entry to be broken into multiple lines, and
% therefore requires a width specification, in this case 5 centimeters.
Eigenstate of \ \ \ & \ \ \  $\hat{\mathrm{Q}}_1$ \ \ \ & \ \ \  $\hat{\mathrm{Q}}_2$ \ \ \  & \ \ \  $\hat{\mathrm{Q}}_3$ \ \ \  & \ \ \  $\hat{\mathrm{Q}}_{\rho}$ \ \ \  & \ \ \  $\hat{\mathrm{\Theta}}_{\rho}$ \ \ \  & \ \ \  $\hat{\lambda}$\ \ \   \\
\hline	% horizontal line to separate headings from data
 $\ket{x_0, y_0}$ &  $x_0$ & $y_0$ &  $-$  &  $\rho_0$  &  $-$  &  $-$ \\
 $\ket{x^{\star}_0,y^{\star}_0}$ &  $x_0$ & $y_0$ &  $-$  &  $-$  &  $\rho_0$  &  $-$ \\
 $\ket{x_0, y_0,\pm}$ &  $x_0$ & $y_0$ &  $-$  &  $-$  &  $-$  &  $\pm1$ \\
 $\ket{\mathring{R}, z_0}$ &  $-$ & $-$ &  $z_0$  &  $R$  &  $-$  &  $-$ \\
 $\ket{\mathring{R}, z_0^{\star}}$ &  $-$ & $-$ &  $z_0$  &  $-$  &  $R$  &  $-$ \\
 $\ket{\mathring{\tilde{R}}, z_0, \pm}$ &  $-$ & $-$ &  $z_0$  &  $-$  &  $-$  &  $\pm1$ \\
$\ket{\tilde{x}_0, \tilde{y}_0, z_0}$ &  $-$ & $-$ &  $z_0$  &  $-$  &  $-$  &  $-$ \\
\hline
\hline
%$^*$ Original authorial velocities
\end{tabular}
%\end{ruledtabular}
\label{eigenvalues}
\end{table}

Photon electric (fibrous) string on a straight line:
\begin{equation}
    \ket{x_0,y_0}(\mathbf{r})=
    \left(\begin{array}{c}
0\\
0\\
\delta(x-x_0)\delta(y-y_0)
\end{array} \right),
\end{equation}
\begin{equation}
\hat{P}\ket{x_0, y_0}=\ket{x_0, y_0}, \ \ \ \ \hat{P}_0\ket{x_0, y_0}=0,
\end{equation}
\begin{equation}
\hat{\mathrm{Q}}_1\ket{x_0, y_0}=x_0\ket{x_0, y_0}, \ \ \ \ \hat{\mathrm{Q}}_2\ket{x_0, y_0}=y_0\ket{x_0, y_0},
\end{equation}
\begin{equation}
\braket{x_1,y_1|x_2,y_2}=\delta(x_1-x_2)\delta(y_1-y_2)\int{dz}.
\end{equation}

Photon magnetic (spiral) string on a straight line:
\begin{equation}
    \ket{x^{\star}_0,y^{\star}_0}(\mathbf{r})=
    \left(\begin{array}{c}
\partial_y\frac{1}{\sqrt{(x-x_0)^2+(y-y_0)^2}}\\
-\partial_x\frac{1}{\sqrt{(x-x_0)^2+(y-y_0)^2}}\\
0
\end{array} \right),
\end{equation}
\begin{equation}
\hat{P}\ket{x^{\star}_0, y_0^{\star}}=\ket{x^{\star}_0, y^{\star}_0}, \ \ \ \ \hat{P}_0\ket{x^{\star}_0, y^{\star}_0}=0,
\end{equation}
\begin{equation}
\hat{\mathrm{Q}}_1\ket{x^{\star}_0, y^{\star}_0}=x_0\ket{x^{\star}_0, y^{\star}_0}, \ \ \ \ \hat{\mathrm{Q}}_2\ket{x^{\star}_0, y^{\star}_0}=y_0\ket{x^{\star}_0, y^{\star}_0}.
\end{equation}
\begin{equation}
\braket{x^{\star}_1,y^{\star}_1|x^{\star}_2,y^{\star}_2}=\delta(x_1-x_2)\delta(y_1-y_2)\int{dz}.
\end{equation}

Photon helical strings on a straight line:
\begin{equation}
\ket{x_0, y_0,\pm }:=
\frac{1}{\sqrt{2}}\ket{x_0, y_0}\pm \frac{1}{\sqrt{2}}\ket{x_0^{\star}, y_0^{\star}},
\end{equation}
\begin{equation}
\hat{P}\ket{x_0, y_0,\pm}=\ket{x_0, y_0,\pm}, \ \ \ \ \hat{P}_0\ket{x_0, y_0,\pm}=0,
\end{equation}
\begin{equation}
\hat{\mathrm{Q}}_1\ket{x_0, y_0}=x_0\ket{x_0, y_0}, \ \ \ \hat{\mathrm{Q}}_2\ket{x_0, y_0,\pm}=y_0\ket{x_0, y_0,\pm},
\end{equation}
\begin{equation}
\hat{\lambda}\ket{x_0, y_0,\pm}=\pm\ket{x_0, y_0,\pm},
\end{equation}
\begin{equation}
\braket{x_1,y_1,\pm|x_2,y_2,\pm}=\delta(x_1-x_2)\delta(y_1-y_2)\int{dz},
\end{equation}
\begin{equation}
\braket{x_1,y_1,\pm|x_2,y_2,\mp}=0.
\end{equation}
Photon electric (fibrous) string on a circle:
\begin{equation}
 \ket{\mathring{R},z_0}(\mathbf{r})=\frac{1}{R}
\left(\begin{array}{c}
-y\ \delta(\sqrt{x^2+y^2}-R)\ \delta(z-z_0)\\
x\ \delta(\sqrt{x^2+y^2}-R)\ \delta(z-z_0)\\
0
 \end{array} \right),
\end{equation}
\begin{equation}
\hat{P}\ket{\mathring{R}, z_0}=\ket{\mathring{R}, z_0}, \ \ \ \ \hat{P}_0\ket{\mathring{R}, z_0}=0,
\end{equation}
\begin{equation}
\hat{\mathrm{Q}}_{\rho}\ket{\mathring{R},z_0}=R\ket{\mathring{R},z_0}, \ \ \ \ \hat{\mathrm{Q}}_3\ket{\mathring{R},z_0}=z_0\ket{\mathring{R},z_0},
\end{equation}
\begin{equation}
\braket{\mathring{R}_1,z_1|\mathring{R}_2,z_2}=2\pi R_1\ \delta(R_1-R_2)\ \delta(z_1-z_2).
\end{equation}
Photon magnetic (spiral) string on a circle:
\begin{equation}
 \ket{\mathring{R},z_0^{\star}}(\mathbf{r})=\frac{2R^2}{\pi}
\left(\begin{array}{c}
\frac{2x(z-z_0)}{\sqrt{(R^2+x^2+y^2+(z-z_0)^2)^2-4R^2(x^2+y^2)}^3}\\
\frac{2y(z-z_0)}{\sqrt{(R^2+x^2+y^2+(z-z_0)^2)^2-4R^2(x^2+y^2)}^3}\\
\frac{R^2-x^2-y^2+(z-z_0)^2}{\sqrt{(R^2+x^2+y^2+(z-z_0)^2)^2-4R^2(x^2+y^2)}^3}
 \end{array} \right),
\end{equation}
\begin{equation}
\hat{P}\ket{\mathring{R}, z_0^{\star}}=\ket{\mathring{R}, z_0^{\star}}, \ \ \ \ \hat{P}_0\ket{\mathring{R}, z_0^{\star}}=0,
\end{equation}
\begin{equation}
\hat{\mathrm{\Theta}}_{\rho}\ket{\mathring{R},z_0^{\star}}=R\ket{\mathring{R},z_0^{\star}}, \ \ \ \ \hat{\mathrm{Q}}_3\ket{\mathring{R},z_0^{\star}}=z_0\ket{\mathring{R},z_0^{\star}},
\end{equation}
\begin{equation}
\braket{\mathring{R}_1,z_1^{\star}|\mathring{R}_2,z_2^{\star}}=2\pi R_1\ \delta(R_1-R_2)\ \delta(z_1-z_2).
\end{equation}
Photon helical strings on a circle:
\begin{equation}
\ket{\mathring{\tilde{R}},z_0,\pm }:=
\frac{1}{\sqrt{2}}\ket{\mathring{R},z_0}\pm \frac{1}{\sqrt{2}}\ket{\mathring{R},z_0^{\star}}, 
\end{equation}
\begin{equation}
\hat{P}\ket{\mathring{\tilde{R}},z_0,\pm}=\ket{\mathring{\tilde{R}},z_0,\pm}, \ \ \ \ \hat{P}_0\ket{\mathring{\tilde{R}},z_0,\pm}=0,
\end{equation}
\begin{equation}
\hat{\mathrm{Q}}_3\ket{\mathring{\tilde{R}},z_0,\pm}=z_0\ket{\mathring{\tilde{R}},z_0,\pm},
\end{equation}
\begin{equation}
\hat{\lambda}\ket{\mathring{\tilde{R}},z_0,\pm}=\pm\ket{\mathring{\tilde{R}},z_0,\pm},
\end{equation}
\begin{equation}
\braket{\mathring{\tilde{R}}_1,z_1,\pm|\mathring{\tilde{R}}_2,z_2,\pm}=2\pi R_1\ \delta(R_1-R_2)\ \delta(z_1-z_2),
\end{equation}
\begin{equation}
\braket{\mathring{\tilde{R}}_1,z_1,\pm|\mathring{\tilde{R}}_2,z_2,\mp}=0.
\end{equation}
Flat photon vortex (infinitely short open magnetic string):
\begin{equation}
    \ket{\tilde{x}_0,\tilde{y}_0,z_0}=\frac{1}{2\pi}
    \left(\begin{array}{c}
\partial_y \Big(\frac{1}{\sqrt{(x-x_0)^2+(y-y_0)^2}}\Big)\delta(z-z_0)\\
-\partial_x \Big(\frac{1}{\sqrt{(x-x_0)^2+(y-y_0)^2}}\Big)\delta(z-z_0)\\
0
\end{array} \right),
\end{equation}
\begin{equation}
\hat{P}\ket{\tilde{x}_0,\tilde{y}_0,z_0}=\ket{\tilde{x}_0,\tilde{y}_0,z_0},\ \ \ \ \hat{P}_0\ket{\tilde{x}_0,\tilde{y}_0,z_0}=0,
\end{equation}
\begin{equation}
\hat{\mathrm{Q}}_3\ket{\tilde{x}_0,\tilde{y}_0,z_0}=z_0\ket{\tilde{x}_0,\tilde{y}_0,z_0},
\end{equation}
\begin{equation}
\braket{\tilde{x}_1,\tilde{y}_1,z_1|\tilde{x}_2,\tilde{y}_2,z_2}=
    \delta(\tilde{x}_1-\tilde{x}_2) \delta(\tilde{y}_1-\tilde{y}_2)\delta(z_1-z_2).
\end{equation}

\bibliographystyle{unsrt}

\end{document}